\documentclass[12pt]{spieman}  % use this instead for A4 paper
\usepackage{amsmath,amsfonts,amssymb,mathrsfs}
\usepackage{graphicx}
\usepackage{setspace}
\usepackage{tocloft}
\usepackage{fancyvrb}
\usepackage{booktabs}
\VerbatimFootnotes
\usepackage{tikz}
\usepackage{xcolor}
\usepackage[capitalise]{cleveref}

\newcommand{\revision}[1]{{#1}} %unmarked version

\newcommand{\R}{\ensuremath{\mathbb{R}}}

\newcommand{\RecSpace}{\ensuremath{X_f}}
\newcommand{\DataSpace}{\ensuremath{Y}}

\newcommand{\noise}{\varepsilon}

\newcommand{\approxInv}[1]{#1^{\dagger}}
\DeclareMathOperator{\ForwardOp}{\ensuremath{\mathcal{A}}}
\DeclareMathOperator{\ForwardOpPseudoInv}{\ensuremath{\approxInv{\ForwardOp}}}
\DeclareMathOperator{\ForwardOpInvLearned}{\ensuremath{\approxInv{\mathcal{A}}_{\param}}}

\DeclareMathOperator{\ForwardOpAdj}{\ensuremath{\ForwardOp^*}}

\DeclareMathOperator*{\argmin}{arg\,min}

\DeclareMathOperator{\Loss}{\ensuremath{L}}

\DeclareMathOperator{\Network}{\ensuremath{\Lambda}}
\newcommand{\param}{\theta}

\newcommand{\signal}{\ensuremath{f}}

\newcommand{\data}{\ensuremath{g}}
\title{Deep Learning in Photoacoustic Tomography: Current approaches and future directions}

\author[a,b]{Andreas Hauptmann}
\author[c]{Ben Cox}
\affil[a]{University of Oulu, Research Unit of Mathematical Sciences, Oulu, Finland}
\affil[b]{University College London, Department of Computer Science, London, United Kingdom}
\affil[c]{University College London, Department of Medical Physics and Biomedical Engineering, London, United Kingdom}

\cftpagenumbersoff{figure}
\cftpagenumbersoff{table} 
\begin{document} 
\maketitle

\begin{abstract}
Biomedical photoacoustic tomography, which can provide high resolution 3D soft tissue images based on the optical absorption, has advanced to the stage at which translation from the laboratory to clinical settings is becoming possible. The need for rapid image formation and the practical restrictions on data acquisition that arise from the constraints of a clinical workflow are presenting new image reconstruction challenges.
There are many classical approaches to image reconstruction, but ameliorating the effects of incomplete or imperfect data through the incorporation of accurate priors is challenging and leads to slow algorithms.
Recently, the application of Deep Learning, or deep neural networks, to this problem has received a great deal of attention. This paper reviews the literature on 
\emph{learned image reconstruction}, summarising the current trends, and explains how these new approaches fit within, and to some extent have arisen from, a framework that encompasses classical reconstruction methods. 
In particular, it shows how these new techniques can be understood from a Bayesian perspective, providing useful insights. 
The paper also provides a concise tutorial demonstration of three prototypical approaches to \emph{learned image reconstruction}. The code and data sets for these demonstrations are available to researchers.
It is anticipated that it is in \emph{in vivo} applications - where data may be sparse, fast imaging critical and priors difficult to construct by hand - that Deep Learning will have the most impact.
With this in mind, the paper concludes with some indications of possible future research directions.
\end{abstract}

% Include a list of up to six keywords after the abstract
\keywords{Photoacoustic tomography, learned image reconstruction, deep learning, neural networks, data-driven methods, \emph{in vivo} imaging}

% Include email contact information for corresponding author
{\noindent \footnotesize Corresponding author: \linkable{andreas.hauptmann@oulu.fi} }

%\begin{spacing}{2}   % use double spacing for rest of manuscript

\tableofcontents

\section{Introduction}
\label{sect:intro}

The potential of biomedical photoacoustic tomography (PAT) to obtain high resolution images based on optical absorption and, moreover, provide images that depend quantitatively on endogenous or exogeneous molecular contrast, has resulted in rapidly growing interest in the modality. For example, the ability to obtain accurate, spatially-resolved, estimates of blood oxygenation would have significant impact both clinically and for pre-clinical applications.

There are two aspects to PAT image reconstruction, an acoustic inversion from the measured acoustic time series to the initial acoustic pressure distribution
\cite{Kuchment:2011mtat,Poudel2019survey}, and a spectroscopic optical inversion to recover optical absorption coefficients or quantities derived from them \cite{Cox2012quantitative}. 
The acoustic inverse problem can be solved exactly in closed form in the ideal circumstance that complete data is available and the medium has a uniform and known sound speed. In most practical scenarios, however, there are divergences from this ideal case, e.g.\ heterogeneities in the sound speed or bandlimited detection over an incomplete set of measurement points, making the acoustic inversion challenging. (The use of linear arrays for \emph{in vivo} imaging is a case in point.) When, in addition, a solution is required to the optical inversion, the image reconstruction task becomes more challenging still as the forward operator is nonlinear.
Iterative model-based approaches have been devised that manage this greater complexity by providing a flexible way to frame the problem and incorporate prior knowledge of the kind of solution expected \cite{Huang:2013fwi,Arridge:2016apat,Boink2018framework}.
However, such approaches, while appealing, are typically computationally intensive and time-consuming, which precludes their use in many applications.

In contrast to purely model-based approaches, data-driven techniques, and in particular Deep Learning (DL), are increasingly widely used for tomographic image reconstruction\cite{Kang2017,Jin2017,Hammernik2018learning,Adler2017,Zhu2018automap,arridge2019solving}. These techniques, which primarily originate from computer vision and are known to excel at segmentation and classification tasks, are frequently treated as `black boxes'. This is widely considered undesirable in biomedical imaging and inverse problems and recent work has started to provide insights into why certain network architectures work well for certain tasks\cite{ye2018deep,haber2017stable,ruthotto2019deep},
and also to provide justifications for the use of DL approaches in the solution of inverse problems including image reconstruction. We will refer to the application of DL \revision{within the image reconstruction pipeline} as {\it learned image reconstruction}.

The rising interest in learned image reconstruction has lead to a transition from classical analytical methods to such data-driven approaches. While much of this work has focused on established imaging modalities such as MRI\cite{Hammernik2018learning,Schlemper2017deep,Hauptmann2019MRM} and CT\cite{Kang2017,Jin2017,Adler2018}, this transition is also clearly discernible in the recent literature on PAT image reconstruction. In this paper we will review the recent work done in this area and place these new approaches into a broader context by drawing connections to classical analytical reconstructions. We also provide a tutorial style introduction to the the use of DL in PAT image reconstruction, including describing and demonstrating several different approaches for learned image reconstruction. Code is available for these examples, free to download, allowing researchers to reproduce them, and providing them with a starting point for their own learned  reconstructions.\footnote{Codes and training/test data for all experiments discussed will be made available at: \url{https://github.com/asHauptmann/PAT_CODES} } 

PAT is a particularly suitable area in which to review these methods for several reasons. There is a very active experimental community interested in a wide range of applications, from data-intensive, large-scale, 3D imaging to 2D high-frame rate uses. This results in a wide variety of different approaches for data collection and presents many different challenges in the reconstruction pipeline, including those of limited data, computationally expensive forward operators, uncertainty in model parameters, and the lack of training data; the latter especially a problem for \emph{in vivo} applications. This leads to a final point, which is that the community is not only in a transition from classical to data-driven approaches, but also in a transition from proof-of-concept studies to applying the techniques in challenging clinical and pre-clinical scenarios. Indeed, these two transitions may prove to be symbiotic: data-driven approaches are rarely needed in proof-of-concept studies in which complete data is available and time is not of the essence, but many of the problems facing \emph{in vivo} use are not easily tackled within a classical framework. We hope that by describing a framework for learned reconstructions, and by presenting an overview of the diverse work done, this review can provide guidance for possible future directions for image reconstruction as PAT transitions from the bench to the clinic.

\subsection{Scope of Tutorial Review}
There are multiple ways in which DL could be used within the context of PAT, so to keep this review to a reasonable length it is necessary to limit its scope. This review will concentrate on DL as applied to tomographic reconstruction in photoacoustics. In other words, the focus will be on using DL networks, sometimes in combination with classical approaches, to reconstruct photoacoustic (PA) images from projections (which here are acoustic time series)\revision{, this includes pre- and post-processing approaches with the intend to improve reconstruction quality.} With this as the focus, there are several applications of DL to PA imaging that must regrettably wait for a future review. First, this review will be limited to PA \textit{tomography} and will not cover the use of DL in relation to PA \textit{microscopy} (PAM), the principal difference being, for our purposes, that in PAT it is necessary to reconstruct the image from a set of measured projections but in PAM the image can be measured directly. 
%Second, this review will not cover work where DL approaches have been used with PAT images but the same work could equally have been performed on images generated using another modality. For example, where DL has been used to segment or classify PAT images, or regions of images, into, say, diseased or healthy.
\revision{Second, this review will not cover work where DL approaches have been used subsequent to a final reconstruction. This includes, for example, applications where DL has been used to segment or classify PAT images, or regions of images, into, say, diseased or healthy.}
Third, this review will not cover the use of DL to make diagnostic judgements, eg. to answer questions such as ``Does this image indicate diabetes, rheumatism, cancer, etc?''.

Papers on DL for PAT reconstruction are currently appearing at a steady rate and we anticipate that trend will continue. This review attempts to cover all relevant papers or preprints appearing up to the end of June 2020. 

\section{Forward and Inverse Problems in Photoacoustic Tomography}\label{sec:PAT}

\subsection{PAT Forward Problems}
\subsubsection{Physics of photoacoustic \revision{signal} generation}
The \emph{photoacoustic effect} is the name given to the phenomenon by which the absorption of an optical pulse generates an acoustic pulse. A light pulse incident on soft biological tissue will be scattered around in the tissue, eventually either leaving the tissue or being absorbed by absorbing molecules in the tissue, known as chromophores (hemoglobin being one of the most important). The energy of the excited chromophore is then converted into heat. This all occurs on a timescale ($\sim$ns) which is much shorter than the timescale required for the tissue to move (for the local mass density to change, $\sim\!\!\!\mu$s), so the heating is isochoric and therefore accompanied by an increase in pressure. Tissue is elastic, so the regions of higher pressure will act as sources of acoustic waves. Because of the difference in timescales, the pressure increase is usually treated as occurring instantaneously, and PA wave generation and propagation is modelled as the initial value problem:
\begin{align}
(\partial_{tt} - c^2\Delta) p(x,t) = 0, 
\quad p(x,0) = f(x), 
\quad \partial_t p(x,0) = 0,
\label{eqn:wave_equation}
\end{align}
where $x \in \R^3$ is the spatial variable, $t \in \R^{\ge 0}$ is time, and $p(x,t)$ is the acoustic pressure. The medium properties to which the acoustic wave is sensitive, sound speed and mass density, will in general vary with position. However, for propagation through soft tissue the variations are often small and are rarely known in advance so the medium is usually treated as \revision{acoustically} homogeneous. (Acoustic absorption, not described by Eq.\ \eqref{eqn:wave_equation},
may also become important in some applications.) The initial condition, $f(x)\geq 0 $, is the termed the  \textit{initial acoustic pressure distribution} and is related to the optical properties of the tissue by the equation
\begin{align}
f(x,\lambda) = \Gamma \mu_a(x,\lambda) \phi(x, \lambda),
\label{eqn:initial_pressure}
\end{align}
where $\lambda$ is the optical wavelength, $\mu_a$ is the optical absorption coefficient (dimensions of reciprocal length), $\phi(x)$ is the optical fluence (dimensions of energy per unit area), and $\Gamma$ is a dimensionless constant which accounts for the efficiency of the acoustic generation (sometimes called the Gr\"uneisen parameter, which it equals in some circumstances). The dependence of $\phi$ on wavelength has been made explicit in Eq.\ \eqref{eqn:initial_pressure}, but $\phi$ also depends on the absorption and scattering throughout the tissue, making Eq.\ \eqref{eqn:initial_pressure} nonlinear in the absorption coefficient $\mu_a$. The positivity of the initial pressure distribution, $f(x)$, arises from the fact that $\mu_a\phi$ is the energy density due to absorption of the light and $\Gamma$ is positive for most materials, especially soft tissue.

\subsubsection{Tissue optics}
\label{subsubsec:tissue_optics}
The nature of the dependence of the fluence $\phi$ on the absorption and scattering is usually modelled in biological tissue using transport theory
\cite{Arridge1999, Welch2011book2ndedn}, ie.\ making the assumption that coherent optical effects can be safely ignored. Under this assumption, the light field is described in terms of energy by the radiance, $\psi(x,t,\hat{s}, \lambda)$, which is the rate of energy flow per unit area per unit solid angle in direction $\hat{s} \in S^2$ at position $x$ at time $t$ (units of power per unit area per unit solid angle). When there are no significant inelastic processes such as fluorescence present, the radiance at each wavelength obeys the following integro-differential equation, known as the \textit{radiative transfer equation}, which can be thought of as a statement of the principle of the conservation of energy:
\begin{align}
\frac{1}{v}\frac{\partial\psi}{\partial t} = q - (\hat{s}\cdot\nabla + \mu_a + \mu_s)\psi + \mu_s\int_{S^2}\vartheta(\hat{s},\hat{s}')\psi(\hat{s}') \,d\hat{s}', 
\label{eqn:rte}
\end{align}
where $v$ is the speed of light, $q$ is a source term, $\mu_s$ is the scattering coefficient, and $\vartheta(\hat{s},\hat{s}')$ is the scattering `phase' function, a probability density function describing the likelihood of a photon travelling in direction $\hat{s}'$ being scattered into the direction $\hat{s}$. The fluence, for a given wavelength, can be found by integrating the radiance at that wavelength over all angles and time:
\begin{align}
\phi(x,\lambda) = \int_{S^2}\int_{\R^{\ge 0}} \psi(x,t,\hat{s},\lambda) \,d\hat{s} \,dt.
\label{eqn:fluence}
\end{align}
The quantity of interest in quantitative PAT is sometimes the absorption coefficient, but more often it is a related quantity. For example, the absorption coefficient is related to the concentrations of the chromophores present by
\cite{Bigio2016book}
\begin{align}
\mu_a(x,\lambda) = \sum_q \alpha_q(\lambda)C_q(x),
\label{eqn:molar_abs}
\end{align}
where $C_q$ is the concentration of the $q^{\text{th}}$ chromophore and $\alpha_q(\lambda)$ is its molar absorption coefficient spectrum. A quantity of considerable clinical interest is blood oxygen saturation\cite{Li2018reviewSO2}, which is related to the concentrations of two particular endogenous chromophores, oxy- and deoxy-hemoglobin, $C_{HbO}$ and $C_{Hb}$ respectively, by
\begin{align}
sO_2(x) = \frac{C_{HbO}(x)}{C_{HbO}(x) + C_{Hb}(x)}.
\label{eqn:saturation}
\end{align}

\subsubsection{Photoacoustic measurements}
\label{subsubsec:Photoacoustic_measurements}

\begin{figure}[tb!]
\centering
\includegraphics[width=0.6\textwidth]{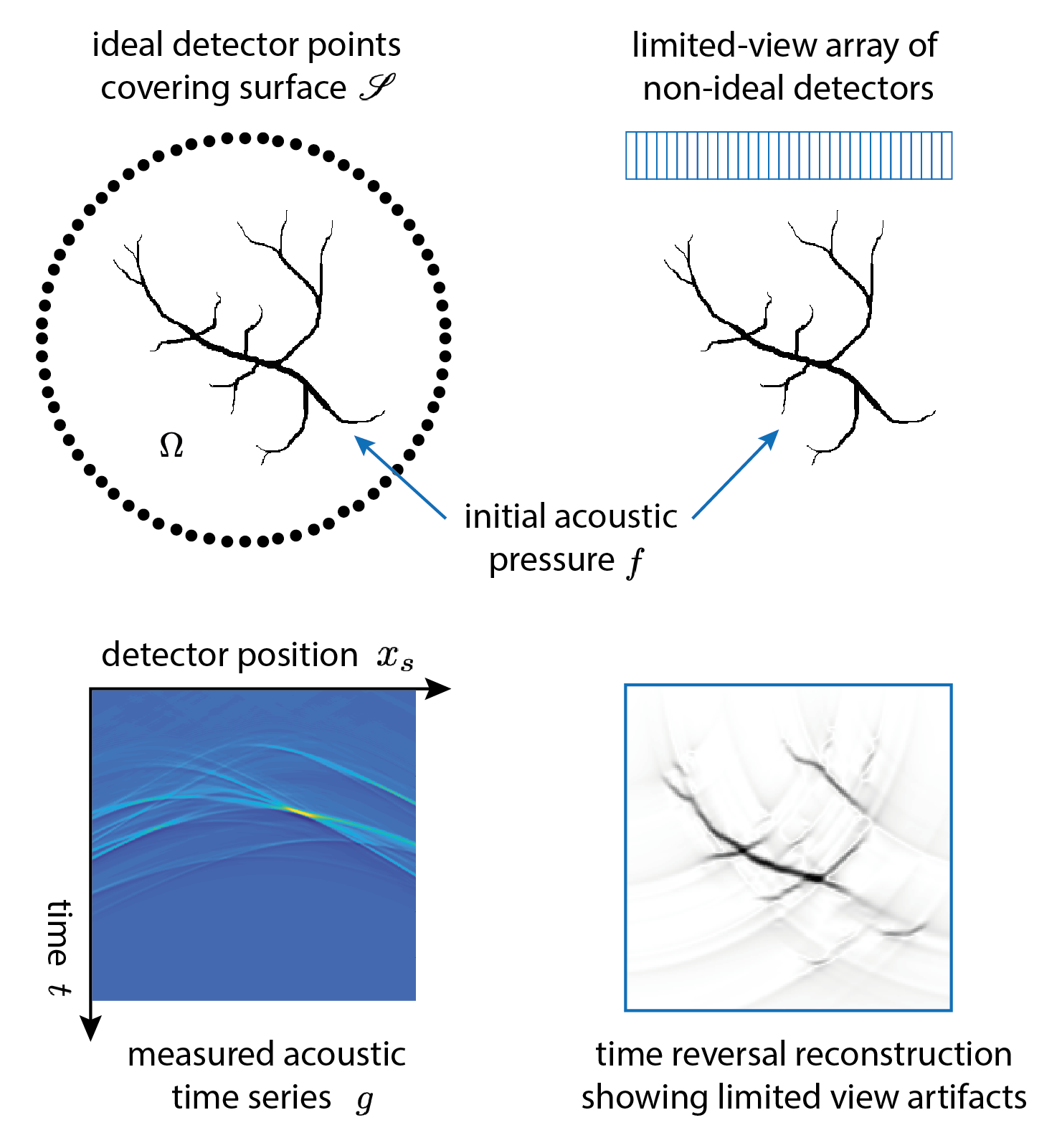}
\caption{\label{fig:measurement_setup} (\emph{Top Left}) Ideal PAT measurement setup with point-like omnidirectional detectors covering the surface $\mathscr{S}$ surrounding the initial pressure distribution $f$. (Here shown in 2D but the array would ideally surround $f$ in 3D.) (\emph{Top Right}) More typical PAT measurement setup using a finite-sized linear array of detectors. This is the setup used for the tutorial in Sec.\ \ref{sec:comparisonMethods}. (\emph{Bottom Left}) The acoustic time series measured by the linear array. (\emph{Bottom Right}) A PAT image reconstructed from these time series using the classical time reversal approach, Sec.\ \ref{subsubsec:time_reversal}, showing arc-like artifacts due to the limited view detection.}
\end{figure}

In PAT, measurements of the PA-generated acoustic waves are made on a surface $\mathscr{S}$ surrounding a region $\Omega \supset \text{supp}(f)$
containing the object to be imaged, $f$ (see Fig.\ \ref{fig:measurement_setup}). Note that $\mathscr{S}$ is not a boundary, ie.\ it is assumed not to affect the acoustic field. The measurement operator $\mathcal{M}$ will typically consist of a filtering operator, $\mathcal{W}$, which accounts for the angle-dependent frequency response of the detectors, and a spatial sampling operator, $\mathcal{S}$, which selects the part of the acoustic field to be detected such that
\begin{align}
g = \mathcal{M} p  + \noise = \mathcal{S}\mathcal{W} p + \noise,
\label{eqn:measurement_operator}
\end{align}
where $\noise$ is additive measurement noise.
(In some imaging systems, eg.\ in those using LED excitation, the signal-to-noise ratio can be very low and it is necessary to average many times.) A variety of different sampling operators have been considered for PAT, including detection at a set of points, $\{ x_s \in \mathscr{S} \}$, for which $\mathscr{S}$ is a simple geometric surface such as a plane
\cite{Koestli:2001f}, cylinder \cite{Xu2002cylinder}, sphere \cite{Finch:2004msph}, 
ellipsoid \cite{haltmeier2014ellipsoids}
or polyhedron \cite{Kunyansky2011polyhedra}, 
measurements of spatial integrals of the acoustic field along planes or lines
\cite{Burgholzer2005integrating}
or patterns \cite{huynh2019:SinglePixelPAT}, 
2D measurements using a ring of detectors focused in a plane \cite{Wang:2003}, and measurements made with a linear array of elements also focused in a plane (as used for conventional ultrasound imaging)
\cite{yin2004linear_array}.

PA signals are by their nature broadband, often more broadband than the ultrasound detectors used to measure them, so the detected frequency range is usually restricted. Furthermore, due to the finite size of real ultrasound detectors, they also filter the spatial wavenumbers. (As the detection area increases, the more directional the detectors become, ie.\ the narrower the acceptance angle.) The filtering operator, $\mathcal{W}$, accounts for both the frequency and wavenumber filtering effects.

When the detectors are ideal, $\mathcal{W} = \mathrm{Id}$, the identity, and neglecting noise, Poisson's solution \cite{Landau1987fluid} to the initial value problem in Eq.\ \eqref{eqn:wave_equation} shows that the relationship between the measured time series $g(x_s,t)$ and the initial acoustic pressure $f(x)$ can be written in the form: 
\begin{align}
\frac{1}{t}\int_0^t g(x_s,t') \,dt' = \frac{1}{4\pi (ct)^2}\int_{|x_s - x| = ct} f(x) \,dA,
\label{eqn:spherical_mean}
\end{align}
where $dA$ is an area element of the surface given by the spherical shell $|x_s - x| = ct$. This shows that the time average of $g$ between times $0$ and $t$ equals the spatial average of the initial pressure $f(x)$ over a spherical shell of radius $ct$ centered at $x_s$. More concisely, we can write $g_{\text{sm}} = \mathscr{R}_{\text{sm}} f(x)$, where $\mathscr{R}_{\text{sm}}$ is the \textit{spherical mean Radon transform}, and $g_{\text{sm}} = t^{-1}\int_0^t g(x_s,t') dt'$ is the \textit{spherical mean data}. 
Some of the literature relevant to PAT image reconstruction considers the data in this form
\cite{Finch:2004msph,Kuchment:2011mtat}.

\subsubsection{Acoustic, optical and spectroscopic operators}
\begin{figure}[htb!]
\centering
\includegraphics[width=\textwidth]{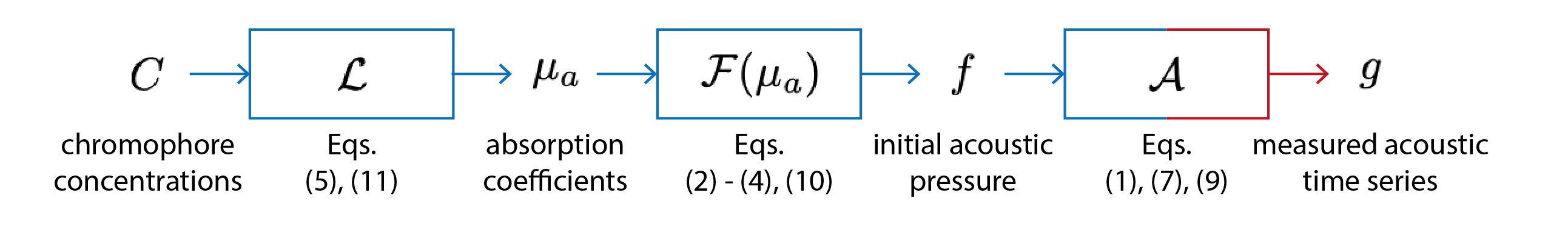}
\caption{\label{fig:forward_operators} The three operators describing the PAT forward problem: spectroscopic $\mathcal{L}$, optical $\mathcal{F}$, and acoustic $\ForwardOp$.
(Blue indicates the image space, $X$, and red the data space $\DataSpace$.)}
\end{figure}

Before discussing PAT inverse problems, it will be helpful to define three operators describing the forward or direct problems (see Fig.\ \ref{fig:forward_operators}). First, the operator $\ForwardOp$, a linear mapping from the initial acoustic pressure distribution $f$ to the measurements $g$ under additive measurement noise $\noise$, which is based on Eqs.\ \eqref{eqn:wave_equation} and \eqref{eqn:measurement_operator}: 
\begin{align}
g = \ForwardOp f + \noise.
\label{eqn:acoustic_operator}
\end{align}
$\ForwardOp$ maps from image space $\RecSpace$ to data space $\DataSpace$.
Second, the operator $\mathcal{F}$, a nonlinear mapping from the absorption coefficient, $\mu_a$, to the initial pressure distribution, $f$, which is based on Eqs.\ \eqref{eqn:initial_pressure}, \eqref{eqn:rte} and \eqref{eqn:fluence}:
\begin{align}
f = \mathcal{F}[\mu_a](\mu_a).
\label{eqn:optical_operator}
\end{align}
$\mathcal{F}$ maps from image space $X_{\mu_a}$ to image space $\RecSpace$.
Finally, a third operator maps chromophore concentrations to absorption coefficients, from $X_{C}$ to $X_{\mu_a}$, 
based on Eq.\ \eqref{eqn:molar_abs}: 
\begin{align}
\mu_a = \mathcal{L} C.
\label{eqn:spectroscopic_operator}
\end{align}

\subsection{PAT Inverse Problems}
\label{sec:inverse_problems}
There are two main inverse problems in PAT, corresponding to the acoustic and optical forward operators described above. First, an acoustic inversion from the measured time series to an estimate of the initial acoustic pressure distribution, ie.\ an estimate of $\ForwardOp^{-1} g$, and second, an optical inversion which attempts to recover quantitatively accurate estimates of optical coefficients (or related properties), eg.\ an estimate of $\mathcal{F}^{-1}(f)$. 
It can be shown \cite{Kuchment:2011mtat} that the acoustic inverse problem is well-posed when sufficient data has been measured, but what constitutes sufficient data? 
The data measured by a closely-spaced array of omnidirectional, broadband, noise-free, point detectors arranged such that all the rays passing through every point in the imaged object reach at least one of the detectors would be sufficient data. For example, if ideal detectors are positioned on the surface of a hemisphere at a spacing of $\lambda_{\text{min}}/2$, where $\lambda_{\text{min}}$ is the shortest wavelength generated (to satisfy the spatial Nyquist criterion), the sound speed is constant everywhere, and the object lies inside the hemisphere's convex hull - the `visible' region \cite{Xu2004Limitedview} - then the acoustic inversion will be well-posed.
Given the stringency of these requirements, it is unsurprising that real experimental settings will often diverge from this ideal, leading to inversions that are no longer well-posed. 
One challenge for the reconstruction, then, relates to dealing with incomplete or imperfect measurement data. There are also challenges relating to the forward operator. These issues are outlined below as it is in tackling these issues that DL may be able to make the most useful contributions.

\subsubsection{Incomplete or imperfect data}
For the acoustic inversion, the data could be incomplete or imperfect for several reasons:
\begin{itemize}
\setlength\itemsep{-0.2em}
    \item \textit{Noise} is present in any real measurement.
    \item \textit{Detector responses} are never perfectly broadband or omni-directional. Compromising on these characteristics is sometimes necessary in order to achieve sufficient detection sensitivity.
    \item \textit{Limited-view detection}, in other words insufficient coverage of the object, perhaps because of the limitations of the available hardware, eg.\ a 2D linear array to image a 3D object, or due to restricted access to the object.    
    \item \textit{Undersampling} in space or time, perhaps in order to achieve faster data acquisition, or due to hardware constraints.
\end{itemize}
In the optical inversion, when considered separately from the acoustic inversion, the input data are images of the initial pressure distribution $f$ obtained from the acoustic inversion. There are two ways in which this data can be deficient:
\begin{itemize}
\setlength\itemsep{-0.2em}
    \item \textit{Artifacts} may be present in the image data due to an imperfect acoustic reconstruction.
    \item \textit{Wavelengths}. The data must contain images at a set of wavelengths chosen such that the spectroscopic aspect of the optical problem, $\mathcal{L}^{-1}$, is well-posed.
\end{itemize}

\subsubsection{Inaccurate forward operators}
\label{subsubsec:inaccurate_forward}
Eqs.\ \eqref{eqn:wave_equation} - \eqref{eqn:fluence} are broadly considered to capture the physical phenomena relevant to PAT, but to solve practical problems they must be implemented as numerical models. There are two ways in which these models can differ from the ideal.
\begin{itemize}
\setlength\itemsep{-0.2em}
    \item \textit{Simplifying approximations} are often made to the forward operators in order to reduce the complexity of the computations necessary to implement them as numerical models. For example, the radiative transfer equation, Eq.\ \eqref{eqn:rte}, is often approximated using a diffusion approximation, and the wave equation, Eq.\ \eqref{eqn:wave_equation}, is sometimes substituted with a simpler model, eg.\ based on rays.
    \item \textit{Inaccurate model inputs}. Although the focus in experimental settings is usually on the accuracy of the measurement data, the accurate determination of the auxiliary parameters on which the forward operators depend is often just as crucial. For example, the acoustic operator $\ForwardOp$ depends on the sound speed and how it varies within the tissue, which is rarely known to a high degree of precision. And the optical operator $\mathcal{F}$ not only requires knowledge of how the tissue was illuminated but also of the tissue's scattering properties, both of which can be hard to determine accurately.
\end{itemize}
This latter problem, the difficulty of accurately measuring the necessary model inputs, in particular the sound speed and the optical scattering distributions, has led some researchers to consider them as additional unknowns in the inverse problem. \cite{Cox2009c,Huang2016joint}.
These inversions have been shown to be less well-posed 
\cite{Tarvainen2012, Stefanov2013instability} than the inversions of $\ForwardOp$ and $\mathcal{F}$, and additional data or constraints are usually required to find meaningful solutions. 

% To facilitate this, variants of the acoustic and optical operators have been proposed which take the unknown parameters as inputs, along with the data. The acoustic variant can be written \cite{Huang2016joint} $\tilde{\ForwardOp}[c] (f, c)$ where $c$ is the sound speed, and the optical operator variant $\tilde{}[\mu_a, \mu_s] (\mu_a, \mu_s)$. (In principle, the phase function could be included in $\tilde{\mathcal{F}}$ too, but usually this is assumed known.) 

\subsubsection{Statistical framework: noise models and priors}
\label{sec:statisticalFramework}
The question naturally arises as to what can be done to improve the image reconstruction when the data is imperfect or the forward model is only known approximately. An approach which sounds like common sense, but in practice can be challenging, is to try to find the reconstruction $f$ that is most probable given the data $g$. This requires a statistical framework\cite{KaipioSomersalo2004}, which also provides a way to incorporate in the reconstruction any other information that is already known about the final image, the data, or the operator, in order to constrain the solution to one that has a higher probability of being correct.
Specifically, we want to find the posterior probability distribution $\pi(f|g)$, \footnote{Here, the notation $\pi(f)$ is used to denote the probability density function of $f$, and $\pi(f|g)$ the conditional probability density of $f$ given $g$.} or some related quantity that characterises the most probable reconstructions. In the Bayesian framework, we can incorporate our prior knowledge about the problem via Bayes' formula
\begin{equation}\label{eqn:Bayes}
		\pi(f|g) \propto \pi(g|f)\pi(f),
\end{equation}
where $\pi(f)$ incorporates prior knowledge of the solution $f$, and $\pi(g|f)$, called the \emph{likelihood}, incorporates the known noise statistics using the forward operator $\ForwardOp$. For example, if the noise in Eq.\ \eqref{eqn:acoustic_operator} is normally distributed with zero mean and variance $\sigma^2$ we can express the likelihood as\cite{KaipioSomersalo2004}
\begin{equation}\label{eqn:likelihood}
		    \pi(g|f) \propto \exp\left(-\frac{1}{2 \sigma^2}\|\ForwardOp f - g \|^2_2\right).
\end{equation}
Even though in many applications we might not be able to explore the full posterior distribution $\pi(f|g)$, the Bayesian framework can provide guidance for the interpretation of specific image reconstruction approaches.  For instance, computing the \emph{maximum a posteriori} (MAP) estimate relates to finding the minimiser in variational approaches, as we will discuss later in Sec.\ \ref{subsubsec:acoustic_variational}.

Classically, both the likelihood and the prior are explicitly modelled and might therefore be limited in their expressibility. Hence a natural question arises in the context of this study: \emph{Can we use learning-based models instead of analytical models to generalise this approach?} 
In particular, two ways in which learning-based methods could be incorporated are:
\begin{itemize}
        \item[i)] Learning a prior $\pi(f)$ that describes the unknown initial acoustic pressure distribution better,
        \item[ii)] Compensating, in the likelihood $\pi(g|f)$, for model uncertainties or complex noise statistics.
\end{itemize}
In DL it is conceptually easier to address the estimation of a prior, as it relates to the training set, as we will discuss in the later sections; it is not so straight-forward to incorporate model uncertainties into the likelihood estimation. 
A useful direction on how to tackle this is given by the well-established approach of Bayesian approximation error modelling\cite{KaipioSomersalo2004,Kaipio2007,arridge2006a}. In this approach, modelling errors in the forward operator $\ForwardOp$ are estimated as normally distributed and explicitly corrected in the likelihood term, Eq.\ \eqref{eqn:likelihood}. This approach has been applied in PAT to compensate for uncertainty in the measurement parameters (model uncertainty)\cite{sahlstrom2020modeling,tarvainen2013}.

\subsubsection{Image reconstructions}
\label{sec:varieties_of_inverse_problem}
While the two inverse problems described above, the acoustic and optical inversions, are the fundamental image reconstruction problems in PAT, variations on them are often used in practice. Common PAT reconstruction problems that appear in the literature are:
\begin{itemize}
\setlength\itemsep{-0.2em}
    \item reconstructing an image of the initial acoustic pressure distribution from the measured time series data, $\ForwardOp^{-1} g$,
    \item reconstructing an image of the optical absorption coefficient from initial acoustic pressure distribution images, $\mathcal{F}^{-1} (f)$,
    \item reconstructing an image of optical absorption coefficient directly from the measured time series data, $(\ForwardOp\mathcal{F})^{-1} (g)$,
    \item reconstructing images of quantities related to optical absorption, eg.\ chromophore concentrations or blood oxygenation, from a multiwavelength set of initial acoustic pressure images, 
    $(\mathcal{F}\mathcal{L})^{-1} (f)$,
    \item reconstructing images of quantities related to optical absorption directly from a multiwavelength  set of time series data, $(\ForwardOp\mathcal{F}\mathcal{L})^{-1} (g)$.
\end{itemize}
Research has already begun on applying DL to several of these tasks; this literature will be reviewed in Sec.\ \ref{sec:reviewPart}. The next two sections will give an overview of the classical approaches to PAT image reconstruction, and a short tutorial on the kinds of DL that are being used for image reconstruction.

\section{Classical Approaches to PAT Image Reconstruction}
\label{sec:PATrecon}
DL can be used to complement or augment current approaches to PAT reconstruction, or replace parts of them. For this reason, as well as to provide context, this section describes several widely used `classical' - ie.\ not learning-based - approaches that have been used for solving the PAT inverse problems.
This section is not intended to be a comprehensive review of classical methods for PAT image reconstruction, for which the literature is large, but for later reference.

\subsection{Acoustic Reconstruction}
Here we consider the \emph{acoustic} inversion of PAT, ie.\ the linear problem of solving Eq.\ \eqref{eqn:acoustic_operator} for $f$, the initial pressure distribution, given $g$, the measurement data. We will denote a generic reconstruction operator, or data-to-image mapping, by $\ForwardOpPseudoInv:\DataSpace\to\RecSpace$ throughout this review. Let us now discuss specific choices for such a mapping.

\subsubsection{Backprojection and beamforming}
\label{subsubsec:backprojection}
Algorithms based on the idea of \emph{backprojection} are widely used in PAT. This terminology comes from X-ray tomography in which the forward operator (the linear ray transform) maps from image to data space by integrating the target along a set of straight lines for each detector, and the backprojection operator maps from data to image space by putting the data back along those straight lines and summing over all detectors. In the X-ray case, these dual operations are also adjoint; the backprojection operator is the adjoint of the forward operator\cite{Natterer1986}.
In PAT the situation is slightly different. The forward operator $\ForwardOp$ maps from image space $\RecSpace$ to data space $\DataSpace$ by integrating through $f$ along a set of spherical shells of radius $t = |x - x_s|/c$ centered on the detector points $x_s$ (see Sec.\ \ref{subsubsec:Photoacoustic_measurements}). Correspondingly, the backprojection operator $\ForwardOp^{\#}$ maps a function of $x_s$ and $t$, from data space $\DataSpace$ to image space $\RecSpace$ by putting the data back onto the same spherical shells with the mapping $t \rightarrow |x - x_s|/c$, and summing over all detector points, $x_s$. For some function $h(x_s,t)$, which might be the measurement data or some function of it, the backprojection operator is
\begin{align}\label{eqn:backprojectionIntegral}
(\ForwardOp^{\#} h)(x) = \int_{\mathscr{S}} \left[h(x_s,t)\right]_{t = |x - x_s|/c} \, d\mathscr{S}(x_s) 
\end{align}
where $d\mathscr{S}$ is an area element on the measurement surface $\mathscr{S}$.
On the other hand, the adjoint operator $\ForwardOpAdj$ is given by \cite{Arridge:2016adj}
\begin{align}
(\ForwardOpAdj g)(x) = \int_{\mathscr{S}} \left[\frac{1}{4\pi |x-x_s|}\frac{\partial g}{\partial t}(x_s, t) \right]_{t = |x - x_s|/c} d\mathscr{S}(x_s),
\end{align}
which is clearly a backprojection, but not of the data $g$ (see also Sec.\ \ref{subsubsec:acoustic_variational}). When the data is processed before backprojection (or sometimes the image is processed after backprojection) the resulting algorithm is often referred to as a \emph{filtered backprojection}. Filtered backprojection formulas for PAT have been found for a variety of measurement surface geometries \cite{Finch:2004msph,Xu2005UBP,Kuchment:2011mtat,Kunyansky2011polyhedra,Haltmeier2015universal}. Perhaps the most well-known, called the `universal backprojection' algorithm\cite{Xu2005UBP}, gives exact reconstructions for detector points covering a spherical, cylindrical or planar measurement surface, and can be written as 
\begin{align}
\label{eqn:UBP_3D}
f(x) = \frac{-2}{\Omega_s c^2} \int_{\mathscr{S}} \left[
\frac{\partial}{\partial t}\left( \frac{g(x_s,t)}{t} \right)
\right]_{t = |x_s - x|/c} \cos(\alpha) \,d\mathscr{S}(x_s),
\end{align}
where $\alpha$ is the angle between the inward normal to $\mathscr{S}$ and the vector $(x - x_s)$, and $\Omega_s$ is the solid angle of $\mathscr{S}$ as seen from a point $x\in\Omega$, eg.\ $\Omega_s = 4\pi$ when $\mathscr{S}$ is a sphere. A 2D version of this has also been derived\cite{burgholzer2007temporal}:
\begin{align}
\label{eqn:UBP_2D}
f_{\text{2D}}(x) = \frac{-4}{\Omega_s c^2} \int_{\mathscr{S}} \left( \int_{|x - x_s|/c}^{\infty} \frac{1}{\sqrt{t^2 - |x - x_s|^2/c^2}} 
\frac{\partial}{\partial t}\left( \frac{g(x_s,t)}{t} \right)
dt \right)
\kappa(x,x_s) \cos(\alpha) \,d\mathscr{S}(x_s),
\end{align}
where the weighting factor $\kappa(x,x_s) = |x - x_s|$ for the universal backprojection algorithm, but has also been treated as a learnable parameter (Sec.\ \ref{subsubsec:extensions}).

Linear array transducers of the kind used in conventional ultrasound imaging are increasingly being used for PAT, with backprojection-type formulas commonly used for image reconstruction. In this context, image reconstruction is sometimes referred to as `beamforming' and the backprojection operation $\ForwardOp^{\#}$ is descriptively dubbed `delay-and-sum'. Linear arrays are typically short, consisting of just 128 bandlimited detection elements focused in a plane, so the image reconstruction is very ill-posed. Many variations of backprojection-type algorithms with different pre- and post-processing steps have been explored to try to maximise the image quality given these severe constraints. 
\cite{park2008beamforming,mozaffarzadeh2017delay_and_sum}. In DL approaches to PAT image reconstruction, backprojection / beamforming-type algorithms have been used widely to map from data space $\DataSpace$ to image space $\RecSpace$ before and after post- and pre-processsing networks respectively (see Secs.\ \ref{subsubsec:Rec_nPost}, \ref{subsec:post-proc} and \ref{subsec:pre-proc}).

\subsubsection{Series solutions}
\label{subsubsec:seriesSol}
The first analytical solution to Eq.\ \eqref{eqn:wave_equation} was found in the form of an infinite series, and more have since been derived
\cite{Koestli:2001f,Xu:2002efdr,Xu2002cylinder,Kunyansky2007series,Kunyansky2012fast}. A formula for the case of detection points lying on a plane is of particular interest because it is in the form of a Fourier transform, which can be computed efficiently using the Fast Fourier Transform \cite{Koestli:2001f}. The solution relies on the fact that any acoustic wavefield, $p(x,t)$, can be written as a sum of travelling plane waves whose temporal frequency, $\omega$, and wavevector, $k = (k_1,k_2,k_3)$, are linked by the dispersion relation $\omega = c|k| = c\sqrt{k_1^2 + k_2^2 +k_3^2}$. The solution takes the form
\begin{align}
f(x) = \mathscr{F}^{-1}_{1,2,3} \{\tilde{f}(k)\},
\quad
\tilde{f}(k_1,k_2,\omega) = B(k_1,k_2,\omega) \mathscr{F}_{1,2}\left\{ \left\{\mathscr{C}_t\left\{
g(x_1,x_2,t)\right\}\right\}\right\},
\end{align}
where $B(k_1,k_2,\omega) = \sqrt{(\omega/c)^2 - k_1^2 - k_2^2}/\omega$, $\mathscr{F}$ and $\mathscr{C}$ are Fourier and Cosine transforms respectively, and $\tilde{f}(k)$ is obtained by algebraic transform from $\tilde{f}(k_1,k_2,\omega)$ using the dispersion relation.
In DL, this method has been used as a component in learned iterative reconstrucions (see Sec.\ 
\ref{subsubsec:learned_iterative_3D}). When used with linear array transducers, this method and its variants are sometimes referred to as `Fourier beamforming'.

\subsubsection{Time reversal}
\label{subsubsec:time_reversal}
Perhaps the most physically intuitive algorithm is based on the concept of time reversal.\cite{Xu2004TR,burgholzer2007,Hristova2008} Consider a measurement surface $\mathscr{S}$ surrounding a region $\Omega \supset \text{supp}(f)$. Imagine the photoacoustically-generated waves propagating outwards and being measured as they pass through the surface $\mathscr{S}$. After a suitably long time $T$ the acoustic field in $\Omega$ will be zero (guaranteed in a 3D homogeneous medium by Huygens' principle\cite{Baker2003Huygens}). If the measured pressure, $g(x_s,t)$, were now reproduced on $\mathscr{S}$ in time-reversed order, starting with $g(x_s,T)$, then the acoustic field in $\Omega$ created by the \emph{in-going} waves would reproduce the out-going wavefield exactly but backwards in time. In particular, the field at $t=0$ would be the initial acoustic pressure distribution $f(x)$. Based on this idea, time reversal image reconstruction uses a numerical acoustic model to solve the following time-varying boundary value problem for the time-reversed field $p_r(x,t_r)$, from time $t_r = 0$ to $T$,
\begin{align}
(\partial_{tt} - c^2\Delta) p_r(x,t_r) = 0, 
\quad p_r(x_s,t_r) = g(x_s,T - t_r), 
\quad p_r = \partial_t p_r(x,0) = 0.
\label{eqn:time_reversal}
\end{align}
The solution $p_r(x,T) = f(x)$ for $x\in\Omega$.

In DL studies, the time reversal approach is sometimes used for comparison with network approaches, but care must be exercised here to ensure a fair comparison. To help elucidate two problems with time reversal, note that the time-varying Dirichlet condition, $p_r(x_s,t_r) = g(x_s,T - t_r)$, is equivalent to reintroducing the measurement data as a source term within a reflective cavity defined by the measurement surface $\mathscr{S}$\cite{Xu2004TR}. First, then, time reversal is not a good choice when using data detected on a sparse array of points because during the time reversal procedure they act like point scatterers. Second, when the true sound speed is spatially-varying but the reconstruction uses a homogeneous sound speed, the reflective effect of the boundary condition can trap artifacts in the image region\cite{cox2009artifact}. In these scenarios, time reversal may not be the best method for comparison.
Furthermore, when the sound speed is spatially-varying, resulting in multiple scattering, the requirement that the acoustic field in $\Omega$ will fall to zero in a finite time $T$ is no longer satisfied. One solution\cite{Stefanov:2009tatvs} is to use the following iterative scheme:
\begin{align}
f^{(n+1)} = f^{(n)} - \ForwardOp^{\text{TR}}(\ForwardOp f^{(n)} -  g),
\label{eqn:iterative_time_reversal}
\end{align}
where $\ForwardOp^{\text{TR}}$ signifies the time-reversal operator.

\subsubsection{Variational approaches}
\label{subsubsec:acoustic_variational}
The iterative time reversal algorithm points to a more general approach to reconstruction as it looks very similar to this gradient descent scheme
\begin{align}
f^{(n+1)} = f^{(n)} - \eta\nabla_{\!f} \mathcal{E} 
= f^{(n)} - \eta\ForwardOpAdj(\ForwardOp f^{(n)} -  g),
\label{eqn:gradient_descent}
\end{align}
which solves the least-squares minimisation problem:
\begin{align}
f^* = \underset{f}{\argmin} \, \mathcal{E}(f), \quad  \mathcal{E}(f) = \tfrac{1}{2}\| \ForwardOp f - g \|^2_2,
\label{eqn:data_consistency}
\end{align}
where $f^*$ denotes the optimal solution.
(The similarities between Eqs.\ \eqref{eqn:gradient_descent} and \eqref{eqn:iterative_time_reversal} become even clearer if we observe that the adjoint operator, $\ForwardOpAdj$, can be implemented numerically in a similar way to the time reversal operator, $\ForwardOp^{\text{TR}}$,  except that the pressure time series are reintroduced to the domain in time-reversed order by \textit{adding} them to the existing field rather than enforcing the pressure at the detector points \cite{Arridge:2016adj}.)
The idea of posing the image reconstruction as a numerical optimisation is appealing \cite{Guo2010compressed,Wang2012modelbased,Huang:2013fwi,Arridge:2016apat,Haltmeier2017iterative,Boink2018framework}, because it provides a very flexible framework both for how the forward operator is defined (eg.\ the sound speed could be spatially-varying) and for tackling ill-posedness in the inverse problem. Eq.\ \eqref{eqn:data_consistency} will have a unique solution when $g$ is a complete set of ideal data. However, if $g$ is incomplete or imperfect then Eq.\ \eqref{eqn:data_consistency} may not have a unique solution, or overfitting (in which the model starts to fit to the noise in the data) may become a concern. Early stopping of the iteration in Eq.\ \eqref{eqn:gradient_descent} is one way to avoid overfitting, but a more general approach to restricting the solution space is to add another term to the functional in Eq.\ \eqref{eqn:data_consistency} that expresses prior information about the kind of solution that is expected, eg.\ non-negativity of solutions and smoothness or sparsity conditions. The problem then becomes
\begin{align}\label{eqn:variationalFormulation}
f^* = \underset{f}{\argmin} \, \mathcal{E}(f), \quad  \mathcal{E}(f) =\tfrac{1}{2}\|\ForwardOp f - g \|^2_2 + \alpha\mathcal{R}(f),
\end{align}
where the regularisation parameter $\alpha$ balances the importance placed on the first term - the data consistency term - and the second term, $\mathcal{R}$, which encodes the prior information about $f$. There is an extensive literature on methods to solve minimisations such as this
\cite{chambolle2016introduction,benning2018modern}.
If the regularisation term $\mathcal{R}(f)$ is differentiable, one could simply employ a gradient descent, Eq.\ \eqref{eqn:gradient_descent} with 
$\alpha \partial \mathcal{R}/\partial f$. 
If not, another approach to computing solutions iteratively is the proximal gradient method, which means computing the iteration:
\begin{align}\label{eqn:proximalGrad}
f^{(n+1)} = \text{prox}_{\mathcal{R},\eta\alpha}\left(f^{(n)} - \eta\ForwardOpAdj(\ForwardOp f^{(n)} -  g)\right)
\end{align}
where $\ForwardOpAdj(\ForwardOp f^{(n)} -  g))$ is the gradient of the data consistency term and $\text{prox}_{\mathcal{R},\eta\alpha}$, 
the proximal operator, takes the updated image estimate and projects it into the constrained set defined by the regularisation, or in other words the space in which the solution is thought to exist. It is formally defined as the minimisation problem
\begin{equation}\label{eqn:proximalOperator}
    \text{prox}_{\mathcal{R},\eta\alpha}(h) = \underset{y}{\argmin}\left\{ \eta\alpha\mathcal{R}(y) + \frac{1}{2}\|h-y\|_2^2 \right\}.
\end{equation}
The formulation as a minimisation problem in Eq.\ \eqref{eqn:variationalFormulation} is directly connected to the Bayesian formulation in Eq.\ \eqref{eqn:Bayes} and corresponds to maximising the posterior distribution $\pi(f|g)$ to find the most likely reconstruction $f$. This represents a point estimator known as the \emph{maximum a posteriori} (MAP) estimate\cite{KaipioSomersalo2004}. In this context, the negative logarithm of the prior distribution directly relates to the regularising term, $-\log \pi(f) \propto \mathcal{R}(f)$.
This general framework provided by the variational approach has inspired several learned iterative approaches to PAT reconstruction (see Secs.\  
\ref{subsubsec:Model-based_learned_iterative_reconstruction} and
\ref{subsec:reviewLearnedIterative}).

\subsubsection{Matrix formulation}
\label{sec:matrixRep}
The acoustic forward operator $\ForwardOp$ is linear and so can in principle be discretised and written as a (large) matrix. When this matrix can actually be explicitly computed, the image reconstruction problem has been reduced to a matrix inversion and all the machinery of linear algebra, and the associated methods of regularisation, can be brought to bear to solve it. This includes the variational approaches above in Sec.\ \ref{subsubsec:acoustic_variational}. For instance, if one considers a quadratic regularisation in Eq.\ \eqref{eqn:variationalFormulation}, such as $\mathcal{R}(f)=\|f\|_2^2$, then the solution can be computed in closed form and is given by
\[
f^* = (\ForwardOpAdj\ForwardOp + \alpha \mathrm{Id})^{-1}\ForwardOpAdj g,
\]
where $\mathrm{Id}$ denotes the identity, and which is sometimes called a Tikhonov-regularised solution. 

There are many methods that can be used to discretise the forward operator, from pseudo-spectral methods \cite{Huang:2013fwi} to semi-analytical approaches \cite{Rosenthal2010fast}. However, whether it is convenient - or even possible - to compute and store $\ForwardOp$ explicitly as a matrix will depend on the number of detectors and the size of the image, and whether sparsity or other structures in the matrix can be exploited\cite{Paltauf2002iterative}. In fact, we will make use of a matrix representation for $\mathcal{A}$ in the tutorial part of this review (Sec.\ \ref{sec:comparisonMethods}), as the problem under consideration is sufficiently small.

\subsection{Optical Reconstructions}
\label{subsec:Optical_Reconstructions}
This section briefly summarises classical approaches to solving the nonlinear Eq.\ \eqref{eqn:optical_operator} for the absorption coefficient or related quantities. From Eq.\ \eqref{eqn:initial_pressure} we can see formally that $\mu_a = \mathcal{F}^{-1}(f) = f/(\Gamma\phi(\mu_a))$. An empirically determined value for $\Gamma$ is sometimes used, or, when the final quantity of interest is a ratio of concentrations, see Eq.\ \eqref{eqn:saturation}, $\Gamma$ is assumed to be constant with wavelength and cancels out. The dependence of the fluence $\phi$ on $\mu_a$, however, needs to be considered carefully.

\subsubsection{Non-iterative approaches}
A simple approach to deal with the dependence of $\phi$ on $\mu_a$, but one with questionable accuracy, is to ignore the dependence and apply the spectroscopic inversion directly to the PA data, $\mathcal{L}^{-1} f$. This is sometimes known as \emph{linear unmixing}. Despite its obvious flaws, this stance has been taken (usually implicitly) in many experimental papers in which the PA spectrum at a point, $f(\lambda)$, has been assumed to be proportional to the absorption spectrum at that point $\mu_a(\lambda)$. The difference between $f(\lambda)$ and $\mu_a(\lambda)$, which linear unmixing ignores, is known as \emph{spectral coloring}. 
A better approach, but still one whose accuracy needs to be demonstrated on a case-by-case basis, is to approximate the fluence using estimated average background absorption and scattering values, and suppose that this fluence remains unchanged by small changes in the optical absorption coefficient. 
In some cases, the fluence distribution can be measured directly using a second imaging modality in addition to PAT \cite{Bauer2011quantitative,Hussain2018photoacoustic}. However, this requires complementary hardware to make the additional measurements, and it is difficult to achieve the same spatial resolution for $\phi$ as for $f$ (or one may as well measure just the fluence distribution and not do PAT at all).

\subsubsection{Fixed-point iterations}
If the scattering is known, then the absorption coefficient can be found using a model of light transport, such as Eq.\ \eqref{eqn:rte} or a suitable approximation, to calculate both the fluence and the absorption coefficient iteratively using the fixed point iteration \cite{Cox2006}:
\begin{align}
\mu_a^{(n+1)}(x,\lambda) = f(x,\lambda) / \Gamma \phi^{(n)}(x, \lambda; \mu_a^{(n)}).
\label{eqn:fixed_point}
\end{align}

\subsubsection{Variational approaches}
As with the acoustic inversion described in Sec.\ \ref{subsubsec:acoustic_variational}, casting the optical inversion as a minimisation problem allows the various constraints and prior information to be included systematically. Here, the inverse problem for the absorption coefficient is stated as
\begin{align}
\mu_a^*(x) = \underset{\mu_a(x)}{\text{argmin}} \, \mathcal{E}(\mu_a), \quad  \mathcal{E}(\mu_a) = \tfrac{1}{2}\| \mathcal{F}[\mu_a](\mu_a) - f \|^2_2 + \alpha\mathcal{R}(\mu_a),
\label{eqn:qpat_variational}
\end{align}
and a similar expression can be written for the oxygenation saturation or other quantities of interest. As, from Eq.\ \eqref{eqn:initial_pressure}, $\mathcal{F}(\mu_a) := \Gamma\mu_a\phi(\mu_a)$, the functional gradient is given by $\nabla_{\mu_a} \mathcal{E} = \Gamma(\mu_a D\phi + \phi)$, where $D\phi$ is the Fr\'{e}chet derivative of $\phi$, the form of which will depend on the particular model of light transport used \cite{Cox2009c,tarvainen2013,Buchmann2019three}.

\section{Tutorial Introduction to Deep Learning for PAT Image Reconstruction}
\label{sec:tutorial}

\subsection{What Role could Deep Learning Play?}
How can DL help to solve the challenges posed by the twin problems of incomplete data and inaccurate forward models outlined in Sec.\ \ref{sec:inverse_problems}?
Or are there other ways in which DL can be used to enhance PAT image reconstruction? There are many areas in which DL could make an impact. For example, a DL network could be used to 
\begin{itemize}
\setlength\itemsep{-0.2em}
    \item correct for missing or corrupted data in the measured time series data (pre-processing),
    \item reconstruct images from incomplete or imperfect data given the forward operator (effectively learning prior information to regularise the solution),
    \item approximate a forward operator (eg.\ when it is difficult to write an accurate and computationally-efficient forward model explicitly),
    \item approximate an inverse operator (even when the data is perfect and the forward operator known this may speed up the image reconstruction),
    \item remove artifacts and noise from reconstructed images (post-processing),
    \item segment images,
    \item classify or label images or regions of images.
\end{itemize}
(As mentioned in the Introduction, the last two points are out of the scope of this review.) 
An important attribute of a DL network is the speed with which it can process an input. For small networks this can be very fast, which may be useful in settings where reconstructions are required on short time scales such as real-time or dynamic imaging. However, the speed of evaluation will depend on the size of the network and the size of the input data. It is also important to keep in mind, that the final reconstruction speed will still depend on how the forward operator is utilised in the processing pipeline.

The motivation to use DL in image reconstruction, which these conceptual advantages provide, can readily be followed by action thanks to the availability of easy-to-use DL tools, such as TensorFlow\cite{tensorflow2015-whitepaper} and PyTorch\cite{pytorch_NEURIPS2019},
which make employing these new methods straightforward. Furthermore, the tendency of the machine/deep learning community to provide open-source algorithms and data accelerates the development of new methods and makes it simpler for researchers to try new approaches. Consequently, we provide the codes accompanying this review along with a basic example of training and test data.

\subsection{Brief Introduction to Deep Learning}
This review concentrates on the application of Deep Learning, by which we mean in particular deep neural networks, to image reconstruction tasks in PAT. Specifically, we will concentrate throughout this section on the acoustic inverse problem, so we are interested in finding a mapping from the measurement data $g\in\DataSpace$ to the initial acoustic pressure $f\in\RecSpace$. The driving incentive is the hope that a reconstruction operator $\ForwardOpInvLearned:\DataSpace\to\RecSpace$ that is parametrised by a set of \emph{learnable} parameters $\param$, such that 
\begin{equation}
\label{eqn:paraInvProb}
f \approx \ForwardOpInvLearned(g), 
\end{equation}
can give better (faster, more accurate) reconstructions than classical approaches. The mapping in Eq.\ \eqref{eqn:paraInvProb} may be a composition of model-based parts, involving a known operator describing the acquisition geometry and physics, and pure learning-based components. Before we can review common approaches and network architectures, we will give a short introduction to DL and the main network components. We will concentrate here on a high-level overview to help develop an intuition for the operations involved. For a more extensive review, see, for instance \cite{lecun2015deep,goodfellow2016deep,schmidhuber2015deep}.

\subsubsection{Deep neural networks}
A deep neural network, denoted here by the nonlinear operator $\Network_\param$, maps an input vector to an output vector. The network consists of several `layers', each of which is a composition of an affine linear function with learnable parameters and a nonlinear function (often referred to as the `activation function' but referred to as a nonlinearity here).  The term Deep Learning refers, roughly, to networks that consist of multiple layers, in contrast to shallow networks consisting of only a few layers. 

\begin{figure}[th!]
\centering
\includegraphics[width=0.5\textwidth]{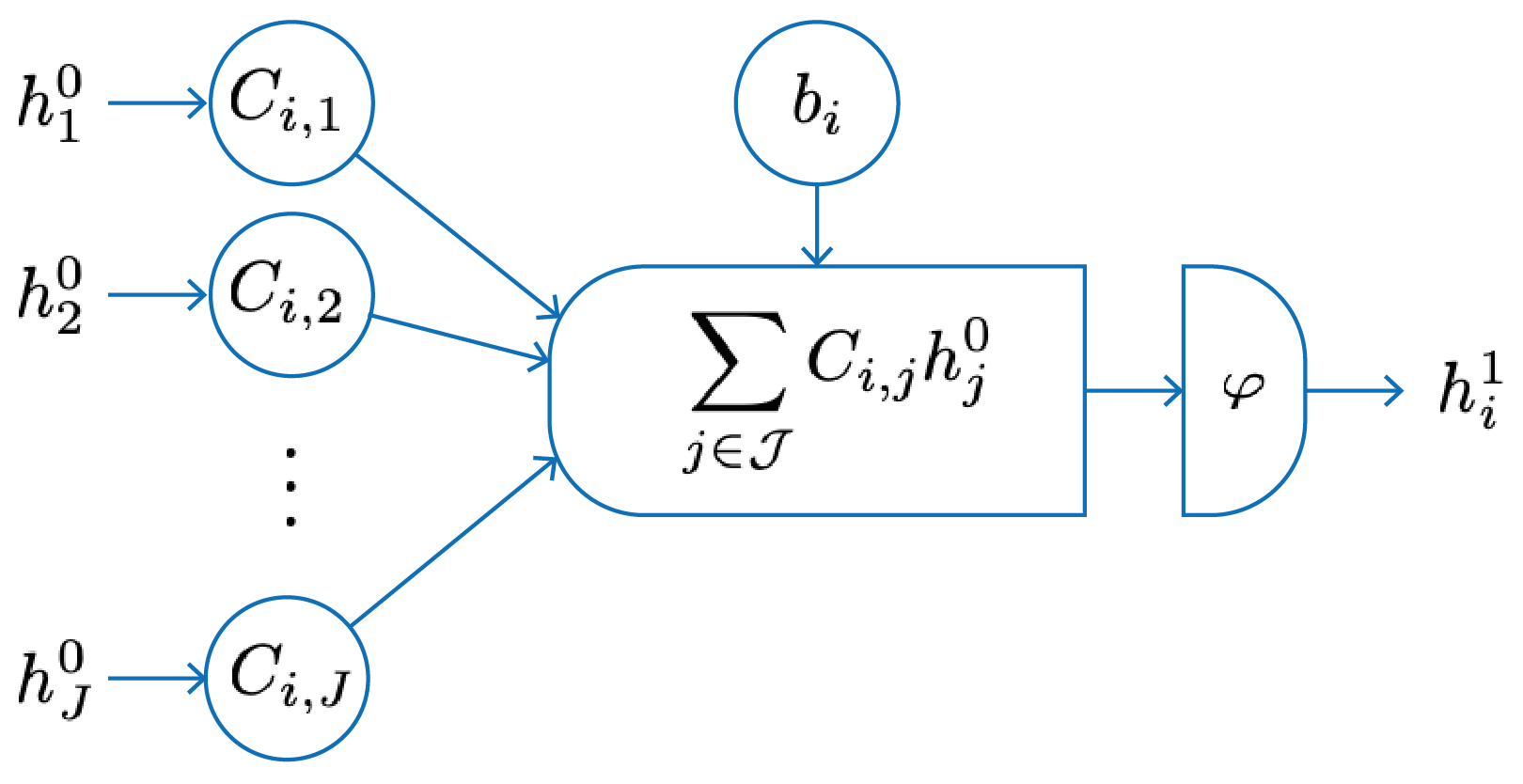}
\caption{\label{fig:ANN} The $i^{\text{th}}$ neuron in one layer of an artificial neural network takes an input vector $h^0$ and computes an output vector $h^1$ according to Eq.\ \eqref{eqn:denseLayer}.}
\end{figure}

Let us now formalise the notion of a layer. Given an input vector $h^0 = \{h^0_j\}_{j=1}^{J} \in \R^J$ where $j\in \mathcal{J}=\{1,\dots,J\}$ and an output vector $h^1=\{h^1_i\}_{i = 1}^{I} \in\R^I $ where $i\in\mathcal{I}=\{1,\dots,I\}$, a linear map given as a matrix $C\in \R^{I\times J}$, a vector $b\in \R^{I}$, and a point-wise nonlinear function $\varphi:\R \to \R$, then one layer $\mathscr{L}$ in a network is given by
\footnote{In the literature, the term `layer' is used somewhat ambiguously. Here, it will be used to refer to both an operation and its output, not just the output. One exception is the \emph{input layer}, which refers to just the input data with no prior operation.}
\begin{equation}\label{eqn:oneLayer}
\mathscr{L}(h^0) = \varphi(C h^0 + b) = h^1.
\end{equation}
The individual neurons in such a neural network are now the mapping to one element of the output vector, see Fig.\ \ref{fig:ANN} for an illustration. If we write this out for the above case, Eq.\ \eqref{eqn:oneLayer}, then the result of the $i$-th neuron is the $i$-th element of the output vector $h^1_i$ and each neuron sums over all input elements of $h_0$ with a common bias $b_i$:
\begin{equation}\label{eqn:denseLayer}
h^1_i=\varphi\left( \sum_{j\in\mathcal{J}} C_{i,j} h^0_j + b_i \right) \hspace{-0.3em} \text{ for each } i\in \mathcal{I}.
\end{equation}
The network type is essentially defined by the linear mappings in each layer, defined by the structure of the matrix $C$.  

\subsubsection{Fully connected layers}
\label{subsubsec:full_connected}
The basic choice for $C$ is a dense matrix, which gives a \textit{fully connected layer} as all the inputs are related to all the outputs. 
We then obtain a simple $L$-layered network by the composition of $L$ fully connected layers. This network can be expressed as the composition of several layers $\mathscr{L}_l$ for $l=1,\dots,L$ to obtain
\begin{equation}\label{eqn:twoLayerDense}
h^L = \Network_\param(h^0)  = (\mathscr{L}_L \circ \mathscr{L}_{L-1} \circ \dots \circ \mathscr{L}_1)(h^0).
\end{equation}
For example, if we write this out for a two-layer network, we get the relation
\begin{equation}
h^2 =  \varphi(C^{2} h^1 +b^2) = \varphi(C^{2} \varphi(C^{1}h^0+b^{1}) +b^2)
\end{equation}
In the general case, the trainable parameter set $\param$ of the network $\Network_\param$ is given by the matrices and the bias vectors, that is $\param = \{C^l,C^{l-1},\dots,C^1,b^l,b^{l-1},\dots ,b^1\}$. 
This basic network architecture, consisting of multiple fully connected layers, is the basis for many deep neural networks.
%\footnote{\revision{Networks consisting of fully connected layers are sometimes called \emph{artificial neural networks}, although this is also used as a more general term.}}
When using fully connected layers in imaging applications, the input, either an image or other measured signals, must be reshaped into a vector for the input layer. 
If one then aims to extract some relevant low-dimensional information from the input, the dimensions of successive layers will be gradually decreased until the desired output dimensions are reached. 
An often used network architecture worth mentioning in this class is termed an \emph{autoencoder}. Here, the input is first \emph{encoded} using a contracting path to extract a low-dimensional representation of relevant features and then subsequently \emph{decoded} using an expanding path to represent a clean version of the input signal. Input $h^0$ and output $h^l$ typically have the same or similar dimensions. 

\subsubsection{Convolutional neural networks}
Often the values in image pixels or voxels are related in some way with those in neighbouring pixels or voxels, eg.\ both may be part of the same image feature. For applications of DL to imaging, therefore, it seems wise to take spatial relations, and especially local relations, into account. A fully connected layer does not explicitly maintain these spatial relations, as all inputs are connected to all outputs without reference to their respective spatial positions. In other words, the linear mapping $C$ in Eq.\ \eqref{eqn:oneLayer} does not have any pre-determined structure. It is possible, however, to think of structures for $C$ that do retain spatial information and can use local features in the input, such as edges, to encode such features more efficiently in the output.
Convolutions, especially with small filters - $3\times3$ say - are a popular and very successful choice for such operations, as these are also translation equivariant\footnote{We say that a function is translation equivariant if translating the input and then applying the function is equivalent to applying the function followed by translation of the output.} 
and hence encode the same local features under translation of the image and are agnostic to the image size. Additionally, localised filters have the advantage of leading to linear mappings with sparse structure that can be efficiently implemented without an explicit matrix representation. 
In this case, instead of learning the whole matrix $C$, one needs to learn only the filter coefficients. Usually multiple such filters are used, each one referred to as a `channel' here. Networks using this idea are called \emph{convolutional neural networks}, or CNNs. 

\begin{figure}[th!]
\centering
\includegraphics[width=0.95\textwidth]{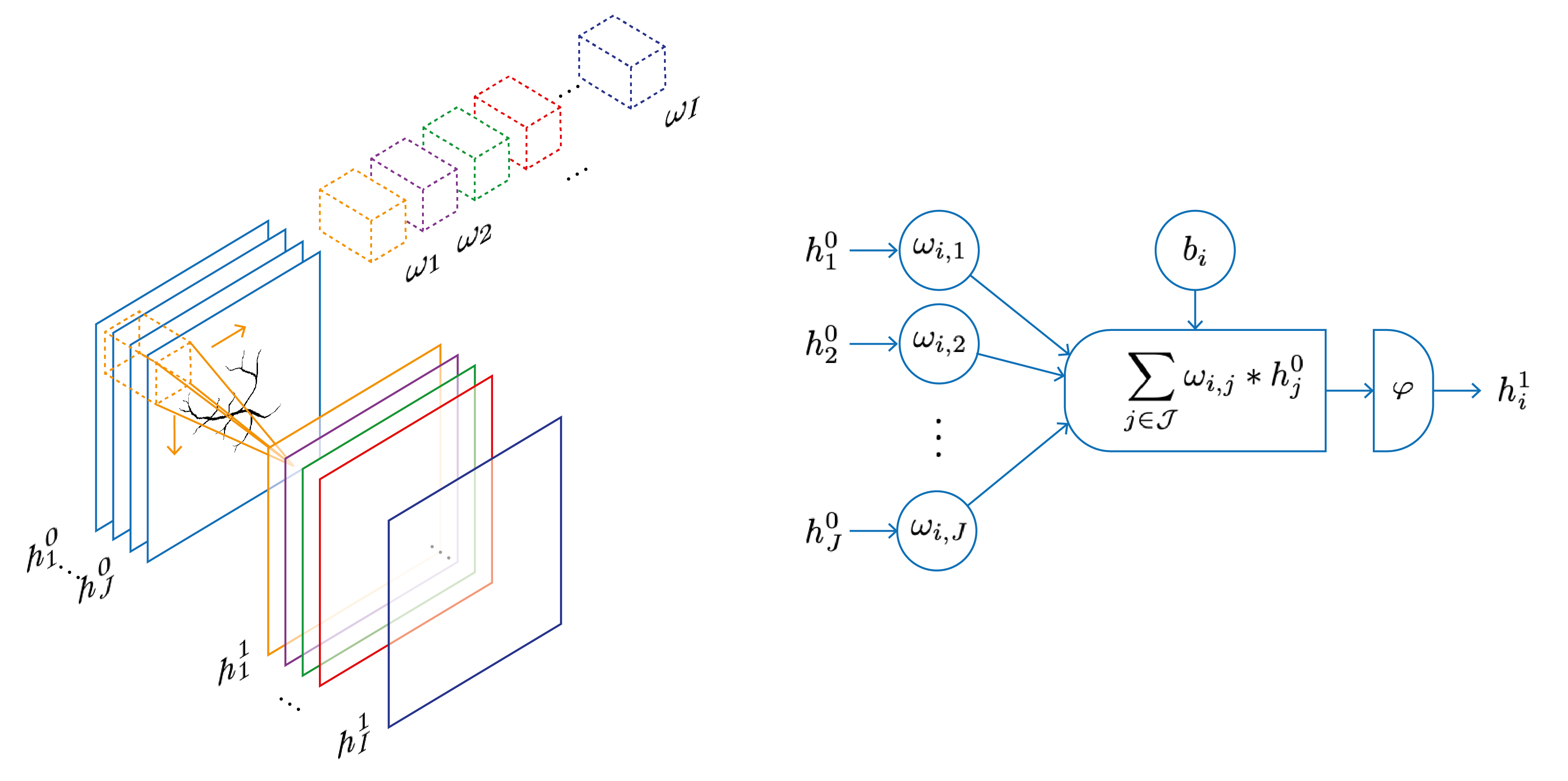}
\caption{\label{fig:CNN} 
One layer of a convolutional neural network (CNN). (\emph{Left}) The input, consisting of $J$ channels $\{h^0_1,\ldots, h^0_J\}$, is convolved with the $I$ filters $\omega_i$, then the nonlinearity $\varphi$ is applied and biases $b_i$ added (not shown) to give the output in $I$ channels, $\{h^1_1,\ldots, h^1_I\}$.
(\emph{Right}) In analogy with Fig.\ \ref{fig:ANN} for the fully connected network, the $i^{\text{th}}$ output channel of a CNN takes an input $h^0\in\R^{m\times m\times J}$ and computes an output $h^1\in\R^{m\times m\times I}$ according to Eq.\ \eqref{eqn:convLayer}.}
\end{figure}

Consider an application to imaging in $\R^2$. The input is either a single or multichannel image $h^0 =\{h^0_j \in\R^{m\times m}\}_{j=1}^J \in\R^{m\times m\times J}$, where $j\in \mathcal{J}=\{1,\dots,J\}$ denotes the input channels, and similarly an output 
$h^1=\{h^1_i \in\R^{m\times m}\}_{i = 1}^I \in\R^{m\times m\times I}$ where $i\in\mathcal{I}=\{1,\dots,I\}$ denotes the output channels.
(These images are square but this can straightforwardly be extended to non-square images.)
The affine linear mapping is then defined by a set of $I$ filters
$\omega_i \in \R^{m_\omega\times m_\omega\times J}$ where $\omega_i = \{ \omega_{i,j} \in \R^{m_\omega\times m_\omega} \}_{j=1}^J$, and biases $b\in\R^I$, where each output channel has one bias. The convolutional layer that maps between the two multichannel images $h^0$ and $h^1$ is then defined for each channel as 
\begin{equation}
\label{eqn:convLayer}
h^1_i = \varphi\left(\sum_{j\in \mathcal{J}} \omega_{i,j}\ast h^0_j + b_i \right)
\hspace{-0.3em} \text{ for each } i\in \mathcal{I},
\end{equation}
where $\ast$ denotes a discrete convolution (see Fig.\ \ref{fig:CNN}).
The set of parameters in this case is given by the coefficients of the filters $\omega_i$ and biases $b_i$. 
Each output channel has one scalar bias and the input and output of each channel are connected by one specific filter. Thus we could consider an analogy here: in a CNN, each channel in the convolutional layer, Eq.\ \eqref{eqn:convLayer}, acts similarly to a neuron in a fully connected layer, Eq.\ \eqref{eqn:denseLayer}, with the filter components $\omega_{i,j}$ analogous to the point-wise weights $C_{i,j}$; compare Figs.\ \ref{fig:ANN} and \ref{fig:CNN}. 
We also note that the convolutional layer, Eq.\ \eqref{eqn:convLayer}, could be written in the general form Eq.\ \eqref{eqn:oneLayer} by vectorising the input and representing the convolution as a (sparse) matrix.

An important feature of CNNs is the fact that each layer maps between multichannel images of the same (or similar) resolution.\footnote{In this paper, in common with much of the image processing literature, we use `image resolution' to refer to the number of pixels or voxels in an image, so an image with $128\times 128$ pixels has twice the image resolution of one with $64\times 64$ pixels. The same term is used in imaging physics with a different meaning, there referring to the smallest resolvable features in an image. For example, a blurry image consisting of a large number of pixels would have a high resolution in the terminology we use here, but a low resolution in the sense that the fine features in the image are not distinguishable.}
Therefore, a CNN is a natural choice to represent data-to-data or image-to-image mappings, rather than mappings between spaces with different dimensions such as data-to-image. Nevertheless, in many applications there are reasons why it might be desirable to downsample input images during the processing (eg. memory constraints, sparser representation, wider receptive field) and hence many architectures include downsampling operations, called pooling layers, which reduce the image resolution using mean or maximum filters, for instance. This will become clearer in the following section on network architectures, specifically in Sec.\ \ref{subsubsec:Rec_nPost}.
By combining convolutional and fully connected layers, we can define the majority of network architectures that are used in the literature for image reconstruction. The specific networks depend on the task for which they are employed and hence we will discuss the particular architectures later in this section. Let us now focus on how the network parameters are learned.

\subsubsection{The learning task}
\label{subsubsec:learning_task}
After defining the network architecture, the parameters of the network need to be determined. This is done by learning them from a set of training data. Before this can be done, we need to define the actual learning task that will determine the network's mapping properties. That is, we want train the network to perform a specific task, such as either reconstructing or denoising an image, or in other applications to perform segmentation or classification. More precisely, given a network $\Network_\param$ we need to find an optimal set of parameters $\param^*$, such that our network fulfils the desired mapping property, ie.\ it does what we want. The training of the network is nothing more than an optimisation problem to find the optimal set of parameters $\param^*$, which can be formulated in various ways as we will summarise shortly.
Specifically, we will consider the reconstruction task of recovering the initial pressure $f$ from the measurement data $g$ given the parametrised reconstruction operator $\ForwardOpInvLearned$, such that Eq.\ \eqref{eqn:paraInvProb} is fulfilled. 

\paragraph{Supervised training}
The first idea that comes to mind is to minimise a distance function between the desired output - the known ground-truth - and the actual output of the network. This leads to \emph{supervised training}, in which the optimisation problem is formulated with knowledge of a desired ground-truth in order to find the parameters of $\ForwardOpInvLearned$.
For the optimisation we need pairs of measurement data $g_\mathfrak{i}$ and corresponding ground-truth $f_\mathfrak{i}$ for $\mathfrak{i}=1,\dots,\mathfrak{I}$. The set of pairs $\{(g_\mathfrak{i},f_\mathfrak{i})\}_{\mathfrak{i}=1}^\mathfrak{I}$ is called the training set. Next we need to define how to measure the closeness of the resulting reconstruction. For that purpose one typically formulates a loss function in the $L^p$-norm, such as
\begin{equation}\label{eqn:lossFunc}
L_\param(f_\mathfrak{i},g_\mathfrak{i}) = \|\ForwardOpInvLearned(g_\mathfrak{i}) - f_\mathfrak{i}\|_p^p.
\end{equation}
The learning task is to find an optimal set of parameters $\param^*$ in the space of possible parameters $\Theta$ that minimises Eq.\ \eqref{eqn:lossFunc} with respect to the given training set
\begin{equation}\label{eqn:paramOptimisation}
\param^* = \arg\min_{\param\in\Theta} \frac{1}{\mathfrak{I}} \sum_{\mathfrak{i}=1}^\mathfrak{I}     
     \Loss_{\param}(\signal_\mathfrak{i},\data_\mathfrak{i}) ,
\end{equation}
In fact, one is not limited to loss functions of the form Eq.\ \eqref{eqn:lossFunc} and depending on the learning task other more suitable choices can be made.
Additionally, one can add regularisation terms to the loss function, either on the output of the network or even on the parameters, for instance requiring sparsity by minimising the $L^1$-norm $\|\param\|_1$, where the 1-norm here acts element-wise as usual.

Finding a set of optimal parameters $\param^*$ as formulated in Eq. \eqref{eqn:paramOptimisation} leads to an optimisation problem and hence can be solved with suitable optimisation techniques. Here, gradient based methods are typically used in DL, where the gradients for the update are computed via backpropagation \cite{rumelhart1986learning,lecun1989backpropagation}. The most common optimisation strategies are stochastic gradient methods, where the stochasticity refers to randomisation in the subset of training samples (batches), such as the popular adaptive moments estimation algorithm \emph{Adam}\cite{kingma2014adam}.

\paragraph{Alternative training regimes}
Although the majority of learned image reconstruction approaches applied to PAT to date have been fully supervised, one current direction within the DL community is the investigation of possible alternative training regimes. In particular, these are concerned with cases in which only a small number of input and ground-truth pairs are available. Such approaches are typically referred to as \emph{semi-supervised} or \emph{self-supervised} training. 
These developments will not be covered extensively in this review, but we will discuss some possible directions on how to move away from fully supervised training in the conclusions.
Roughly speaking, what these approaches have in common is that instead of requiring closeness to a known ground-truth for all data pairs, we define an auxiliary measure on the goodness of reconstructions. For instance, one could think of a data consistency term, $\|\ForwardOp\ForwardOpInvLearned (g) - g \|_2^2$, that is used in a similar way to the concept of cycle consistency in the computer vision community \cite{zhu2017unpaired}. Related directions use the concept of adversarial networks, in which a discriminator is used to evaluate how well reconstructions resemble `realistic' ones during the training procedure. 
\newline

In summary, regardless of the chosen training regime, defining the learning task leads to an optimisation problem, where we aim to find an optimal set of parameters for the network architecture with respect to a chosen measure and training set.

\subsection{Architectures for Learned Reconstruction}
\label{sec:architectures}

\begin{figure}[b!]
\centering
\includegraphics[width=0.75\textwidth]{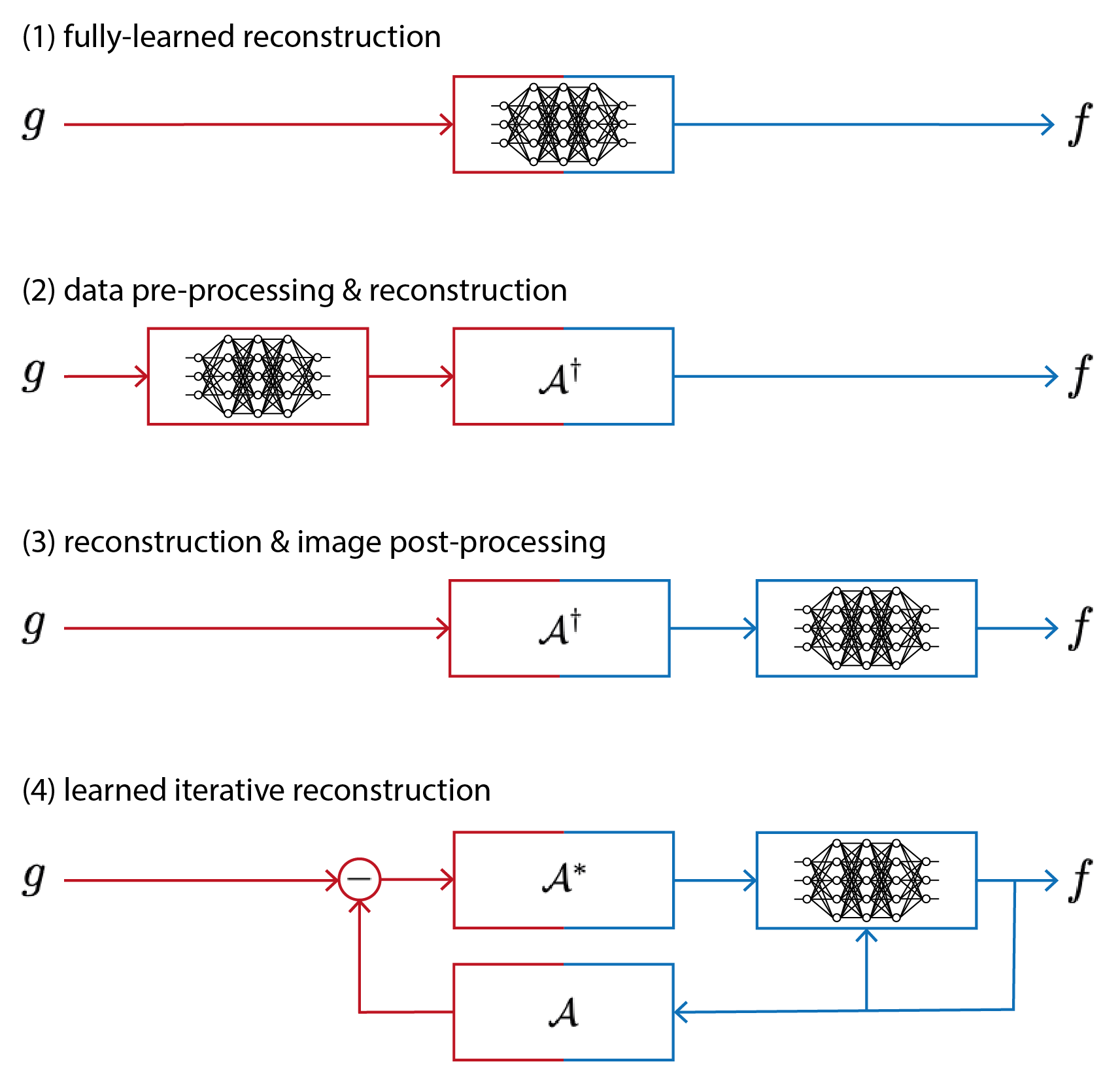}
\caption{\label{fig:DL_recon_approaches} Four different approaches to using a DL step (a network) within a PAT image reconstruction framework, ie.\ four types of learned reconstruction operator  $\ForwardOpInvLearned$.
(1) Fully-learned $\ForwardOpInvLearned = \Network_\param$, Eq.\ \eqref{eqn:fully_learned}. 
(2) Data pre-processing \& reconstruction 
$\ForwardOpInvLearned = 
\ForwardOpPseudoInv \circ \Network_\param$.
(3) Reconstruction \& image post-processing 
$\ForwardOpInvLearned = \Network_\param \circ \ForwardOpPseudoInv $, Eq.\ \eqref{eqn:postProc}.
(4) A learned iterative reconstruction based on gradient descent, Eq.\ \eqref{eqn:classicLGS}; see Fig.\ \ref{fig:learned_primal_dual} for another example of a learned iterative reconstruction scheme.
Red indicates the data space $\DataSpace$ and blue the image space, $\RecSpace$.}
\end{figure}

The reconstruction task in PAT can be addressed in various ways, as outlined in Sec.\ \ref{sec:PATrecon}, and since learning-based reconstruction algorithms are often inspired by these classical methods there is a wide range of possible approaches. In an attempt to classify learned reconstructions, we could divide the possible approaches into three classes by the number of times the physical model, the forward operator $\ForwardOp$ or a related operator, is involved in the reconstruction process: never, once, and multiple times. Four common strategies that are directly related to classical schemes are illustrated schematically in Fig.\ \ref{fig:DL_recon_approaches}. (The middle two strategies in Fig.\ \ref{fig:DL_recon_approaches} fall into the same class in this classification.)
In the following, we discuss these three classes of approach on a conceptual level, giving one example of a standard architecture for each. As mentioned above, we will concentrate here on the acoustic reconstruction problem, Eq.\ \eqref{eqn:paraInvProb}; extensions and applications to the optical reconstruction problem will be discussed in the literature review in Sec.\ \ref{subsec:qpat_review}.

\subsubsection{The fully-learned approach}
\label{sec:fullyLearned}
In the fully-learned approach, the whole learned reconstruction operator $\ForwardOpInvLearned$ is given by one network architecture, ie.\ 
\begin{equation}
\label{eqn:fully_learned}
\ForwardOpInvLearned := \Network_\param,
\end{equation}
where $\Network_\param:\DataSpace\to\RecSpace$.
At first sight, such fully-learned approaches seem promising as they eliminate the need for a potentially expensive reconstruction operator. However, the `no free lunch' concept applies here, as this improved reconstruction speed comes with a major limitation, which will be discussed below. First, though, we discuss the potential advantages. 
The forward operator $\ForwardOp: \RecSpace\rightarrow \DataSpace$ is non-local in nature. For instance, a point source $f\in \RecSpace$ has a spatially global effect on the measurement data $g\in \DataSpace$ (although it is localised in time). Similar non-locality of data-image relations are observed in most tomographic inverse problems. A fully connected layer has filter coefficients connecting each input to each output, and they can all be different, so it can cope with non-locality in the data-image relation, and can represent any linear mapping. In particular, the linear forward operator $\ForwardOp$ could be learned by a fully connected network. (It could even be learned by one fully connected layer with no nonlinearity, although that would just be $\ForwardOp$ represented as a dense matrix, which could be computed directly rather than learned.) Also, an inverse mapping such as the backprojection $\ForwardOp^\#$ in Eq. \eqref{eqn:backprojectionIntegral} can be learned by a dense layer, and in particular by a composition of dense layers with nonlinearities. In a CNN, on the other hand, a layer acts only locally, meaning an output pixel is only related to nearby input pixels. A fully connected network might therefore seem, at first glance, a better choice than a CNN for this task. (Although some ability to learn non-localities can be regained by using multi-scale CNNs such as the U-Net, as described below.) Another potential advantage of a fully learned approach, depending on the particular architecture, is that it can provide reconstructions quickly, with low latency, as no explicit model evaluation is required. 

The use of a fully connected network, however, has a major limitation similar to the problem faced by matrix representations of operators for high-dimensional problems, in that we need to learn a dense matrix of size $M\times T$, where $M$ is the total number of pixels, or voxels, and $T$ the product of the number of detectors and the number of sampling points in time. Let us for example consider a three-dimensional setting with $m\times m\times m = M$ voxels and $m\times m \times \mathbf{t} = T$ measurement points, where $m = 64$ and $\mathbf{t}=128$. Then a single dense layer, mapping between data and image space, represented in single precision (32 bit) would occupy $\sim\!\!500$GB. Thus, reasonable applications of this approach are limited in practice to two-dimensional problems. Also, the large number of learnable parameters necessitates a large training set for the training procedure to avoid overfitting to the training samples.
Additionally, as the fully connected layer associates each point in the input with the output nodes, the trained network depends specifically on consistent dimensions in the data space as well as image space, and hence the acquisition geometry. For PAT this means that one needs to train a separate network if the measurement setup changes, such as the number or location of the sensors, or the time-sampling points, or if there is a change in the sound speed distribution, for example.

\begin{figure}[th!]
\centering
\includegraphics[width=\textwidth]{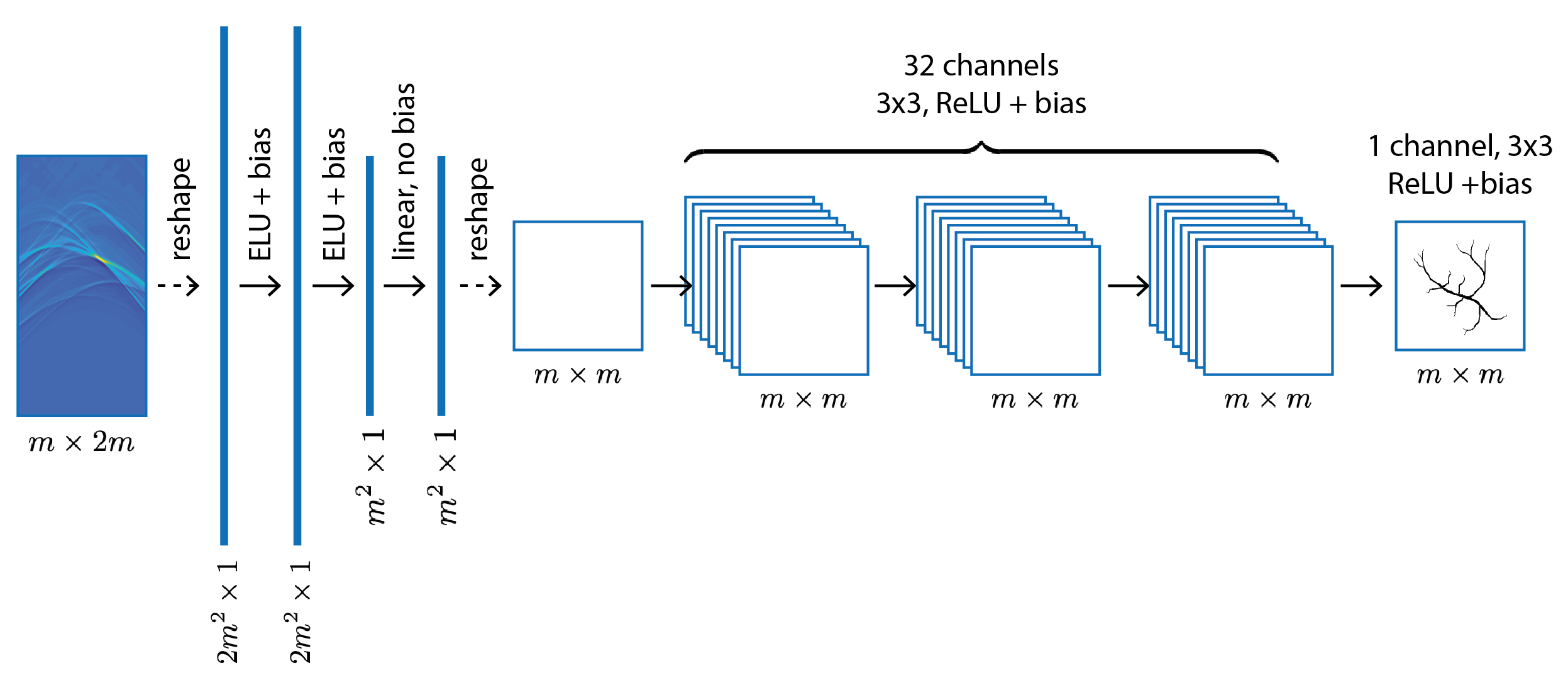}
\caption{\label{fig:AUTOMAP} 
A fully connected network similar to the AUTOMAP architecture\cite{Zhu2018automap}.
Three dense layers with ELU nonlinearity and bias are followed by a small CNN of 3 layers with 32 channels followed by a final CNN layer with 1 channel for the output. The ReLU nonlinearity on the final layer imposes a non-negativity constraint.}
\end{figure}

One could append a small CNN to the fully connected layers to exploit spatial features in the output from the fully connected layers to produce the final reconstructions. This thought leads to the architecture known as AUTOMAP\cite{Zhu2018automap}, originally devised for magnetic resonance imaging. A version of this kind of network is shown in Fig.\ \ref{fig:AUTOMAP}. In our case, the input to the network is given by the time-series of measured acoustic pressure. For the application of the fully connected layers, the input must be flattened or vectorised, ie.\ reshaped into a vector, before being passed to the network. The vector output of the fully connected layers is reshaped into an image and post-processed by a small convolutional network to produce the final output (Fig.\ \ref{fig:AUTOMAP}). This is just one architecture that uses fully connected layers and there are many variations on this theme, some of which are discussed in Sec.\ \ref{subsubsec:Convolutional_approaches}. 
This network can be thought of as a learned regularised backprojection operator (the fully connected layers) followed by a post-processing network to improve the image (the convolutional layers). This way of seeing the network leads us directly to the next approach, in which a classical backprojection operation is first performed with knowledge of the physical model and the network acts on the output in image space.

\subsubsection{Reconstruction and post-processing}
\label{subsubsec:Rec_nPost}

A major limitation of the fully-learned approach is the inflexibility with regard to the acquisition geometry and acoustic properties, ie.\ each network is specific to a fixed arrangement of the detectors and the sound speed. This can be overcome by using an explicitly model-based (classical) reconstruction from measured time-series to image space as an initial reconstruction step. This allows for potentially higher image resolutions to be used, as the memory burden of the fully connected layers has been removed, and potentially facilitates efficient initial reconstructions using approximate and computationally cheaper models.
In other words, if we substitute the fully connected part in Fig.\ \ref{fig:AUTOMAP} with an explicitly known reconstruction operator $\ForwardOpPseudoInv$, we arrive at the approach of an initial analytical reconstruction followed by a learned post-processing step. More precisely, let $\ForwardOpPseudoInv \colon \DataSpace \to \RecSpace$ be an analytically known reconstruction operator, that is ideally known to be robust (small changes in the input give small changes in the output). For example, $\ForwardOpPseudoInv$ could be $\ForwardOp^{\#}$ or $\ForwardOpAdj$ or another approximation to $\ForwardOp^{-1}$. Then one can train a CNN to remove the reconstruction artifacts that arise from using $\ForwardOpPseudoInv$ \cite{Kang2017,Jin2017,antholzer2019deep}. 
In the PAT case these artifacts can range from blurred out edges and noise to more severe undersampling and limited-view artifacts. The learned \emph{inverse mapping} is now given as   
\begin{equation}\label{eqn:postProc}
 \ForwardOpInvLearned =\Network_\param \circ \ForwardOpPseudoInv,
\end{equation}
where the network $\Network_\param:\RecSpace\to\RecSpace$ maps between the same space.
The main advantage in this approach lies in the analytical knowledge of the reconstruction operator, and so the network can be designed to focus instead on exploiting the structure in reconstruction artifacts in order to remove them.  Computationally, the evaluation time of the neural network is usually negligible and reconstruction times are typically limited by the complexity of the reconstruction operator. It is important to notice that the evaluation of the reconstruction operator can be decoupled from the training process and just used when creating the training data, and hence this approach is also advantageous in the training phase if the reconstruction operator is expensive to evaluate.

\begin{figure*}[th!]
\centering
\includegraphics[width=\textwidth]{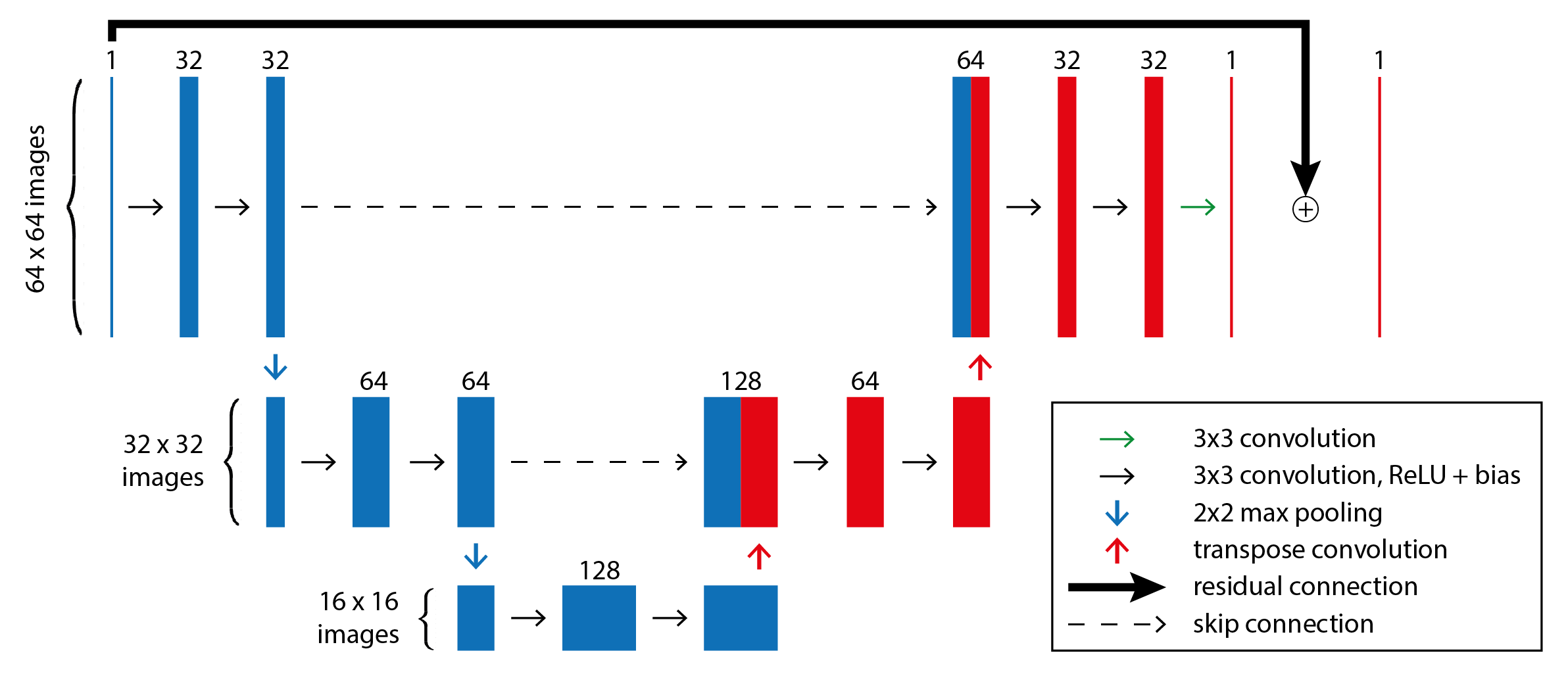}
\caption{\label{fig:UNet} A residual U-Net with 3 scales with two convolutional layers at each scale in the encoding and decoding paths, and concatenating skip connections. The number of channels in each layer is shown above it.}
\end{figure*}

For learned post-processing, typically one employs a high-capacity and particularly expressive network, ie.\ one with many layers and learnable parameters, that are capable of learning complicated image priors. The most prominent architectures for this application are based on the U-Net \cite{Ronneberger2015}, which can be roughly described as a multi-scale convolutional autoencoder. More precisely, instead of applying convolutions only on the full resolution image, the network includes down-sampling layers that reduce the image size in order to extract larger spatial features. The extracted coarse features are then subsequently upsampled to construct the final image. Intuitively, this process can be related to the principle of multi-resolution analysis, such as the wavelet decomposition \cite{daubechies1992ten,mallat1989theory}, where the input image is decomposed into a fine-to-coarse basis.
For image reconstruction tasks, instead of passing the reconstructed image $f_0  = \ForwardOpPseudoInv g$ directly through a network to produce the output image
\begin{equation}
f = \Network_\param (f_0) = \Network_\param (\ForwardOpPseudoInv g),
\end{equation}
the learning task is typically reformulated as a residual problem
\begin{equation}
\label{eqn:residualLearning}
f = f_0 + \Network_\param(f_0),
\end{equation}
in which a correction to the initial image is learned. This is motivated by the notion that the network can be used to identify noise and artifacts to remove from the image. Such networks are often referred to as residual networks, such as a residual U-Net \cite{Jin2017}, for which a basic architecture (with 3 scales\footnote{The classic U-Net architecture\cite{Ronneberger2015} has 5 scales; we use fewer here due to the small image size in the experiments.}) is illustrated in Fig.\ \ref{fig:UNet}. In the encoder part of the network, the left side, in each scale we employ two convolutional layers, followed by a downsampling of factor 2. This downsampling is done by a max-pooling operation, which takes the maximum value in a window of $2\times 2$ and reduces the image size. The numbers on top of each bar indicate the number of channels and, as can be seen, the number of channels is increased as the resolution is decreased. For the decoder part, we follow a similar approach of using 2 convolutions in each scale followed by an upsampling by factor two with a transposed convolutional layer. The final result is then added to the input via the residual connection. 
A particular design choice in the U-Net is the use of skip connections that connect the encoder and decoder parts at each scale by a concatenation. The reason for using these skip connections is two-fold. Computationally, they stabilise the training procedure by avoiding the problem of vanishing gradients in very deep networks. Additionally, the skip connections help to preserve the finer structures in the higher resolution scales.
It is interesting that even though CNNs are translation equivariant, the U-Net is able to learn local dependencies due to the decomposition into the coarser scales and the resulting large receptive field. 
Thus the post-processing approach proves powerful even in applications with strong local dependencies, such as limited-view problems. On the downside, such large capacity networks tend to overfit to the training data if training data is scarce, but still need considerably fewer training samples than the fully-learned approach. We will discuss this further in the experimental part in Sec.\ \ref{sec:comparisonMethods}. Additionally, the output depends solely on the quality of the initial reconstruction and the capability of the network to correct for these shortcomings. Therefore, without further modifications, we cannot guarantee that the reconstructed image is optimally consistent with the data - one possibility to overcome this will be discussed next.

\subsubsection{Model-based learned iterative reconstruction}
\label{subsubsec:Model-based_learned_iterative_reconstruction}
In order to improve data consistency of the reconstructions, one could use the forward operator multiple times in the reconstruction procedure and not only for an initial reconstruction. We call such approaches \emph{learned iterative} schemes, as neural networks are interlaced with evaluations of the forward operator $\ForwardOp$, its adjoint $\ForwardOpAdj$, and possibly other hand-crafted explicitly known operators. 
Typically, such learned iterative schemes outperform other learned reconstruction approaches in reconstruction quality  \cite{Adler2017,Hammernik2018learning,hauptmann2018model,Adler2018}, but come with a higher computational complexity. We also observe that this allows for the use of smaller networks, as the reduced network capacity is compensated for by providing more informative inputs to the network. 
We will introduce the concept with a simple learned gradient-like scheme \cite{Adler2017,Putzky2017recurrent}.
For instance, minimising the data consistency term  $\mathcal{D}(\signal;\data) = \frac{1}{2} \bigl\|\ForwardOp \signal - \data \bigr\|_2^2$ in a gradient descent scheme, as in Eq.\ \eqref{eqn:data_consistency}, could be formulated as a network with the updates
\begin{equation}\label{eqn:GradDescAsNet}
\Network_\param(\signal, \nabla_f \mathcal{D}(\signal;\data)) := \signal - \param \nabla_f \mathcal{D}(\signal;\data) = \signal - \param \ForwardOpAdj(\ForwardOp\signal - \data). 
\end{equation}
Comparing with Eq.\ \eqref{eqn:gradient_descent} we see that the only learnable parameter of the network is the step length $\param \in \R$. 
Extending this idea, we can devise a learned gradient scheme by using a CNN $\Network_\param:\RecSpace\times\RecSpace\to \RecSpace$ to compute the update in Eq.\ \eqref{eqn:GradDescAsNet}, and iterate the process such that
\begin{equation}\label{eqn:classicLGS}
	\signal^{(n+1)} = \Network_{\param_n}\bigl(\signal^{(n)}, \ForwardOpAdj(\ForwardOp \signal^{(n)} - \data)\bigr),\  n=0,\ldots, N-1.
\end{equation}
Here, each network $\Network_{\param_n}$ has its own set of parameters. 
The iterative process in Eq.\ \eqref{eqn:classicLGS} then defines a reconstruction operator when stopped after $N$ iterates:
\begin{equation}\label{eqn:LGS_reconOp}
    \ForwardOpInvLearned(\data) := \signal^{(N)}
   \quad\text{where $\param = (\param_0,\ldots,\param_{N-1})$}, 
\end{equation}
with an initialisation, such as the adjoint of the measurement data $\signal^{(0)}=\ForwardOpAdj g$. The initialisation is essential in this approach as it maps from data space $\DataSpace$ to image space $\RecSpace$ whereas the networks only map from $\RecSpace$ to $\RecSpace$. Each network $\Network_{\param_{n-1}}$ is a \emph{learned updating operator} for the $n$\textsuperscript{th} iterate and we can see a conceptual similarity of Eq.\ \eqref{eqn:classicLGS} to the proximal gradient descent scheme in Eq.\ \eqref{eqn:proximalGrad}, which provides a way to interpret the learned updating operator similar to a proximal mapping.
Such learned iterative approaches are also known as \emph{model-based} learned reconstructions, as the learned reconstruction operator $\ForwardOpInvLearned$ repeatedly uses the explicit forward and adjoint operators. Clearly, this leads to an increased complexity depending on the number of operator evaluations required, but the additional knowledge supplied to the networks allows the use of smaller architectures to achieve similar, or even superior, reconstruction quality compared to the previously discussed approaches.

\begin{figure}[bt!]
\centering
\includegraphics[width=0.9\textwidth]{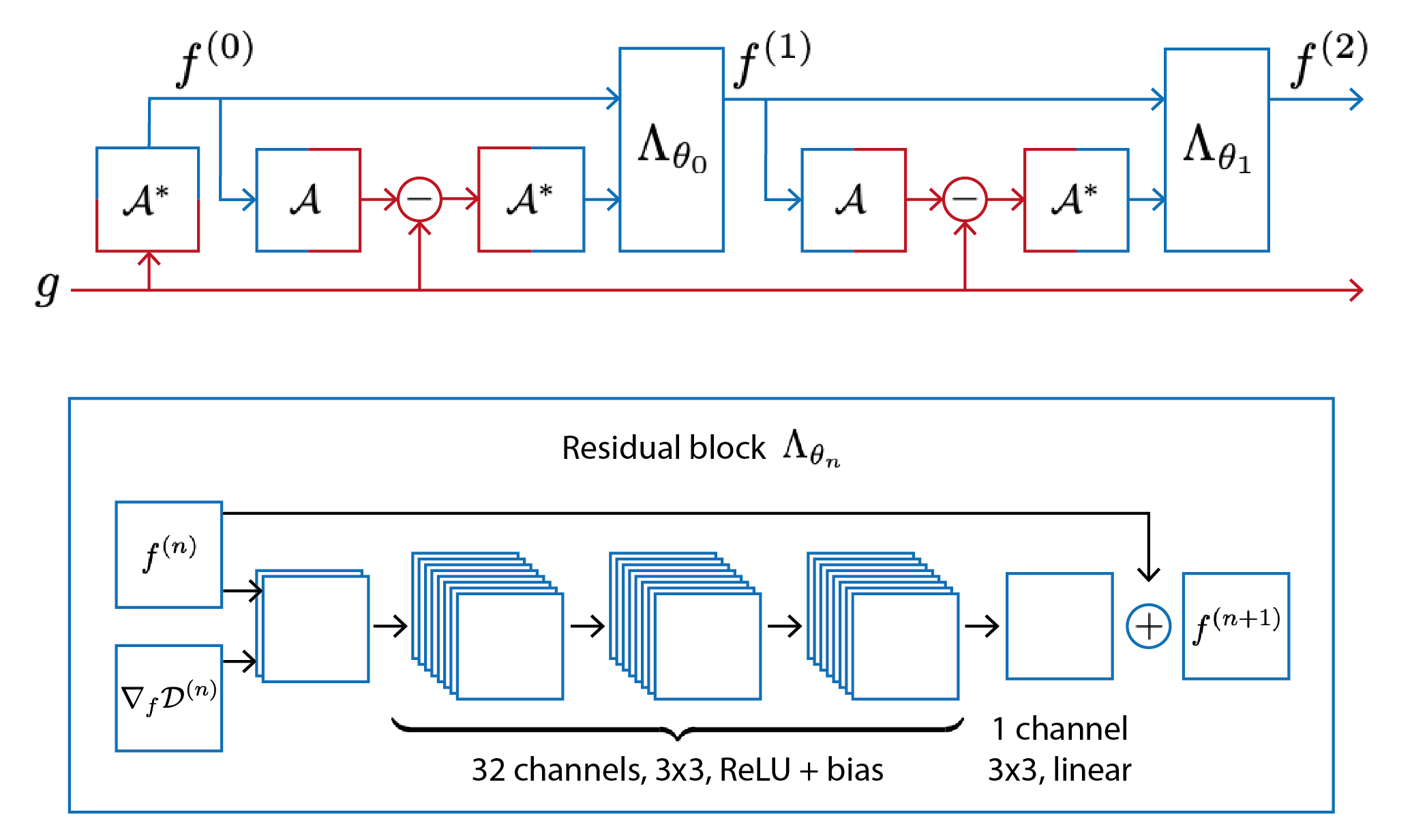}
\caption{\label{fig:residualLearnedIterative} (\emph{Top}) Unrolled network for two iterations of a learned iterative reconstruction as given by Eq.\ \eqref{eqn:residualLearnedIterative}. 
(Red indicates the data space, $\DataSpace$, and blue the image space, $\RecSpace$.)
(\emph{Bottom}) The architecture for the residual blocks $\Network_{\param_n}$, consisting of 3 convolutional layers of 32 channels with ReLU nonlinearity, followed by a linear convolutional layer with 1 channel to give the update for the next iterate. The network parameters $\param_n$ are different for each block (each iteration $n$). $\nabla_f\mathcal{D}^{(n)}$ denotes the gradient of the data consistency $\nabla_f \mathcal{D}(\signal^{(n)};\data) = \ForwardOpAdj(\ForwardOp \signal^{(n)} - g)$.}
\end{figure}

A basic network architecture for this task is illustrated in Fig.\ \ref{fig:residualLearnedIterative}.
The whole unrolled reconstruction is shown, for two iterations, in Fig.\ 
\ref{fig:residualLearnedIterative} \emph{(Top)}
and the architecture of the \emph{residual blocks}, based on the residual network ResNet\cite{he2016deep}, is shown in Fig.\ \ref{fig:residualLearnedIterative} \emph{(Bottom)}.
At each iteration, the current reconstruction, $\signal^{(n)}$, and the corresponding gradient of the data consistency term, $\nabla_f \mathcal{D}(\signal^{(n)};\data) = \ForwardOpAdj(\ForwardOp \signal^{(n)} - g)$, are concatenated into a two-channel input to the learned updating operator $\Network_{\param_n}$. This input layer is followed by 3 convolutional layers with 32 channels, $3\times 3$ filters, ReLU nonlinearity and bias.
The final layer (3x3, linear, no bias) reduces the 32 channels to a single residual update to be added to $\signal^{(n)}$ such that we can rewrite the learned update equation in Eq.\ \eqref{eqn:classicLGS} as
\begin{equation}
\label{eqn:residualLearnedIterative}
	\signal^{(n+1)} = f^{(n)} + \Network_{\param_n}\bigl(\signal^{(n)}, \ForwardOpAdj(\ForwardOp\signal^{(n)} - \data)\bigr).
\end{equation}
The last convolutional layer does not use a nonlinearity as the residual updates need to be able to be both positive and negative. After each residual block the intermediate result $\signal^{(n+1)}$ is used to compute the new gradient $\nabla_f \mathcal{D}(\signal^{(n+1)};\data)$, which is then passed on to the next residual block. 

By using smaller networks than in the post-processing approach, with both the current iterate and the gradient information as inputs, the networks rely less on prior knowledge from the training data and rather learn a desirable combination of both inputs. In fact, the gradient of the data consistency contains information on where the image needs improvement to fit the observed data. Just as importantly, smaller networks are less prone to overfitting and so require less training data. This aspect is further emphasised by a recent study that showed using explicitly known operators in the network architecture does indeed reduce the training error\cite{maier2019learning}. 
As the operator is used repeatedly in the reconstruction process, this also allows for some flexibility in acquisition geometry that the network can be applied to. Many extensions of the basic learned gradient scheme have been proposed in the literature and applications will be discussed in Sec.\ \ref{subsubsec:LPD}.

\subsection{Generating Training Data}
\label{subsec:training_data}
The training set will define the features learned by the network. This essentially defines the probability distribution describing our images of interest, in other words the prior of possible images $\pi(f)$ in the Bayesian framework, Eq.\ \eqref{eqn:Bayes}. This directly addresses the first point raised in Sec.\ \ref{sec:statisticalFramework} of how to learn a better prior. For many biomedical applications it is difficult to handcraft informative priors that represent structures of interest, eg.\ blood vessels, and so the alternative approach of learning a prior from a set of sample images is appealing. The choice of the training data set then becomes highly important as it defines the prior distribution.
A suitable training set will have two primary features: good representation of the relevant structures, and enough variety to represent the image distribution. 
In established medical imaging modalities, such as magnetic resonance imaging, one can use large databases of highly sampled gold standard reconstructions as ground-truth images. In PAT, such a database is not currently available, and, furthermore, many scanner geometries are not able even in principle to collect complete data. There are therefore essentially two options for creating a training set: simulate synthetic data as realistically as possible, or define a high-quality (albeit imperfect) reconstruction using a classical inversion method as the ground-truth.
It's important to emphasise that the training set, together with the training regime, determines the reconstruction quality one can expect. For instance, in a fully supervised setting with only reconstructions from classical inversion methods as the ground-truth, the network would not be expected to provide better reconstruction quality than the classical approach, although it may be able to compute the images more quickly. On the other hand, if augmented training sets or semi-supervised approaches are employed, more complicated priors might be learned and classical methods may be outperformed in reconstruction quality.

\paragraph{Synthetic training data}
The first step in the creation of synthetic training data is to define the ground-truth images $f_\mathfrak{i}\in \RecSpace$ for $\mathfrak{i}=1,\dots,\mathfrak{I}$. From these ground-truth images we can then simulate the corresponding synthetic measurement data $g_\mathfrak{i}\in \DataSpace$ according to Eq.\ \eqref{eqn:acoustic_operator}. Note that this includes the simulation of measurement noise. The pairs of ground-truth image and synthetic measurement data then define the training set $\{ (g_\mathfrak{i},f_\mathfrak{i})\in \DataSpace\times \RecSpace \}_{\mathfrak{i}=1}^{\mathfrak{I}} $. 
In PAT we are often interested in imaging vasculature, so we need a way to create a large enough set of images with relevant vessel structures. A standard way to obtain such structures is to use other imaging modalities that provide images or volumes of vessels and then to segment them to extract the relevant vessels as a ground-truth image. For the following experiments we designed two datasets from different image databases. The first set was created from a set of lung CT scans\footnote{ELCAP Public Lung Image Database: \url{http://www.via.cornell.edu/lungdb.html}} via vessel segmentation and projection to two dimensions, and the second set was taken from retina scans\footnote{DRIVE: Digital Retinal Images for Vessel Extraction: \url{https://drive.grand-challenge.org/}} with a segmentation provided. As Fig.\ \ref{fig:vesselImages} shows, these two sets have very different characteristics, one having piece-wise constant features the other smoother features, in other words the prior distributions $\pi(f)$ are different. As we will see, this difference will have a major impact on the reconstructions obtained depending on how the two training sets are used in the training and testing.
(In this particular case, as an alternative to the segmentation of vessel structures from other modalities, one could imagine creating ground-truth images by, for example, using vessel growing algorithms\cite{scianna2013review} to create a large set of synthetic training data.)

{\newcommand{\showpic}[2]{%
\begin{tikzpicture}%[spy using outlines={circle, magnification=3, 
%size=2.5cm, connect spies}]%
\draw (0,0) node [anchor=south] {\phantom{f}#1\phantom{g}};%
\draw (0,0) node [anchor=north] {\includegraphics[width=0.3\linewidth]{images/#2_true_5.png}};%
%\spy on (-1.075cm,-1.475cm) in node at (-.2cm, -4cm);
\end{tikzpicture}\hspace*{-2mm}%
}

\begin{figure}[ht!]
\centering
\showpic{Piece-wise phantom}{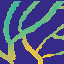}%
\hspace{1cm}
\showpic{Smooth phantom}{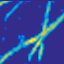}%

\caption{\label{fig:vesselImages} Example images for the two data sets used in the experimental section. Here the left phantoms promote piece-wise constant or linear features, whereas the right dataset promotes smoother features.}
\end{figure}}

% \begin{figure}[ht!]
%     \centering
%     \includegraphics[width =0.3\linewidth]{images/ves_true_5.png}
%     \includegraphics[width =0.3\linewidth]{images/seg_true_12.png}
%     \label{fig:vesselImages}
% \end{figure}

\paragraph{Experimental training data}
An alternative to synthetic training data is to use measured  data for the creation of the training set, ie.\ start with measurement data $g_\mathfrak{i}\in \DataSpace$ and create a reference reconstruction $f_\mathfrak{i}\in \RecSpace$. In this scenario, ideally one would have complete measurement data available that can be used to create a high quality reference reconstruction, for instance with a variational approach, Sec.\ \ref{subsubsec:acoustic_variational}. Then one can either train a network on the pairs of $(g_\mathfrak{i},f_\mathfrak{i})$ to speed up reconstruction times, or one can retrospectively undersample the measurement data to obtain $\widetilde{g}_\mathfrak{i}$ and train with pairs $(\widetilde{g}_\mathfrak{i},f_\mathfrak{i})$ to improve reconstructions from undersampled measurements. In our experience we have found that in the application to real measurement data it is essential to include some experimental data in the training procedure, as structures and noise can vary significantly from synthetic to experimental data. 

\paragraph{Transfer training}
A third option is to combine synthetic and experimental data. This is usually a good idea if one does not have sufficient measurement data available. Here, one can exploit a concept known as \emph{transfer training} or update training. We refer the reader to two discussions on the topic\cite{erhan2010does,yosinski2014transferable}. In our case, the underlying idea is to perform pre-training with a large set of synthetic training data that represents a good prior for the targets we are interested in. Then, after the first training phase on the synthetic data, we can update the obtained parameters with a shorter training on the limited set of available measurement data. This fine-tunes the network parameters to the characteristics of the experimental data, for instance by adjusting threshold capabilities to the noise level. This update is typically done with a reduced learning rate. Sometimes, rather than updating all parameters of the network, the majority are fixed and only the first and/or last layers are updated with the new data.  

Finally, we remark that here self-supervised training regimes, as mentioned in Sec.\ \ref{subsubsec:learning_task}, might be promising in the transition to experimental measurement data, although this area has not yet been widely explored.

\subsection{Comparison of Learned Image Reconstruction Approaches}
\label{sec:comparisonMethods}
\begin{figure}[ht!]
    \centering
    \includegraphics[width=0.75\linewidth]{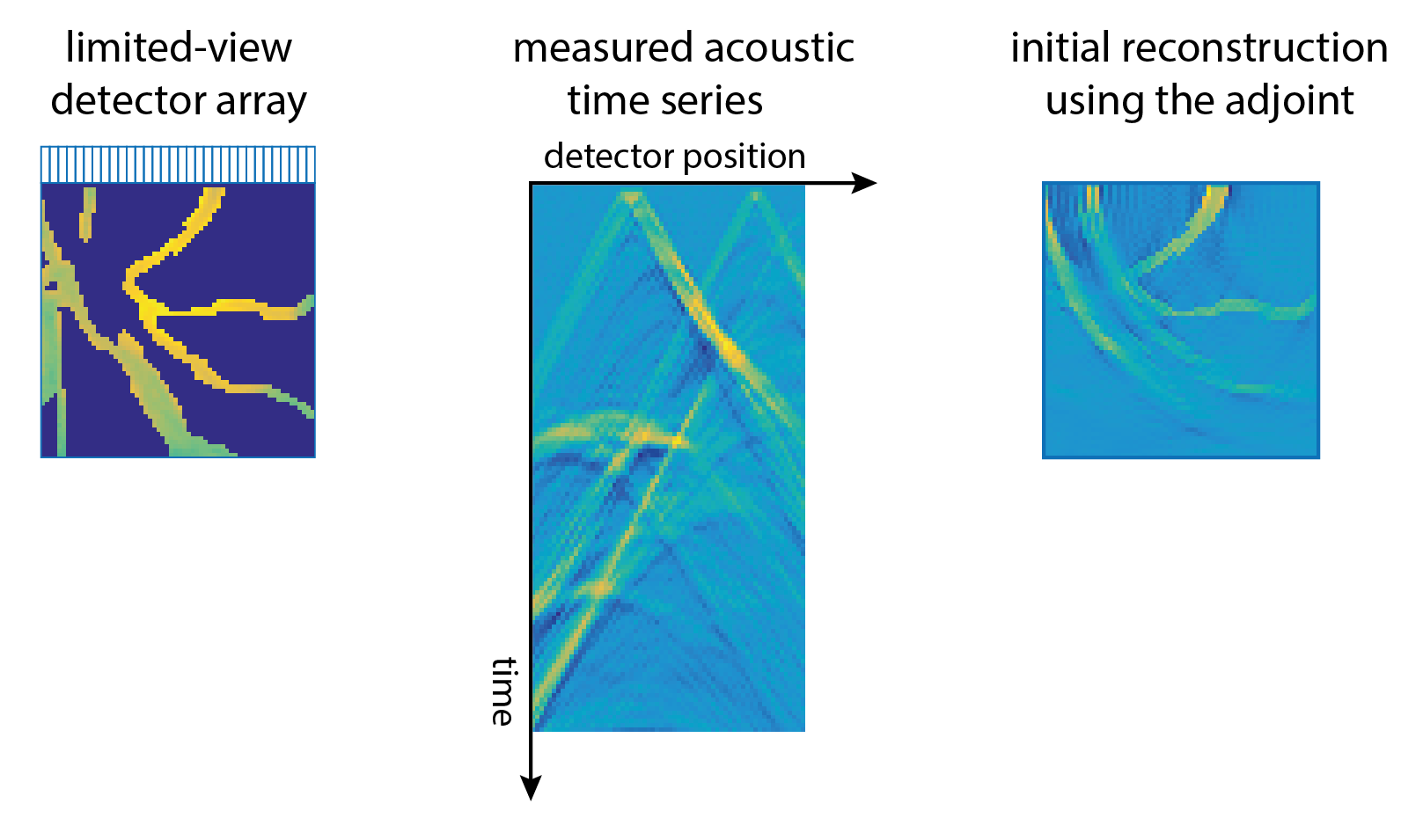}
    \caption{Illustration of the experimental setup for the examples. (\emph{Left}) We consider here a limited view geometry in two dimensions with a line detector on the top of the domain. (\emph{Center}) The corresponding time series measurements. (\emph{Right}) The initial reconstruction obtained by application of the adjoint to the measurements.}
    \label{fig:geometry}
\end{figure}

In this section, we use the synthetic data sets introduced above to examine the performance of the three learned image reconstruction approaches described in Sec.\ \ref{sec:architectures} with respect to accuracy and robustness.
We consider a 2D limited-view scenario with a line detector at the top of the domain, as illustrated in Fig.\ \ref{fig:geometry}. For simplicity, we create a matrix representation of the acoustic forward model as described in Sec.\ \ref{sec:matrixRep} by sampling the forward operator with k-Wave\cite{treeby2010kWave}. Since this section serves in part as a tutorial, we will describe the individual steps required to set up the experiments in detail.

\paragraph{Experimental design}
Here we describe the steps necessary to train and evaluate the `reconstruction and post-processing' reconstruction approach as outlined in Sec.\ \ref{subsubsec:Rec_nPost}. This will cover all the concepts needed to set up the examples for the other learned reconstruction approaches.
\begin{enumerate}
    \item \emph{Data acquisition geometry and definition of the forward operator:} The essential first step is the definition of the imaging setup under consideration, which also defines the forward operator $\ForwardOp$. Here we chose a limited-view planar acquisition geometry in a two-dimensional domain, see Fig.\ \ref{fig:geometry}, and we use k-Wave\cite{treeby2010kWave} for the PA time series simulation in MATLAB. For flexibility in the training data creation and reconstruction, we use a matrix representation of the forward operator by sampling each pixel in the image domain; this is done in the script: {\verb createForwMat.m }
    The resulting matrix is saved to disk\footnote{To enable readability by Python, we added the flag {\Verb'-v7.3' } in MATLAB when saving the mat-file. Additionally, when loading the matrix in Python, it will be transposed, so we transpose the matrix before saving.} so it can be loaded within the learning framework in Python for data creation and reconstruction in the following.
    
    \item \emph{Training data creation:} 
    %For this example, we consider the post-processing approach of cleaning and improving an initial reconstruction. 
    First we need to choose the set of ground-truth images $\{f_\mathfrak{i}\}_{\mathfrak{i}=1}^{\mathfrak{I}}$ we want to use for the training, ie.\ one or both of the sets described above and shown in Fig.\ \ref{fig:vesselImages}. Then we create the corresponding synthetic measurement data by using the matrix form of $\ForwardOp$ to create $g_\mathfrak{i} = \ForwardOp f_\mathfrak{i} + \noise_\mathfrak{i}$, where $\noise_\mathfrak{i}$ is normally distributed noise added to the measurements with standard deviation of 1\% of the \revision{maximum} measurement amplitude.    
    We then create the initial reconstruction using the adjoint, such that $f^\mathrm{rec}_\mathfrak{i} = \ForwardOpAdj g_\mathfrak{i}$. Recall that in the matrix representation, the adjoint corresponds simply to the transpose. 
    The training set $\{(f^\mathrm{rec}_\mathfrak{i},f_\mathfrak{i})\}_{\mathfrak{i}=1}^{\mathfrak{I}}$ for supervised training has now been generated. Test and validation sets can be created in the same way. These preparations are done in the script: {\verb callNetwork.py }
    \noindent
    For the other experiments, with the fully-learned approach and learned iterative schemes, we can just use the generated data as input and hence the set is given by $\{(g_\mathfrak{i},f_\mathfrak{i})\}_{\mathfrak{i}=1}^{\mathfrak{I}}$.
    
    \item \emph{Network selection and training regime:}
    Given the training set, we can now set up the network and define the training regime. All the relevant functions can be imported from the supplied script:  {\verb PATnets.py }
    For this case we choose a classic residual U-Net\footnote{The code package provides a set of standard architectures that can be called instead of the U-Net.} for the network $\Network_\param$, and as a loss function the classic squared $\ell^2$-norm for supervised training. Then the optimisation problem reads as
    \begin{equation}
    \param^* = \argmin_{\param\in\Theta} \frac{1}{\mathfrak{I}}\sum_{\mathfrak{i}=1}^\mathfrak{I}\|\Network_\param (f^\mathrm{rec}_\mathfrak{i}) - f_\mathfrak{i}\|_2^2.
    \end{equation}
    The optimisation is performed with the Adam algorithm, initial learning rate $10^{-4}$, batch size 4, and a total of $5\cdot10^4$ iterations.
    
    \item \emph{Training supervision and evaluation:}
    During the training, we need to ensure both that our cost function is minimised and converges, and also that the learned parameters generalise to other samples not contained in the training set. To achieve this we use the visualisation support provided by tensorboard\footnote{We refer to the online instructions at: \url{https://www.tensorflow.org/tensorboard}}, which can then be called locally from a web browser to provide real time supervision of the training procedure. After the training, the optimal set of parameters $\param^*$ can be saved for later evaluation, or a direct evaluation can be performed. For evaluation, we load the test set and record the average reconstruction quality.
\end{enumerate}

\paragraph{Reconstructions: robustness and generalisation}
In this Section we will evaluate the three reconstruction approaches described in Sec.\ \ref{sec:architectures}, fully-learned, post-processing, and learned iterative reconstruction, following the four-step process outlined above. In particular, we will examine how these methods compare in generalisability with respect to changes in the data sets they are trained on. For that purpose we will consider three scenarios for the training and test sets:
\begin{itemize}
    \item[i.)] \emph{Consistent sets:}
    Trained on the retina data (Fig.\ \ref{fig:vesselImages}b) \revision{of 1000 samples} and tested on a separate but consistent test set from the retina data \revision{with 151 samples}. This corresponds to a scenario where the priors are the same, $\pi_{\rm test}(f)=\pi_{\rm train}(f)$.
    \item[ii.)] \emph{Different test set:} Trained on the retina data (Fig.\ \ref{fig:vesselImages}b) and tested on the lung CT set (Fig.\ \ref{fig:vesselImages}a) \revision{with 151 samples}. This corresponds to a scenario where the priors are different, $\pi_{\rm test}(f)\neq \pi_{\rm train}(f)$.
    \item[iii.)] \emph{Combined set:} Trained on both data sets \revision{with a total of 3760 samples} and tested on a separate combined test set \revision{with 308 samples}. Here the priors are consistent, but more complicated than in (i). We emphasise that this training set is also larger.
\end{itemize}
We trained the three reconstruction approaches for each scenario using the same training regime, as outlined in Sec.\ \ref{sec:comparisonMethods}, with minor tuning as necessary to ensure the parameters are close to optimal. This ensured the results were representative and allowed useful conclusions to be drawn. Nevertheless, as we will see below, not all of these architectures are conceptually the right choice for the scenarios under consideration and it was not possible to improve the performance significantly through further parameter tuning. Note that the fully-learned approach uses a regulariser of the learned parameters $\|\theta\|_1$ to reduce overfitting.

{\newcommand{\showpic}[2]{%
\begin{tikzpicture}%[spy using outlines={circle, magnification=3, 
%size=2.5cm, connect spies}]%
\draw (0,0) node [anchor=south] {\phantom{f}#1\phantom{g}};%
\draw (0,0) node [anchor=north] {\includegraphics[width=0.22\linewidth]{images/seg_#2_12.png}};%
%\spy on (-1.075cm,-1.475cm) in node at (-.2cm, -4cm);
\end{tikzpicture}\hspace*{-2mm}%
}

\begin{figure}[ht!]
\centering
\showpic{Phantom}{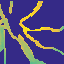}%
\hfill
\showpic{Fully-learned}{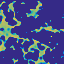}%
\hfill
\showpic{Post-processing}{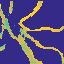}%
\hfill
\showpic{Learned iterative}{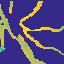}%
\caption{ \label{fig:reconsCase1} Reconstructions obtained for test case i.) with consistent priors $\pi_{\rm test}(f)=\pi_{\rm train}(f)$. Trained and tested on the piece-wise constant phantoms. The fully-learned approach does not perform satisfactorily due to strong overfitting to the training data whereas the other two approaches are able to produce \revision{ quantitatively and qualitatively superior results, but still exhibit errors in the reconstruction.}}
\end{figure}}

% \begin{figure}[ht!]
%     \centering
%     \includegraphics[height =0.2\linewidth]{images/seg_true_12.png}
%     \includegraphics[height =0.2\linewidth]{images/seg_rec_FL_12.png}
%     \includegraphics[height =0.2\linewidth]{images/seg_rec_Unet_12.png}
%     \includegraphics[height =0.2\linewidth]{images/seg_rec_LGS_12.png}
%     \caption{Placeholder: Some recons for the first case: Phantom, Fully learned, U-Net, LGS}
%     \label{fig:reconsCase1}
% \end{figure}

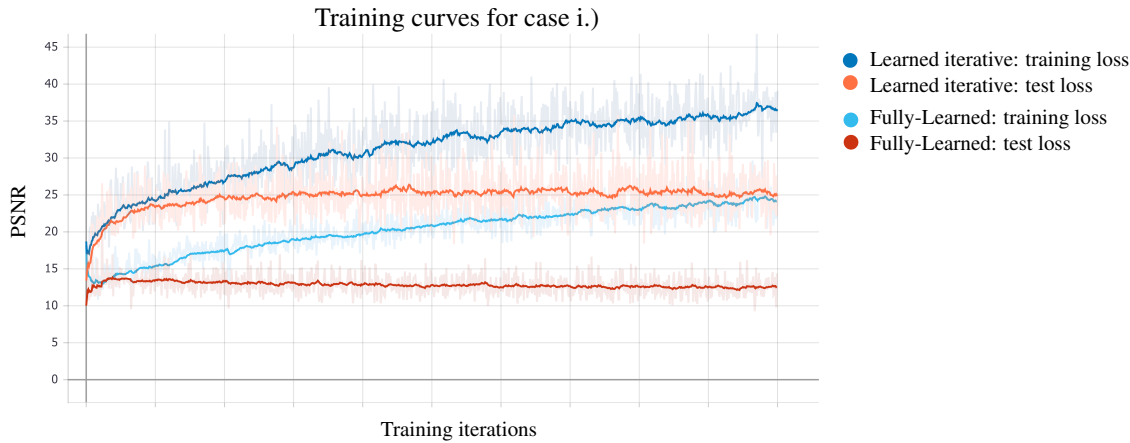
\begin{figure}[ht!]
    \centering
    \begin{picture}(375,155)
    \put(0,10){\includegraphics[width=300pt]{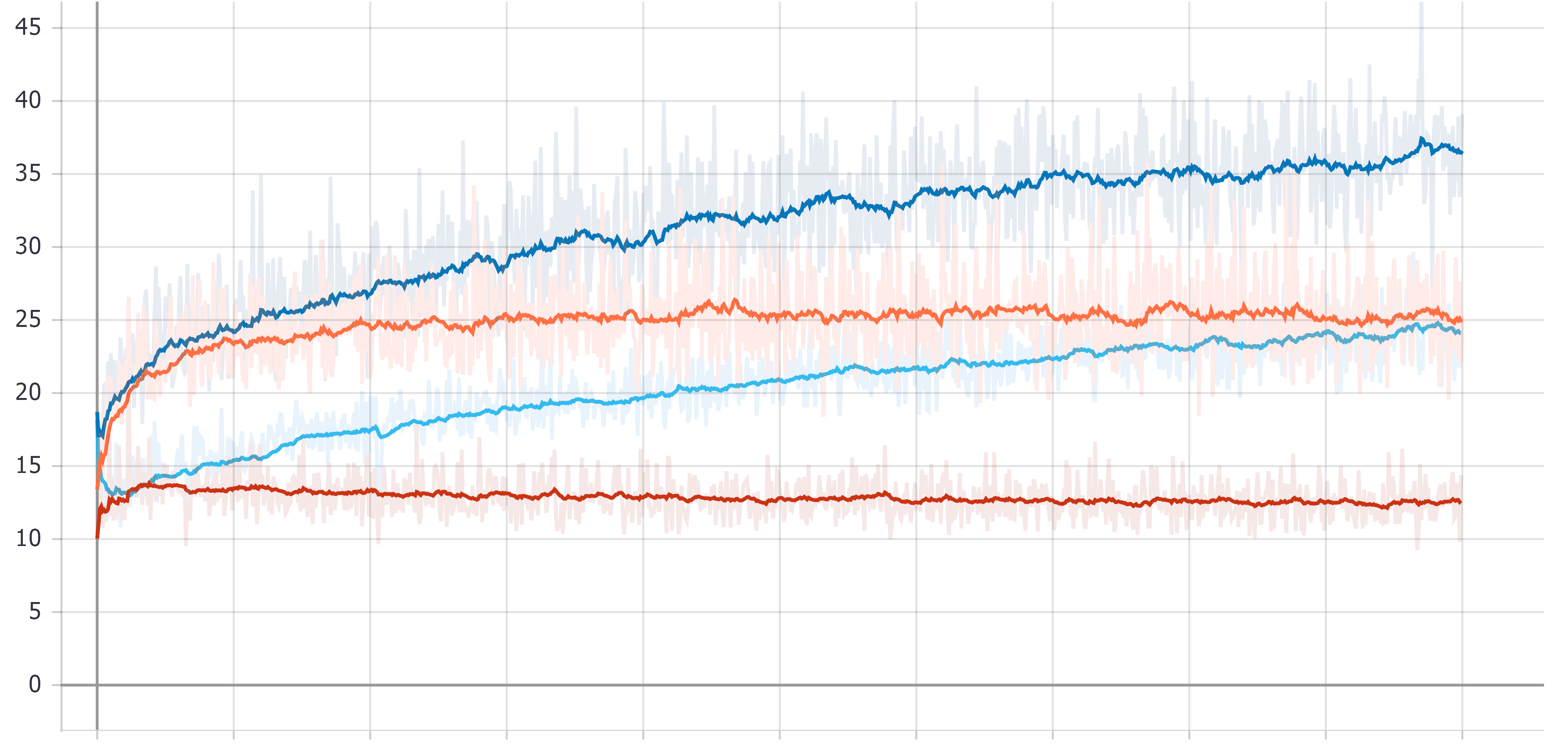}}
    \put(300,100){\includegraphics[height=50pt]{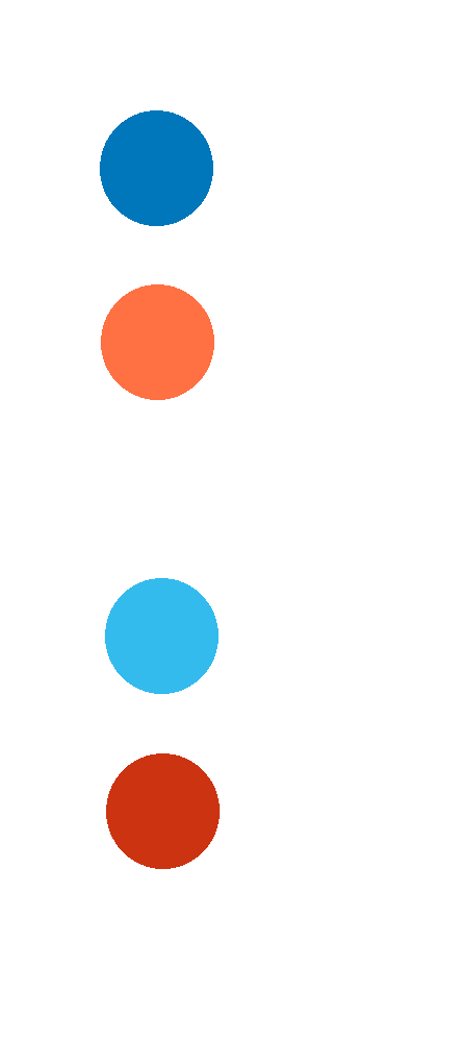}}
     \put(315,140){\scriptsize{Learned iterative: training loss}}
    \put(315,130){\scriptsize{Learned iterative: test loss}}
    
    \put(315,118){\scriptsize{Fully-Learned: training loss}}
    \put(315,108){\scriptsize{Fully-Learned: test loss}}
    
    \put(-10,75){\rotatebox{90}{\scriptsize{PSNR}}}
    \put(130,0){\scriptsize{Training iterations}}
    
    \put(105,155){\footnotesize{Training curves for case i.)}}
    \end{picture}
   
    \caption{Training curves for the fully-learned approach and learned gradient descent in comparison for case i.), exported from the tracking tool tensorboard. While both approaches have a tendency to overfit the training data during training, the fully-learned approach does suffer more compared to the learned iterative reconstruction.}
    \label{fig:FL_train}
\end{figure}

Let us first discuss the obtained reconstructions from a visual perspective. The results obtained for the first case i.) are displayed in Fig.\ \ref{fig:reconsCase1}. Most striking here is the result obtained by the fully-learned approach, which clearly falls short in reconstruction quality compared to the other two approaches. We observe that this is primarily due to the limited size of the training data and hence the network strongly overfitting, even though we use regularisation to reduce this. This is clear from the training error plots shown in Fig.\ \ref{fig:FL_train}. 
In contrast, the other two approaches correctly learned some form of representation of the prior $\pi_{\rm train}$ from the training data. Consequently, the reconstructions for case i.) are visually close to the ground-truth. In particular, we can see a good reconstruction quality close to the detector, but on the boundary where limited-view artifacts are stronger the reconstructions lose quality.

{\newcommand{\showpic}[2]{%
\begin{tikzpicture}%[spy using outlines={circle, magnification=3, 
%size=2.5cm, connect spies}]%
\draw (0,0) node [anchor=south] {\phantom{f}#1\phantom{g}};%
\draw (0,0) node [anchor=north] {\includegraphics[width=0.22\linewidth]{images/ves_#2_5.png}};%
%\spy on (-1.075cm,-1.475cm) in node at (-.2cm, -4cm);
\end{tikzpicture}\hspace*{-2mm}%
}

\begin{figure}[ht!]
\centering
\showpic{Phantom}{true}%
\hfill
\showpic{Fully-learned}{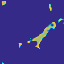}%
\hfill
\showpic{Post-processing}{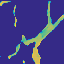}%
\hfill
\showpic{Learned iterative}{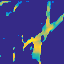}%
\caption{ \label{fig:reconsCase2} Reconstructions obtained for test case ii.) with inconsistent priors $\pi_{\rm test}(f)\neq\pi_{\rm train}(f)$. Trained on the piece-wise constant phantoms and tested on smooth phantoms. All methods struggle to produce satisfactory results and one can see that the piece-wise constant prior from the training data is reproduced by each method.}
\end{figure}}

% \begin{figure}[ht!]
%     \centering
%     \includegraphics[height =0.2\linewidth]{images/ves_true_5.png}
%     \includegraphics[height =0.2\linewidth]{images/ves_rec_FL_5.png}
%     \includegraphics[height =0.2\linewidth]{images/ves_rec_Unet_5.png}
%     \includegraphics[height =0.2\linewidth]{images/ves_rec_LGS_5.png}
    
%     \caption{Placeholder: Some recons for the second case: Phantom, Fully learned, U-Net, LGS }
%     \label{fig:reconsCase2}
% \end{figure}

For the second case ii.), the results are shown in Fig.\ \ref{fig:reconsCase2}. It is clear, on the first sight, that the networks produce results according to the learned piece-wise prior from the training data, as one would expect. Additionally, all the algorithms show a deterioration in reconstruction quality from the consistent case i.). For the post-processing and learned iterative scheme features close to the detector are to some extent correctly reconstructed, but they struggle further away. The fully-learned approach, due again to strong overfitting, produces a result with very limited resemblance to the ground-truth. One interesting feature is that the networks, and especially the learned iterative scheme, tend to smear out features where there is uncertainty in the reconstruction.

{\newcommand{\showpic}[2]{%
\begin{tikzpicture}%[spy using outlines={circle, magnification=3, 
%size=2.5cm, connect spies}]%
\draw (0,0) node [anchor=south] {\phantom{f}#1\phantom{g}};%
\draw (0,0) node [anchor=north] {\includegraphics[width=0.22\linewidth]{images/comb_#2_5.png}};%
%\spy on (-1.075cm,-1.475cm) in node at (-.2cm, -4cm);
\end{tikzpicture}\hspace*{-2mm}%
}

\begin{figure}[ht!]
\centering
\showpic{Phantom}{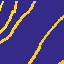}%
\hfill
\showpic{Fully-learned}{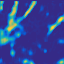}%
\hfill
\showpic{Post-processing}{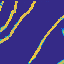}%
\hfill
\showpic{Learned iterative}{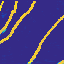}%
\caption{ \label{fig:reconsCase3} Reconstructions obtained for test case iii.) with consistent priors $\pi_{\rm test}(f)=\pi_{\rm train}(f)$. Trained and tested on combined phantoms with piece-wise constant as well as smooth features. The reconstruction quality of the fully-learned approach improved slightly compared to the other test cases due to the larger training set, but it is clearly outperformed by both methods that use the model in the reconstruction pipeline.}
\end{figure}}

% \begin{figure}[ht!]
%     \centering
%     \includegraphics[height =0.2\linewidth]{images/comb_true_5.png}
%     \includegraphics[height =0.2\linewidth]{images/comb_rec_FL_5.png}
%     \includegraphics[height =0.2\linewidth]{images/comb_rec_Unet_5.png}
%     \includegraphics[height =0.2\linewidth]{images/comb_rec_LGS_5.png}
    
%     \caption{Placeholder: Some recons for the third case: Phantom, Fully learned, post-processing U-Net, LGS }
%     \label{fig:reconsCase3}
% \end{figure}

In the final case iii.) the training samples are combined and so the size of the training data set is increased. The results are shown in Fig.\ \ref{fig:reconsCase3}. There is a clear improvement over the second case, as the test data is consistent with the mixed prior, and both approaches that use a model in the reconstruction do fairly well in reconstructing the target. There is a slight influence of the mixed prior visible in the results, as the reconstructions for the piece-wise constant phantom exhibit some smoother features related to the smooth phantoms. Lastly, the fully-learned approach seems to struggle with the mixed priors, but the reconstruction is still arguably closer to the ground-truth than in the other cases, as the increased training data reduced the overfitting. Nevertheless, the result is still not satisfactory.

\begin{table*}[ht!] 
\small
  \caption{Quantitative values for the three test cases in SSIM and PSNR}
  \begin{footnotesize}
  \begin{center}
    \begin{tabular}{l|c|c|c|c|c|c}
     %\multicolumn{5}{c}{\sc Reconstructed Average Values}\\
    %\hline
    &  \multicolumn{2}{c|}{case i.)} & \multicolumn{2}{c|}{case ii.)} & \multicolumn{2}{c}{case iii.)} \\
 & SSIM &  PSNR & SSIM &  PSNR & SSIM &  PSNR \\
 \toprule 
Fully-learned & 0.624\,$\pm$0.181  & 13.34\, $\pm$3.54 & 0.491\,$\pm$0.182  & 16.00\, $\pm$2.19 &  0.592\,$\pm$0.170  & 17.34\, $\pm$3.95  \\
Post-processing & 0.946\,$\pm$0.042  & 21.57\, $\pm$5.85 & 0.570\,$\pm$0.185  & 16.56\,$\pm$2.23  & 0.902\,$\pm$0.133  & 23.22\,$\pm$5.07  \\
Learned iterative & 0.983\,$\pm$0.025  & 28.76\, $\pm$8.10 & 0.679\,$\pm$0.165  & 18.28\, $\pm$2.35 & 0.949\,$\pm$0.089  & 28.04\, $\pm$5.82 \\
    \bottomrule
    \end{tabular}%
  \label{table:Quant}%
  \end{center}
  \end{footnotesize}
\end{table*}%

These observations are supported by the quantitative values shown in Table \ref{table:Quant}, which shows the mean and standard deviation over the whole test data for each case. We computed the peak signal-to-noise ratio (PSNR), which is a logarithmically relative root mean squared error and hence related to the quantity we minimised in the training. Additionally, we computed the structural similarity index measure (SSIM) as an indication of the perceived similarity in the reconstructions. We can see that the post-processing and learned iterative schemes perform better in this test, but with a strong deterioration when changing the prior distribution for the test data, as is also seen visually for case ii.). For case iii.) neither method using a model showed a strong improvement over case i.), in fact both methods deteriorate in terms of SSIM as the priors are not perfectly reproduced, although PSNR is either stable or improves for the post-processing. For the fully-learned approach, PSNR improved considerably with the larger training size, although SSIM slightly deteriorated, most likely due to the difficulty in reproducing the prior correctly. This is further an indicator of the large quantity of data needed for the fully learned approach to work well. Similar observations are made by Baguer et al.\cite{baguer2020computed}: in their overview they show that a fully learned approach only performs well with large amounts of data, which explains in parts the poor performance in this limited data setting.

\section{Deep Learning in PAT - Literature Review}
\label{sec:reviewPart}

The majority of the journal articles in which DL techniques have been applied to PAT image reconstruction are concerned with the acoustic part of the reconstruction and there are fewer papers tackling the optical part. Most of the subsections below will therefore focus on the acoustic reconstruction. The papers concerned with DL approaches applied to the optical reconstructions will be reviewed in Sec.\ \ref{subsec:qpat_review}. 
We also draw attention to related reviews on the matter of optical imaging and/or learned image reconstruction\cite{arridge2019solving,an2020application,zhang2019brief,sivasubramanian2020deep}.

\subsection{Post-Processing}
\label{subsec:post-proc}
% \begin{itemize}
%     \item Haltmeier group: Antholzer (mainly post-proc) \cite{antholzer2018deep,antholzer2019deep,antholzer2018photoacoustic} Schwab (learned backproj) \cite{schwab2018real,schwab2019learned}
%     \item Basic U-Net and extensions: Dense \cite{guan2019fully}, fancy in vivo \cite{davoudi2019deep}, comparison to fully-learned \cite{waibel2018reconstruction}, leaky t \cite{shan2019accelerated}
%     \item Temporal with recurrent \cite{anas2018enabling}
% \end{itemize}

Early approaches to learned image reconstruction concentrated on the 
reconstruction-and-post-processing approach as outlined in Sec.\ \ref{subsubsec:Rec_nPost}. The work by Antholzer et al.\cite{antholzer2018deep,antholzer2019deep,antholzer2018photoacoustic} investigated the approach of using filtered backprojection 
(Sec.\ \ref{subsubsec:backprojection}) to reconstruct an initial image and then train a U-Net, with 5 scales, to do post-processing. This was in a sparse and limited-view data setting, and followed the residual learning approach\cite{Jin2017} given by Eq.\ \eqref{eqn:residualLearning}. Similar to our observations in Sec.\ \ref{sec:comparisonMethods}, the authors report that consistent training and test data, ie. $\pi_{\rm test}(f)\approx\pi_{\rm train}(f)$,
is crucial for optimal performance of the trained network\cite{antholzer2019deep}; this seems to be more so in the case of limited-view detection geometries. This observation was confirmed and clearly demonstrated in the study by Guan et al.\cite{guan2019fully}, who proposed a dense U-Net to ameliorate this negative effect. 
Other extensions have been proposed too: using a leaky ReLU nonlinearity\cite{deng2019machine}, or using the first iterate of a model-based iterative approach (Sec.\ \ref{subsubsec:acoustic_variational}) instead of a backprojection-type reconstruction\cite{shan2019accelerated}. 
In\cite{awasthi2019pa}, the authors propose combining a reconstruction obtained with the adjoint with the first iterate of an iterative algorithm in a learned fusion process.

In comparison to other approaches, U-Net-based networks generally performed better than other architectures, eg.\ compared to a simple 3 layer CNN\cite{antholzer2018deep}, VGG\cite{deng2019machine}, and compared to applying U-Net directly to the measurement data $g$\cite{waibel2018reconstruction}, especially with respect to robustness. It is interesting that Antholzer et al.\cite{antholzer2018photoacoustic} compare their results to a classic $\ell^1$-regularisation approach for compressed sensing and report that when the system matrix is randomly sampled, and hence undersampling artifacts change as well, the classical variational approach clearly outperforms the network-based post-processing approach. This enforces the observation that consistent training and test data is needed for this approach to be successful.

\subsubsection{Application to in-vivo imaging}
In an extensive study, the U-Net-based post-processing approach was successfully applied to \emph{in vivo} measurements\cite{davoudi2019deep} and showed clear improvements over backprojection-based algorithms when the data was undersampled or detected over a partial aperture (limited-view problem).
Hariri et al.\cite{hariri2020deep} showed that this approach can improve \emph{in vivo} imaging when using low-fluence sources. The observation of improved visual performance for \emph{in vivo} applications was also reported in other studies\cite{farnia2020high,zhang2020new}.

\subsubsection{Extensions of the post-processing approach}
\label{subsubsec:extensions}
The primary problem with the post-processing approach is that the result depends on a network that is determined only by the information content of the training data and not the physics of the problem. To tackle this, Antholzer et al.\cite{antholzer2018deep,antholzer2019deep} propose a nullspace projection to ensure data consistency after post-processing. In other words, only components in the nullspace $\mathbf{N}(\ForwardOp)$ of the forward operator $\ForwardOp$ are added to the reconstruction and as such do not change the data consistency term, 
$\| \ForwardOp f - g  \|_2^2$.
The solution therefore takes the form
\begin{equation}
f = f_0 + \mathcal{P}_{\mathbf{N}(\ForwardOp)}\Network_\param (f_0),
\end{equation}
where $\mathcal{P}_{\mathbf{N}(\ForwardOp)}$ denotes the orthogonal projection to the null space.
In Schwab et al.\cite{schwab2018real,schwab2019learned}, the authors combine post-processing by a U-Net with a learning-based filter in the backprojection step ($\kappa$ in Eq.\ \eqref{eqn:UBP_2D}) to improve initial reconstructions from limited-view measurements.

Recently, LED-based excitation systems have become popular but because of their low power output many averages (thousands) are required to improve the signal-to-noise ratio. The resulting long-duration measurements are sensitive to motion artifacts. To compensate for this, Anas et al.\cite{anas2018enabling} proposed using a recurrent neural network, a convolutional LSTM network\cite{hochreiter1997long,xingjian2015convolutional}, to exploit the temporal dependencies in the noisy measurements. They report a considerable improvement over single-frame post-processing. In our opinion, this explicit consideration of the temporal aspect with recurrent units is more promising for low power systems than just post-processing with a U-Net\cite{singh2020deep}.
With a similar motivation to \revision{expand} on the information before post-processing, Kim et al.\cite{kim2020deep} propose to use the delay part of delay-and-sum but without taking the sum (Eq.\ \eqref{eqn:backprojectionIntegral} without the integral). The resulting three-dimensional input is then processed and collapsed by a U-net to produce the final reconstruction. 

\subsubsection{Beyond fully-supervised training regimes}
A possibility to provide an uncertainty estimation for reconstructed images by the post-processing approach was investigated by Godefroy et al.\cite{godefroy2020solving}. The authors propose to train a U-Net with Monte Carlo (MC) dropout to provide reconstructions and an uncertainty estimate. Here, a set of images is sampled with the MC dropout procedure, which provides a reconstruction (the mean of these images) and a standard deviation indicating instabilities in the reconstruction. 

Finally, we observe that the approaches here were all trained in a supervised manner by minimising an explicit loss function given by the $\ell^1$ or $\ell^2$ error. In a recent study, Vu et al.\cite{vu2020generative} explore the possibility of using a generative adversarial network (GAN) to process the image. In this setting, the U-Net is interpreted as the generator producing a clean PA image and the discriminator acts as the loss function evaluating reconstruction quality. GAN-based approaches lead the way to applications where no paired training data is available.

\subsection{Pre-Processing}
\label{subsec:pre-proc}
In a similar manner to the previous approach of using a network for post-processing reconstructions, one can instead focus the learning task on the data side and then use a classical reconstruction algorithm (Sec.\ \ref{sec:PATrecon}) to obtain the PA image; see Fig.\ \ref{fig:DL_recon_approaches}. In this sense we reformulate the learned post-processing reconstruction operator in Eq.\ \eqref{eqn:postProc} to its analogue for learned pre-processing as
\begin{equation}
 \ForwardOpInvLearned = \ForwardOpPseudoInv \circ \Network_\param.
\end{equation}
Here the network $\Network_\param$ can act as a denoising and artifact removal step on the data side to make the inversion step easier (essentially it changes the learning task from an inversion step to a de-noising step).

\subsubsection{Artifact removal for source localisation}
\label{sec:preprocartifactremoval}
Defining a clear purpose for an application enables the formulation of task specific processing algorithms, for instance in the case of tracking applications as explored in the work by Allman et al.\cite{allman2018photoacoustic,allman2018exploring,allman2017machine}. Here the aim is to localise a point-like source and to this end it is essential to distinguish clearly the true signal from noise and artifacts. The authors propose to use an object detection and classification approach to separate artifacts from the true signal. Their approach is based on a network architecture known as Faster R-CNN\cite{ren2015faster} that produces a classification between signal and artifact, a confidence score and locations as a bounding box. After a subsequent artifact removal step the final PA image is reconstructed using beamforming (Sec.\ \ref{subsubsec:backprojection}).
The authors show that their networks for accurate source location trained on simulated data can be transferred successfully to experimental data\cite{allman2018photoacoustic}, as well as \emph{ex vivo} and \emph{in vivo}\cite{allman2018deep} measurements.

\subsubsection{Sampling and bandwidth enhancement}
The PAT reconstruction problem is well-posed if perfect measurement data is available (see Sec.\ \ref{sec:inverse_problems}). One approach to pre-processing is therefore to aim to produce ideal data for the inversion from the non-ideal measurement data. This was investigated in the work by Awasthi et al.\cite{awasthi2020sinogram,awasthi2020deep}. The authors considered a sparse data (but full-view) scenario with limited bandwidth detectors and trained a network to produce high quality data from the degraded input. In particular, the network attempted to upsample the data from 100 detectors to 200, to denoise it, and to increase the bandwidth. The improved data was then reconstructed by filtered backprojection (Sec.\ \ref{subsubsec:backprojection}). Two architectures were used for $\Network_\param$: a simple 7-layer CNN\cite{awasthi2020sinogram} and a U-Net-based architecture\cite{awasthi2020deep}. In general the U-Net architecture performed better, but it is interesting that for low noise the simple CNN architecture was highly competitive. Translation to \emph{in vivo} measurements without retraining was successful for both methods\cite{awasthi2020sinogram,awasthi2020deep}.

Conceptually, such a pre-processing approach can be understood as learning a representation of the likelihood $\pi(g|f)$ conditioned with a training set for the images $f$. Nevertheless, the reconstruction quality is essentially limited by the goodness of the pre-processed measurement data and hence we believe this approach is only viable in fairly simple measurement scenarios, such as the tracking applications discussed above\cite{allman2017machine,allman2018photoacoustic}.
%in Eq. \eqref{eqn:likelihood} 

\subsection{Fully-Learned}\label{subsec:fullyLearned}
When considering a fully-learned reconstruction, it's important to keep in mind that the measurement data $g \in \DataSpace$ lies in a different spatio-temporal space than the reconstructed images $f\in\RecSpace$ and as such a mapping between the spaces $\DataSpace\rightarrow \RecSpace$ needs to be constructed. In Sec.\ \ref{sec:fullyLearned} we discussed the non-local nature of the mapping, and that in principle a fully connected layer can account for this. While the mapping may therefore be done by a fully connected layer, we nevertheless clearly saw in Sec.\ \ref{sec:comparisonMethods} that with a limited amount of data the fully connected layers are hard to train to achieve high quality reconstructions. Additionally, we observed that the CNN following the fully connected layers did most of the visual `heavy-lifting' for the final reconstruction.
This observation is in line with what has been reported in the literature, as discussed below in Sec.\ \ref{subsubsec:Convolutional_approaches}. Following this idea, Shang et al.\cite{shang2020two} proposed a two-step approach, where first a fully connected layer is trained to transform measurements into the image space, and then a U-Net is trained to process the result while the weights of the fully connected layer are fixed.

\subsubsection{Convolutional approaches}
\label{subsubsec:Convolutional_approaches}
Even though there is no clear theoretical justification to use a CNN directly to transform a spatio-temporal signal from $\DataSpace$ into an image in $\RecSpace$, as they learn spatially-invariant mappings, many studies in fact explore this scenario. The strength of convolutional-based networks lies in their capability to exploit local relations in the data, and as such can deal efficiently with noise in the input. The issue of spatial invariance can be overcome by using multiple pooling layers to increase the receptive field of the network, and the representation on the coarse scales effectively encodes the locality of the information.
This implies that large multi-scale networks are needed to transform the signal into the sought-after PAT image effectively. In an early study by Waibel et al.\cite{waibel2018reconstruction, grohl2018confidence} it was shown that using an asymmetric U-Net to reconstruct the PA image directly from raw sensor data is feasible in a limited-view setting. In comparison to a post-processing approach using a U-Net it was competitive in terms of mean reconstruction error, but exhibited a higher variance in reconstruction error. To overcome this, various solutions have been investigated in the literature, including enlarging the network to increase the capacity\cite{lan2019deep,lan2019reconstruct}. Others propose to introduce a pre-processing step to provide more informative input to the network, either by a hand-crafted interpolation\cite{guan2020limited} or even learned pre-processing with a separate CNN\cite{tong2020domain}. Note that in the latter case the transformation after the pre-processing is in fact done by a dense layer and hence is the closest to the AUTOMAP architecture discussed in Sec.\ \ref{sec:fullyLearned}. In both cases, the pre-processing step seems to be essential to provide an input, reduced in dimensionality, to the network performing the transform to the image space. Additionally, the authors in\cite{tong2020domain} motivate the pre-processing architecture based on the universal backprojection, Eq. \eqref{eqn:UBP_3D}, and provide time-series and as well as the time-derivative to the network. Lan et al. \cite{lan2020real} reduce 120 time series to 1 by summing them with delays, then feed this single time series into a LSTM network followed by a fully connected layer and a subsequent CNN to form the reconstructed image.

Following the discussion in Sec.\ \ref{sec:preprocartifactremoval}, there are situations in which the full reconstruction problem can be simplified to the case where only a source location must be found. This can be achieved by, for example, using a feature detection network\cite{reiter2017machine} or first forming a reconstructed image using an extended U-Net then converting to a numerical value for the source location\cite{johnstonbaugh2019novel,johnstonbaugh2020deep}.

\subsubsection{Discussion of fully-learned approaches}
In summary, the more advanced fully-learned approaches seem to provide a slight improvement over reconstruction followed by post-processing with a U-Net. However, the fully-learned approach does not explicitly include the acquisition geometry and sound speed in the inversion procedure.
While this generality might conceivably be useful, it means that for the network to be robust to changes in these experimental parameters, the training data must account for the full range over which they might vary.
As we see it, the fully-learned approach might therefore be useful in cases where a measurement device is available with corresponding data-image pairs, $(g,f)$, to be used as training data, but the acquisition geometry and other underlying parameters needed for reconstruction are not known. (ie.\ If it is a `black box' with examples of known inputs and outputs but the parameters implicit in $\ForwardOp$ are not known.) A fully-learned approach would then provide a way to improve the imaging pipeline without having to go through the potentially difficult procedure of determining the instrument characteristics.
Finally, following our observation in Sec.\ \ref{sec:comparisonMethods}, the fully-learned approach needs substantially more training data than other approaches that involve $\ForwardOp$ explicitly. This might constitute a major limitation when transitioning to experimental measurement data, where data availability is inherently scarce. Nevertheless, pre-processing approaches, as in\cite{guan2020limited,tong2020domain}, are potentially promising in reducing the hunger for training data.

\subsection{Learned Iterative Reconstructions}
\label{subsec:reviewLearnedIterative}
Learned iterative schemes, as described in Sec.\ \ref{subsubsec:Model-based_learned_iterative_reconstruction}, are model-based reconstructions that use known forward and adjoint models within a learned update. Given the reconstruction operator $\ForwardOpInvLearned$ in Eq.\ \eqref{eqn:LGS_reconOp}, defined by the iterates in Eq.\ \eqref{eqn:classicLGS}, we can formulate the training task in an end-to-end manner. That means, given paired training data $(g_\mathfrak{i},f_\mathfrak{i})\in \DataSpace\times \RecSpace$, then an optimal parameter $\param^*$ is found by solving the optimisation problem in Eq.\ \eqref{eqn:paramOptimisation},
where the loss function is given as
\begin{equation}
    	\Loss_{\param}(\signal,\data) :=\| \ForwardOpInvLearned(\data)  - \signal \|_2^2
    	\quad\text{for $(\signal,\data) \in \RecSpace \times \DataSpace$}.
\end{equation} 
Computing the gradient of the loss function with respect to $\param$ requires performing back-propagation through all of the unrolled iterates $n=0,\ldots, N-1$. This requires storage as well as evaluation of forward and adjoint in each training step for each iterate and hence can be computationally burdensome, and so has mostly been demonstrated in 2D imaging scenarios.  

In Shan et al.\cite{shan2019simultaneous} the basic learned iterative reconstruction approach has been applied with an extension to simultaneuously reconstruct sound speed as well, which constitutes a learned version of \cite{matthews2018parameterized}. Following the illustration in Fig.\ \ref{fig:residualLearnedIterative}, the authors suggested to also add a residual connection updating a sound speed estimate together with the reconstruction.

\subsubsection{Learned primal dual in 2D}
\label{subsubsec:LPD}

\begin{figure}[th!]
\centering
\includegraphics[width=0.8\textwidth]{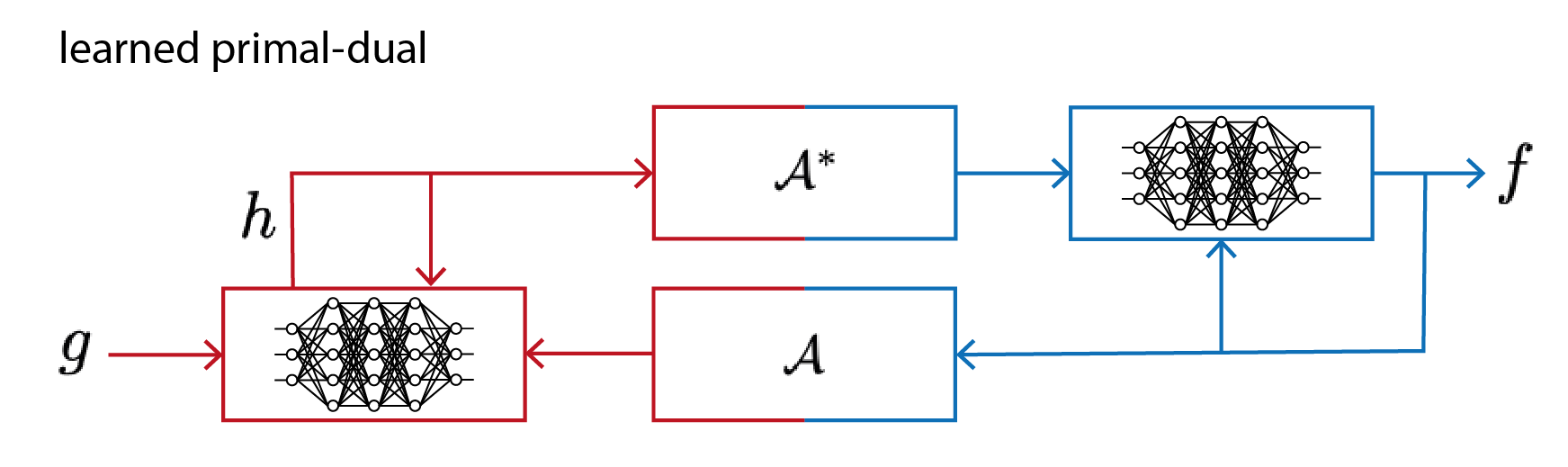}
\caption{\label{fig:learned_primal_dual}
Schematic of the \emph{learned primal-dual} reconstruction scheme, a learned iterative reconstruction based on the primal-dual hybrid gradient algorithm, Eq.\ \eqref{eqn:learned_primal_dual}; see also Fig.\ \ref{fig:DL_recon_approaches}. Red indicates the data space $\DataSpace$ and blue the image space, $\RecSpace$.}
\end{figure}

For reconstructions in PAT, the work by Boink et al. \cite{boink2018sensitivity,boink2019partially,boink2019robustness} has demonstrated the robustness of these learned iterative schemes to a number of \emph{in silico} phantoms as well as in an experimental study. The authors consider an extension to the learned gradient schemes introduced above called \emph{learned primal dual}\cite{Adler2018} (LPD) based on the successful primal-dual hybrid gradient method\cite{Chambolle2011} (also known as the Chambolle-Pock algorithm). The LPD method can be formulated in a similar manner to Eq.\ \eqref{eqn:classicLGS} by learning updating operators in the primal space $\RecSpace$ and the dual space $\DataSpace$
    \begin{align}
     \label{eqn:learned_primal_dual}
     	h^{(n+1)} =&\ \Gamma_\param\bigl(h^{(n)}, \ForwardOp\signal^{(n)}, \data \bigr),\\
     	\signal^{(n+1)} =&\ \Network_\param\bigl(\signal^{(n)}, \ForwardOpAdj h^{(n+1)}\bigr).
     \end{align}
In this case the network $\Gamma_\param$ operates in data space $\DataSpace$, whereas the network $\Network_\param$ operates in image space $\RecSpace$. See also the illustration in Fig.\ \ref{fig:learned_primal_dual}, in which it is clearly seen to be an extension to the learned iterative scheme in Fig.\ \ref{fig:DL_recon_approaches} (Bottom).
In their work \cite{boink2018sensitivity,boink2019partially,boink2019robustness}, the authors examine the robustness of LPD with respect to changes in the target, including the contrast, background, structural changes, and noise level. They found that if the network is trained only on the basic training data it generalises fairly well with respect to noise (1dB degradation in PSNR) and structural changes (3dB), but is most sensitive to changes in background (7dB) and contrast (11dB)\cite{boink2019partially}. Additionally, the authors combine their learned reconstruction with a joint segmentation that is learned with the same network as an additional output, and is shown to provide increased robustness compared to a reconstruction by filtered backprojection and segmentation with U-Net.

\subsubsection{Learned iterative reconstructions in 3D}
\label{subsubsec:learned_iterative_3D}
As already indicated, learned iterative reconstruction methods are ideally (and typically in 2D) trained in an end-to-end manner. While this can provide an optimal set of network parameters, if suitable optimisation procedures have been used, it also comes with two computational challenges. First, the memory footprint of storing and manipulating the network tends to be large and exceeds single GPU configurations making it necessary to use costly (and often less readily available) multi-GPU clusters. More significantly, however, during training the loss function must be evaluated several times, and each of these involves evaluating the forward and adjoint operators for each iterate. This quickly leads to unreasonable training times for three dimensional images, especially when considering large volume sizes, ie.\ many voxels, and accurate forward models. 

To overcome this limitation, Hauptmann et al.\cite{hauptmann2018model} proposed greedy training for learned gradient schemes for 3D PAT. That is, instead of looking for a reconstruction operator that is optimal end-to-end, only iterate-wise optimality is required. For the learned gradient scheme in Eq.\ \eqref{eqn:classicLGS} this amounts to the following loss function for the $n$\textsuperscript{th} unrolled iterate:
\begin{equation}\label{eqn:greedyLoss}
    	\Loss_{\param_n}(\signal^{(n)},\data) =\| \Network_{\param_n}\bigl(\signal^{(n)}, \ForwardOpAdj(\ForwardOp\signal^{(n)} - \data)\bigr) - \signal \|_2^2
\end{equation}
given the output of the previous iterate $\signal^{(n)} := \Network_{\param_{n-1}}\bigl(\signal^{(n-1)}, \ForwardOpAdj(\ForwardOp \signal^{(n-1)} - \data)\bigr)$ and initialisation $\signal^{(0)}=\ForwardOpAdj \data$. It is important to note that, as only iterate-wise optimality is required and the parameters $\param_n$ are not jointly minimised over all iterates, such a greedy scheme constitutes an upper bound on the minimised loss function for end-to-end networks.
Nevertheless, this renders the training procedure feasible since training can be separated from the evaluation of the model; the gradient of the data consistency term  $\ForwardOpAdj(\ForwardOp \signal^{(n)} - \data)$ used in Eq.\ \eqref{eqn:greedyLoss} can be computed before the parameter optimisation is performed. In their study\cite{hauptmann2018model}, the authors show that in this way a learned iterative reconstruction algorithm can be trained for realistic 3D volumes of size $240\times 240\times 80$ in a limited-view acquisition geometry. The results suggest that improved reconstructions can be obtained compared to both post-processing with a U-Net and iterative reconstruction with total variation regularisation. Application to \emph{in vivo} measurement data was presented after transfer training was performed, as outlined in Sec.\ \ref{subsec:training_data}. 
As the authors use an accurate, full-wave, solver for the forward and adjoint operators, reconstruction times were still slow in the order of minutes, but with an 4$\times$ speed-up compared to iterative reconstruction with total variation.

In a follow-up study\cite{hauptmann2018approximate}, the authors considered the use of a faster but approximate forward model to overcome the slow reconstruction times. Here the fast k-space method discussed in Sec.\ \ref{subsubsec:seriesSol} was used for the inverse as well as the forward propagation model, but as the forward model includes a singularity \cite{Cox:2005fwdkspace} this results in an approximate gradient only. Following the greedy training scheme, Eq.\ \eqref{eqn:greedyLoss}, the networks learned to reduce the resulting artifacts to produce a useful update. Using this fast approximate forward model the authors achieve a reconstruction time in the order of seconds, more precisely an 8$\times$ speed-up compared to their previous learned approach\cite{hauptmann2018model}, and 32$\times$ compared to iterative reconstruction with total variation. Results are presented for \emph{in vivo} measurements of a human target. 

Finally, Yang et al.\cite{yang2019accelerated} extend the previous study using an approximate model\cite{hauptmann2018approximate} by using recurrent inference machines\cite{Putzky2017recurrent,lonning2018recurrent} for the network architecture. This way the authors are able to improve reconstruction results for \emph{in silico} experiments in 2D by 2dB in PSNR. 
In conclusion, learned iterative approaches seem to provide an improvement in reconstruction quality compared to other learned reconstructions discussed in this review, but come with the major limitation of reconstruction speed due to the repeated application of the forward model and its adjoint.

\subsection{Hybrid Approaches}
From the previous sections it is apparent that most approaches, while having clear advantages, come with their own shortcomings. To try to mitigate these, a few groups have investigated hybrid approaches. For instance, to overcome the missing model dependence in the fully-learned approach the work by Lan et al.\cite{lan2019hybrid,lan2019ki,lan2019net} proposes augmenting the end-to-end approaches\cite{lan2019deep,lan2019reconstruct} by additionally feeding the network a reconstructed image, either directly into the network at a suitable location\cite{lan2019net} or with a separate processing branch\cite{lan2019hybrid,lan2019ki}.

\subsubsection{Augmented analytical approaches}
Another route is to incorporate learned methods into classical inversion approaches more explicitly than the learned iterative approaches in Sec.\ \ref{subsec:reviewLearnedIterative}. For instance by formulating a variational problem with a learned regulariser in the variational formulation of Eq.\ \eqref{eqn:variationalFormulation}, such that the functional to be minimised becomes
\begin{equation}
\mathcal{E}(f) =\tfrac{1}{2}\|\ForwardOp f - g \|^2_2 + \alpha\Network_\param (f),
\end{equation}
and explicit minimisation of $\mathcal{E}(f)$ can be performed in an iterative algorithm. This approach has been proposed as the NETT framework\cite{li2020nett} and applied to PAT\cite{antholzer2019nett}. The strength of this approach is in the emphasis on the model in the data consistency term and convergence guarantees under certain conditions\cite{li2020nett}, but time consuming iterative minimisation with the explicit forward and adjoint models is still needed, similar to the learned gradient schemes.
Another possibility for an augmented analytical approach is presented by Schwab et al.\cite{schwab2019deepSVD}, who consider a data-driven extension of the truncated singular value decomposition, where the network is trained to produce the singular vectors corresponding to small singular values to improve reconstruction quality.
We emphasise that such augmented analytical approaches are especially important where reconstruction convergence guarantees are needed, such as in critical clinical applications, but they seem to fall short in visual performance compared to the most advanced learned reconstruction approaches.

\subsection{Optical Inversions}
\label{subsec:qpat_review}
There is not, to date, a large literature using DL to tackle the optical inversions in PAT image reconstruction (see Secs.\ \ref{sec:varieties_of_inverse_problem} and \ref{subsec:Optical_Reconstructions}).
What there is all assumes the acoustic inversion has already been solved, which is to say the initial acoustic pressure distribution $f$ is either given as the basic measured quantity or has already been estimated by solving $\ForwardOp^{-1} g$. The inverse problems subsequently tackled fall largely into two classes: solving $\mathcal{F}^{-1}(f)$ to estimate optical absorption coefficients, or solving $(\mathcal{FL})^{-1}(f)$ to estimate chromophore concentrations or, more often than not, blood oxygen saturation, sO$_2$. 
The primary task of the networks in these cases is to account for the effect of the fluence, which is felt in two related ways: voxelwise it makes the PA spectra different from the absorption spectra (spectral coloring), and spatially the PAT image is no longer linearly related to the absorption coefficient distribution. These are related because the absorption coefficient at one voxel can affect the PAT image at another through the fluence. While this non-locality of the operator $\mathcal{F}$ can be strong, eg.\ a large absorber close to the light source may `shadow' a large part of the image region, for small absorbers the effect can be quite localised. The first application of machine learning to this problem \cite{kirchner2018context} used `fluence contribution maps' that made this assumption. In the DL approaches discussed below the use of U-Net-type architectures is common, and it is known that their multi-scale nature can help mitigate the spatial-invariance implicit in CNNs (see Sec.\ \ref{subsubsec:Convolutional_approaches}).

\subsubsection{U-Net-based optical inversions}
\label{subsubsec:U-Net-based}
Cai et al.\ \cite{cai2018end}, in an early contribution, used a variation on the U-Net, named the ResU-Net, to obtain estimates of sO$_2$ and a contrast agent from 2D multiwavelength PAT images. In this architecture, all the convolutional stages of a standard U-Net are replaced by residual blocks.\cite{he2016deep} In a similar approach, Yang et al.\ \cite{yang2019quantitative} proposed the another U-Net variant, DR2U-Net, the principle difference being that the residual blocks contain recurrent loops. Both these networks were shown to outperform linear unmixing $\mathcal{L}^{-1} f$ - which ignores the effect of the fluence - in simple \emph{in silico} tests.

Chen et al.\ \cite{chen2020deep} trained a U-Net to recover a 2D optical absorption coefficient distribution from a single-wavelength 2D PAT image. The loss term included a TV regulariser. The network was initially trained and tested with simple simulated examples and then demonstrated on 2D experimentally measured data. The measured training set was augmented by rotating the images in steps of 1 degree. The one result shown is promising, but the geometric simplicity and similarity of the training and test cases means the general applicability of the network remains unclear. 

Exploiting the fact the U-Net was designed for segmentation of biomedical images\cite{Ronneberger2015}, Luke et al.\ \cite{luke2019net} combine two U-Nets, one for segmentation and one for estimating blood sO$_2$, into a single `O-Net' with common input and output layers. The network input consists of two 2D slices from two 3D images obtained at different wavelengths, and the output is two 2D images: a segmentation and a map of sO$_2$. The network gives promising results on simulated data - it is shown to work better than linear unmixing - but the digital phantoms are simple geometric shapes. 
To overcome this concern, Bench et al.\ \cite{bench2020towards} perform a similar inversion but using 3D multiwavelength training images generated from vessel-like phantoms within a multi-layered tissue. These images also contained limited-view artifacts from the acoustic reconstruction, and therefore incorporated many aspects that would be present in real \emph{in vivo} data. In these simulations, the vessel sO$_2$ estimates were accurate to within 1\% on average, with a standard deviation of 6.3\%.

Yang et al.\cite{yang2019eda} also use more realistic simulated data based on a 3D digital breast phantom, using a 3D light model, and acoustically processing 2D slices as input to the network to mimic the limited-view measurements made by a linear array transducer.
Their network architecture, called an EDA-Net, uses the idea of `iterative deep aggregation'\cite{yu2018deep} to enhance the basic U-Net. In this architecture, every skip-connection is replaced with multiple nodes at the same scale, each of which is fed from below by (nonlinear) upsampling. This network was shown to perform slightly better than ResU-Net and U-Net++\cite{zhou2018unetpp} and much better than linear unmixing. 

In a detailed study, Gr\"ohl et al.\ \cite{grohl2018confidence} use U-Nets to estimate the absorption coefficient in various ways. In two fluence-estimation approaches, asymmetric and symmetric U-Nets were used to estimate the fluence map, $\phi$, from time series data $g$, and from initial pressure distribution $f$ respectively. (This was subsequently divided out of $f$ to estimate $\mu_a$.) Also, a one-step approach was described in which an asymmetric U-Net was used to estimate $\mu_a$ directly from limited-view and limited bandwidth time series data, ie.\ solving $(\mathcal{AF})^{-1}g$ directly. This one-step approach fared worse than the fluence estimation approaches in the \emph{in silico} tests, but the comparison is perhaps unfair. Unlike the fluence estimation approaches, which just have to learn a mapping from one image space $\RecSpace$ to another $X_{\mu_a}$, this inversion requires the network to also learn the mapping from $\DataSpace$ to $\RecSpace$ from incomplete data.

\subsubsection{Learned uncertainty estimation}
\label{subsubsec:learnedUQ}
All the U-Net variants in Sec.\ \ref{subsubsec:U-Net-based} above have been shown to give a degree of accuracy when demonstrated on simulated data (some more realistic than others), that if repeatable with \emph{in vivo} data would be useful in applications. Moving to \emph{in vivo} data, however, is a challenge, as discussed below in Sec.\ \ref{subsubsec:optical_training_data}. 
One of the difficulties with translating sO$_2$ estimation techniques, for example, to a clinical setting is knowing how much confidence one should have in the estimates. This problem is tackled by 
Gr\"ohl et al.\ \cite{grohl2018confidence} who trained a U-Net to act as an error-estimating network, using \{(PAT image,  error image)\} pairs, to give an estimate of the uncertainty in the $\mu_a$ estimates. The uncertainty correlated well with the actual error in the images in this \emph{in silico} study.
The use of a meta-network to observe the performance of a given estimator and output confidence levels for its estimates is very interesting given the difficulties inherent to translating quantitative PAT algorithms to \emph{in vivo} cases.

\subsubsection{Learned spectral unmixing}
In contrast to the U-Net-based approaches discussed above, which exploit the spatial information about the fluence that is present in the PA images, pixelwise approaches attempt to solve the optical inversion using the spectral data alone.

Durairaj et al.\ \cite{Durairaj2020} propose a two-stage autoencoder architecture (Sec.\ \ref{subsubsec:full_connected}) to estimate chromophore concentrations and molar absorption spectra simultaneously. One potentially significant advantage of this approach is that autoencoder networks by their nature do not require ground-truth data for the training. 
As discussed in Sec.\ \ref{subsubsec:full_connected},
by having a smaller hidden layer than input and output layers, autoencoders aim to find a compressed representation of the input. Durairaj et al.\ chose the hidden layer to have many dimensions as there are chromophores contributing to the data, in the hope that the values at the hidden layer are estimates of the chromophore concentrations (endmember abundances in their terminology) and the network weights are estimates of the molar absorption spectra (endmember spectra). Because this approach aims to solve the ill-posed problem of finding both the concentrations and spectra simultaneously, it requires strong prior information. As well as a positivity condition, which is well-justified, they impose the condition that the chromophore concentrations sum to one. However, this is unrealistic as there will also be non-absorbing molecules present in real tissue. Furthermore, it is not clear how this approach can account for the effect of the fluence on the PAT images and therefore unclear the extent to which the approach outlined in this preliminary simulation study will be useful in practice.

A different approach to learned spectral unmixing was taken by Gr\"ohl et al.\ \cite{grohl2019estimation} who used a fully connected network with 8 hidden layers to convert pixelwise PAT spectra into estimates of sO$_2$. The training data was taken from 2D simulated PAT images of vessels, and when the network was tested with simulated data it gave promising results. 
With some bravado, this network was then tested \emph{in vivo} on images of a porcine brain and human forearm, and in the case of the pig brain `seems to provide physiologically more plausible estimations' than linear unmixing.

\subsubsection{Training data}
\label{subsubsec:optical_training_data}
\revision{As a concluding remark on this section, we note that several} classical approaches to quantitative PAT have been demonstrated over the past decade (see \cite{Cox2011,Vogt2019SO2} and their references and citations) but it has proved difficult to translate these methods to work convincingly with measurement data obtained \emph{in vivo}, largely due to the challenge of obtaining all the auxiliary input parameters with sufficient accuracy under experimental conditions. DL holds the promise of overcoming this problem by learning the model, thereby not requiring auxiliary inputs, but a new difficulty arises: obtaining a large collection of experimentally measured \emph{in vivo} data with a known ground-truth to use for the training. As discussed in Sec.\ \ref{subsec:training_data}, there are two approaches: simulating the data or reconstructing ground-truth images using a `gold standard' classical reconstruction technique. The papers discussed in this section have used the former approach of simulating the data, typically using a Monte Carlo method such as MCX\cite{fang2009monte} for modelling the light propagation and collocation method such as k-Wave for modelling the acoustic propagation.\cite{treeby2010kWave} 
The degree to which the simulations are realistic will determine how well a network trained with this data will work on data measured \emph{in vivo}, and therefore will determine the confidence with which any conclusions can be drawn from a study using such an approach.
In conclusion then, the use of DL to tackle the quantitative PAT problem appear to hold promise but the translation to practical, \emph{in vivo}, cases remains a significant challenge.

\section{Conclusions and Future Directions}
The diversity of the work that has been done on learned image reconstruction in PAT in just the last few years, and the increasing rate at which it is being produced, suggests that the field will continue to develop for some time. In particular, we notice that already a move has begun  from straightforward proof-of-concept applications of Deep Learning to more sophisticated approaches. Nevertheless, there are many issues that remain to be addressed. For instance, on the one hand there are model agnostic reconstruction pipelines using fully learned approaches that get a lot of attention due to low latency. On the other hand, as described above, there are learned reconstructions that use a physical model in combination with a network, which have been shown to be more stable and require less training data, but are (considerably) slower in providing a reconstruction. This is in part because accurate numerical models of the physics are often slow compared to networks. Therefore a major question remains: \emph{Is it possible to obtain network speed without sacrificing the stability and accuracy that comes from explicitly incorporating a model?}

Another challenge, that hangs over learned image reconstructions with all biomedical applications, is how to ensure oddities (like a tumour) appear accurately in the image even though nothing quite like them was in the training data. In other words, how do we ensure the distribution of the training data matches that of the imaged target? And if it doesn't, will there be problems, as suggested by results from the tutorial, Sec.\ \ref{sec:comparisonMethods}? Could this problem be ameliorated by ensuring additional constraints, such as data consistency?

To conclude this review, we describe a few current research directions that address these questions, either by considering new training regimes or by combining physical models with neural networks in different ways. 

\paragraph{$\bullet$ Data consistency is important}
Many approaches are still missing a data-consistency term and hence the reconstructions obtained might look realistic but there is no way to assess their correctness. As we have discussed, there are a few approaches which do consider such data consistency during the reconstruction and hence provide a possible direction for further developments, such as the null space approaches discussed in Sec.\  \ref{subsubsec:extensions} or learned iterative reconstructions in Sec.\ \ref{subsec:reviewLearnedIterative}.
Another possible way to tackle this limitation is by using networks that consider uncertainty or provide an uncertainty estimate on top of the reconstruction. First steps in this direction have been taken for PAT \cite{godefroy2020solving}, see also Sec.\ \ref{subsubsec:learnedUQ}, but there is also rising interest in other fields to incorporate such uncertainty estimates into a learned reconstruction framework \cite{adler2019deepPosterior,schlemper2018bayesian,denker2020conditional}, which could be taken as inspiration. 

\paragraph{$\bullet$ Lack of \emph{in vivo} training data} 
For experimental scenarios, especially \emph{in vivo}, using simulated training data is risky because it is hard to ensure the training set distribution matches that of the target.

As the majority of the algorithms discussed above used fully supervised training, these approaches are primarily limited by the available ground-truth data. As this is seldom a viable option when developing imaging pipelines for \emph{in vivo} applications, it may be that different training regimes will be needed, such as semi-supervised approaches, as discussed in Sec.\ \ref{subsubsec:learning_task}. For instance by including a data consistency in the transfer to experimental data (also known as cycle consistency) or discriminator (GAN) based approaches\cite{vu2020generative}.

Another possibility might be to consider the framework of \emph{physics-informed neural networks}\cite{raissi2019physics}, in which the physical model, given by a partial differential equation, is incorporated directly into the loss function. In this case, rather than the network needing to learn the whole physical operator from the data, as in the fully learned cases presented above, the network learns much of the physics by virtue of the terms in the loss function.

\paragraph{$\bullet$ 3D nature of PAT}
The high computationally complexity caused by the inherently three-dimensional nature of PAT is another challenge for learned approaches, as computational models tend to be time-consuming and simply storing the data requires large amounts of memory. 
Possible methods to overcome this have been discussed in some recent papers, for instance by using invertible networks\cite{etmann2020iunets,putzky2019invert}, which do not require the storage of intermediate states in the network to compute the gradients for training. Another idea of how to scale learned iterative schemes to 3D is by computing the forward model on multiple lower resolutions in the reconstruction process\cite{hauptmann2020multi}.

\paragraph{$\bullet$ Model Augmentation and Correction}
\label{subsec:augmentedModels}
The learned schemes that use a model in conjunction with a network are typically slow, and also face the additional problem of uncertainty in model parameters, especially the sound speed and, for the optical inversion, the scattering (see Sec.\ \ref{subsubsec:inaccurate_forward}).

There may be advantages, therefore, of considering different ways to incorporate some of the behaviour of the model equation directly into the network. 
For instance, by designing or constraining networks based on the discretisation of the forward model - similar work has already been done for diffusion equations
\cite{ruthotto2019deep,arridge2019networks}. This way, it is possible to explicitly embed the properties of the model into the network architecture, with a computationally more efficient (network-based) solver.

Another possibility is to use approximate models that are faster or easier to compute in place of the true (expensive) model, and train a network to learn a correction\cite{lunz2020learned,smyl2020learning}. The error may arise from an efficient, but inaccurate, numerical discretisation of the correct model\cite{ hauptmann2018approximate,siahkoohi2019neural} or because the accurate model has been replaced with a more-easily solvable approximation\cite{smyl2020learning}. We believe that this direction could be particularly fruitful for PAT as model information is essential to provide stability and robustness in the inversion, but we need to overcome the two major limitations: computational speed and the inherent uncertainty in the model parameters. Nevertheless, these improvements come with a major increase in training times for such networks.

\paragraph{$\bullet$ Trade-offs and choices}
There are so many options that trade-offs and choices will need to be made in practice. This is not a problem \emph{per se}, but rather an opportunity.
There are many possible ways in which a network can be incorporated into the reconstruction pipeline, and the approach that will be best suited to a particular application will depend on the nature of the application.
It is the responsibility of the designer of the image reconstruction algorithm to consider the trade-offs and constraints - eg.\ Is reconstruction speed or a data-consistency guarantee more important? Does the algorithm needs to be able to work well with more than one hardware system? What hardware is available for the computations? - and construct the algorithm accordingly. This plethora of choice is good, because it gives sufficient flexibility for properly crafted, well-thought-through algorithms to be designed to be optimal for specific tasks. The key to realising that is developing an understanding of the strengths and weaknesses of particular architectures and approaches. We are only at the beginning of this journey, but we hope this paper has illuminated at least a little of the way along the path.

% \disclosures 
%\subsection*{Disclosures}
% Conflicts of interest should be declared under a separate header. If the authors have no relevant financial interests in the manuscript and no other potential conflicts of interest to disclose, a statement to this effect should also be included in the manuscript.

\acknowledgments 
The authors would like to express their thanks to Paul Beard and UCL's Photoacoustic Imaging Group for many helpful discussions on all aspects of PAT over many years, and also to Simon Arridge, Jonas Adler, Sebastian Lunz, Felix Lucka, Marta Betcke, Bradley Treeby, Antonio Stanziola, Ashkan Javaherian, Ciaran Bench and UCL's Biomedical Ultrasound Group for very useful and informative discussions on inverse problems, image reconstruction, Deep Learning and acoustic modelling. This work was partly funded by the European Union’s Horizon 2020 research and innovation program H2020 ICT 2016-2017 under Grant agreement No. 732411, which is an initiative of the Photonics Public Private Partnership, and partly by the Academy of Finland Project 312123 (Finnish Centre of Excellence in Inverse Modelling and Imaging, 2018--2025) and the CMIC-EPSRC platform grant (EP/M020533/1).

%%%%% References %%%%%
%\Andreas{References alphabetical for now: Easier to check for duplicates}

\bibliography{review_PATpart,PAT,PATCS,Inverse_Problems_refs}   % bibliography data in report.bib

\begin{thebibliography}{100}

\bibitem{Kuchment:2011mtat}
P.~Kuchment and L.~Kunyansky, {\em Mathematics of Photoacoustic and
  Thermoacoustic Tomography}, 817--865.
\newblock Springer New York, New York, NY  (2011).

\bibitem{Poudel2019survey}
J.~Poudel, Y.~Lou, and M.~A. Anastasio, ``A survey of computational frameworks
  for solving the acoustic inverse problem in three-dimensional photoacoustic
  computed tomography,'' {\em Physics in Medicine \& Biology} {\bf 64}(14),
  14TR01  (2019).

\bibitem{Cox2012quantitative}
B.~T. Cox, J.~G. Laufer, P.~C. Beard, {\em et~al.}, ``Quantitative
  spectroscopic photoacoustic imaging: a review,'' {\em Journal of biomedical
  optics} {\bf 17}(6), 061202  (2012).

\bibitem{Huang:2013fwi}
C.~Huang, K.~Wang, L.~Nie, {\em et~al.}, ``{Full-Wave Iterative Image
  Reconstruction in Photoacoustic Tomography With Acoustically Inhomogeneous
  Media},'' {\em IEEE Transactions on Medical Imaging} {\bf 32}, 1097--1110
  (2013).

\bibitem{Arridge:2016apat}
S.~Arridge, P.~Beard, M.~Betcke, {\em et~al.}, ``Accelerated high-resolution
  photoacoustic tomography via compressed sensing.,'' {\em Physics in medicine
  and biology} {\bf 61}(24), 8908--8940  (2016).

\bibitem{Boink2018framework}
Y.~E. Boink, M.~J. Lagerwerf, W.~Steenbergen, {\em et~al.}, ``A framework for
  directional and higher-order reconstruction in photoacoustic tomography,''
  {\em Physics in Medicine \& Biology} {\bf 63}(4), 045018  (2018).

\bibitem{Kang2017}
E.~Kang, J.~Min, and J.~C. Ye, ``A deep convolutional neural network using
  directional wavelets for low-dose {X}-ray {CT} reconstruction,'' {\em Medical
  Physics} {\bf 44}(10)  (2017).

\bibitem{Jin2017}
K.~H. Jin, M.~T. McCann, E.~Froustey, {\em et~al.}, ``Deep convolutional neural
  network for inverse problems in imaging,'' {\em IEEE Transactions on Image
  Processing} {\bf 26}(9), 4509--4522  (2017).

\bibitem{Hammernik2018learning}
K.~Hammernik, T.~Klatzer, E.~Kobler, {\em et~al.}, ``Learning a variational
  network for reconstruction of accelerated {MRI} data,'' {\em Magnetic
  resonance in medicine} {\bf 79}(6), 3055--3071  (2018).

\bibitem{Adler2017}
J.~Adler and O.~{\"O}ktem, ``Solving ill-posed inverse problems using iterative
  deep neural networks,'' {\em Inverse Problems} {\bf 33}(12), 124007  (2017).

\bibitem{Zhu2018automap}
B.~Zhu, J.~Z. Liu, S.~F. Cauley, {\em et~al.}, ``Image reconstruction by
  domain-transform manifold learning,'' {\em Nature} {\bf 555}(7697), 487
  (2018).

\bibitem{arridge2019solving}
S.~Arridge, P.~Maass, O.~{\"O}ktem, {\em et~al.}, ``Solving inverse problems
  using data-driven models,'' {\em Acta Numerica} {\bf 28}, 1--174  (2019).

\bibitem{ye2018deep}
J.~C. Ye, Y.~Han, and E.~Cha, ``Deep convolutional framelets: A general deep
  learning framework for inverse problems,'' {\em SIAM Journal on Imaging
  Sciences} {\bf 11}(2), 991--1048  (2018).

\bibitem{haber2017stable}
E.~Haber and L.~Ruthotto, ``Stable architectures for deep neural networks,''
  {\em Inverse Problems} {\bf 34}(1), 014004  (2017).

\bibitem{ruthotto2019deep}
L.~Ruthotto and E.~Haber, ``Deep neural networks motivated by partial
  differential equations,'' {\em Journal of Mathematical Imaging and Vision} ,
  1--13  (2019).

\bibitem{Schlemper2017deep}
J.~Schlemper, J.~Caballero, J.~V. Hajnal, {\em et~al.}, ``A deep cascade of
  convolutional neural networks for {MR} image reconstruction,'' in {\em
  International Conference on Information Processing in Medical Imaging},
  647--658, Springer  (2017).

\bibitem{Hauptmann2019MRM}
A.~Hauptmann, S.~Arridge, F.~Lucka, {\em et~al.}, ``Real-time cardiovascular
  {MR} with spatio-temporal artifact suppression using deep learning--proof of
  concept in congenital heart disease,'' {\em Magnetic resonance in medicine}
  {\bf 81}(2), 1143--1156  (2019).

\bibitem{Adler2018}
J.~Adler and O.~{\"O}ktem, ``Learned primal-dual reconstruction,'' {\em IEEE
  transactions on medical imaging} {\bf 37}(6), 1322--1332  (2018).

\bibitem{Arridge1999}
S.~R. Arridge, ``Optical tomography in medical imaging,'' {\em Inverse
  Problems} {\bf 15}(2), R41--R93  (1999).

\bibitem{Welch2011book2ndedn}
A.~J. Welch, M.~J. Van~Gemert, {\em et~al.}, {\em Optical-thermal response of
  laser-irradiated tissue}, vol.~2, Springer  (2011).

\bibitem{Bigio2016book}
I.~J. Bigio and S.~Fantini, {\em Quantitative biomedical optics: theory,
  methods, and applications}, Cambridge University Press  (2016).

\bibitem{Li2018reviewSO2}
M.~Li, Y.~Tang, and J.~Yao, ``Photoacoustic tomography of blood oxygenation: A
  mini review,'' {\em Photoacoustics} {\bf 10}, 65--73  (2018).

\bibitem{Koestli:2001f}
K.~P. K\"ostli, M.~Frenz, H.~Bebie, {\em et~al.}, ``{Temporal backward
  projection of optoacoustic pressure transients using Fourier transform
  methods},'' {\em Phys Med Biol} {\bf 46}, 1863--1872  (2001).

\bibitem{Xu2002cylinder}
Y.~Xu, M.~Xu, and L.~V. Wang, ``Exact frequency-domain reconstruction for
  thermoacoustic tomography. ii. cylindrical geometry,'' {\em IEEE transactions
  on medical imaging} {\bf 21}(7), 829--833  (2002).

\bibitem{Finch:2004msph}
D.~Finch and R.~Sarah K.~Patch, ``Determining a function from its mean values
  over a family of spheres,'' {\em SIAM Journal on Mathematical Analysis} {\bf
  35}, 1213--1240  (2004).

\bibitem{haltmeier2014ellipsoids}
M.~Haltmeier, ``Exact reconstruction formula for the spherical mean radon
  transform on ellipsoids,'' {\em Inverse Problems} {\bf 30}(10), 105006
  (2014).

\bibitem{Kunyansky2011polyhedra}
L.~Kunyansky, ``Reconstruction of a function from its spherical (circular)
  means with the centers lying on the surface of certain polygons and
  polyhedra,'' {\em Inverse problems} {\bf 27}(2), 025012  (2011).

\bibitem{Burgholzer2005integrating}
P.~Burgholzer, C.~Hofer, G.~Paltauf, {\em et~al.}, ``Thermoacoustic tomography
  with integrating area and line detectors,'' {\em IEEE transactions on
  ultrasonics, ferroelectrics, and frequency control} {\bf 52}(9), 1577--1583
  (2005).

\bibitem{huynh2019:SinglePixelPAT}
N.~Huynh, F.~Lucka, E.~Z. Zhang, {\em et~al.}, ``Single-pixel camera
  photoacoustic tomography,'' {\em Journal of biomedical optics} {\bf 24}(12),
  121907  (2019).

\bibitem{Wang:2003}
X.~Wang, Y.~Pang, G.~Ku, {\em et~al.}, ``{Noninvasive laser-induced
  photoacoustic tomography for structural and functional in vivo imaging of the
  brain},'' {\em Nat. Biotechnol.} {\bf 21}(7), 803--806  (2003).

\bibitem{yin2004linear_array}
B.~Yin, D.~Xing, Y.~Wang, {\em et~al.}, ``Fast photoacoustic imaging system
  based on 320-element linear transducer array,'' {\em Physics in Medicine \&
  Biology} {\bf 49}(7), 1339  (2004).

\bibitem{Landau1987fluid}
L.~Landau and E.~Lifshitz, {\em Fluid mechanics, vol. 6, 2nd edn.},
  Butterworth-Heineman  (1987).

\bibitem{Xu2004Limitedview}
Y.~Xu, L.~V. Wang, G.~Ambartsoumian, {\em et~al.}, ``Reconstructions in
  limited-view thermoacoustic tomography,'' {\em Medical physics} {\bf 31}(4),
  724--733  (2004).

\bibitem{Cox2009c}
B.~T. Cox, S.~R. Arridge, and P.~C. Beard, ``Estimating chromophore
  distributions from multiwavelength photoacoustic images,'' {\em JOSA A} {\bf
  26}(2), 443--455  (2009).

\bibitem{Huang2016joint}
C.~Huang, K.~Wang, R.~W. Schoonover, {\em et~al.}, ``Joint reconstruction of
  absorbed optical energy density and sound speed distributions in
  photoacoustic computed tomography: a numerical investigation,'' {\em IEEE
  transactions on computational imaging} {\bf 2}(2), 136--149  (2016).

\bibitem{Tarvainen2012}
T.~Tarvainen, B.~T. Cox, J.~Kaipio, {\em et~al.}, ``Reconstructing absorption
  and scattering distributions in quantitative photoacoustic tomography,'' {\em
  Inverse Problems} {\bf 28}(8), 084009  (2012).

\bibitem{Stefanov2013instability}
P.~Stefanov and G.~Uhlmann, ``Instability of the linearized problem in
  multiwave tomography of recovery both the source and the speed,'' {\em
  Inverse Problems \& Imaging} {\bf 7}(4), 1367  (2013).

\bibitem{KaipioSomersalo2004}
J.~Kaipio and E.~Somersalo, {\em Statistical and Computational Inverse
  Problems}, vol.~160 of {\em Applied Mathematical Sciences}, Springer Verlag
  (2004).

\bibitem{Kaipio2007}
J.~Kaipio and E.~Somersalo, ``Statistical inverse problems: {D}iscretization,
  model reduction and inverse crimes,'' {\em Journal of Computational and
  Applied Mathematics} {\bf 198}, 493--504  (2007).

\bibitem{arridge2006a}
S.~R. Arridge, J.~P. Kaipio, V.~Kolehmainen, {\em et~al.}, ``{Approximation
  Errors and Model Reduction with an Application in Optical Diffusion
  Tomography},'' {\em Inverse Problems} {\bf 22}(1), 175--196  (2006).

\bibitem{sahlstrom2020modeling}
T.~Sahlstr{\"o}m, A.~Pulkkinen, J.~Tick, {\em et~al.}, ``Modeling of errors due
  to uncertainties in ultrasound sensor locations in photoacoustic
  tomography,'' {\em IEEE Transactions on Medical Imaging}   (2020).

\bibitem{tarvainen2013}
T.~Tarvainen, A.~Pulkkinen, B.~T. Cox, {\em et~al.}, ``Bayesian image
  reconstruction in quantitative photoacoustic tomography,'' {\em IEEE
  Transactions on Medical Imaging} {\bf 32}(12), 2287--2298  (2013).

\bibitem{Natterer1986}
F.~Natterer, {\em The mathematics of computerized tomography}, vol.~32, John
  Wiley \& Sons, Chichester, USA, and B. G. Teubner, Stuttgart, Germany
  (1986).

\bibitem{Arridge:2016adj}
S.~{Arridge}, M.~{Betcke}, B.~{Cox}, {\em et~al.}, ``{On the Adjoint Operator
  in Photoacoustic Tomography},'' {\em ArXiv e-prints}   (2016).

\bibitem{Xu2005UBP}
M.~Xu and L.~V. Wang, ``Universal back-projection algorithm for photoacoustic
  computed tomography,'' {\em Physical Review E} {\bf 71}(1), 016706  (2005).

\bibitem{Haltmeier2015universal}
M.~Haltmeier and S.~Pereverzyev~Jr, ``The universal back-projection formula for
  spherical means and the wave equation on certain quadric hypersurfaces,''
  {\em Journal of Mathematical Analysis and Applications} {\bf 429}(1),
  366--382  (2015).

\bibitem{burgholzer2007temporal}
P.~Burgholzer, J.~Bauer-Marschallinger, H.~Gr{\"u}n, {\em et~al.}, ``Temporal
  back-projection algorithms for photoacoustic tomography with integrating line
  detectors,'' {\em Inverse Problems} {\bf 23}(6), S65  (2007).

\bibitem{park2008beamforming}
S.~Park, A.~B. Karpiouk, S.~R. Aglyamov, {\em et~al.}, ``Adaptive beamforming
  for photoacoustic imaging,'' {\em Optics letters} {\bf 33}(12), 1291--1293
  (2008).

\bibitem{mozaffarzadeh2017delay_and_sum}
M.~Mozaffarzadeh, A.~Mahloojifar, M.~Orooji, {\em et~al.}, ``Double-stage delay
  multiply and sum beamforming algorithm: Application to linear-array
  photoacoustic imaging,'' {\em IEEE Transactions on Biomedical Engineering}
  {\bf 65}(1), 31--42  (2017).

\bibitem{Xu:2002efdr}
Y.~Xu, D.~Feng, and L.~V. Wang, ``{Exact frequency-domain reconstruction for
  thermoacoustic tomography. I. Planar geometry},'' {\em IEEE Transactions on
  Medical Imaging} {\bf 21}, 823--828  (2002).

\bibitem{Kunyansky2007series}
L.~A. Kunyansky, ``A series solution and a fast algorithm for the inversion of
  the spherical mean radon transform,'' {\em Inverse Problems} {\bf 23}(6), S11
   (2007).

\bibitem{Kunyansky2012fast}
L.~Kunyansky, ``Fast reconstruction algorithms for the thermoacoustic
  tomography in certain domains with cylindrical or spherical symmetries,''
  {\em Inverse Problems and Imaging} {\bf 6}(1), 111--131  (2012).

\bibitem{Xu2004TR}
Y.~Xu and L.~V. Wang, ``Time reversal and its application to tomography with
  diffracting sources,'' {\em Physical review letters} {\bf 92}(3), 033902
  (2004).

\bibitem{burgholzer2007}
P.~Burgholzer, G.~J. Matt, M.~Haltmeier, {\em et~al.}, ``{Exact and
  approximative imaging methods for photoacoustic tomography using an arbitrary
  detection surface},'' {\em Phy. Rev. E} {\bf 75}(4), 046706  (2007).

\bibitem{Hristova2008}
Y.~Hristova, P.~Kuchment, and L.~Nguyen, ``{Reconstruction and time reversal in
  thermoacoustic tomography in acoustically homogeneous and inhomogeneous
  media},'' {\em Inverse Problems} {\bf 24}, 055006  (2008).

\bibitem{Baker2003Huygens}
B.~Baker and E.~Copson, {\em The mathematical theory of Huygens' principle},
  vol.~329, American Mathematical Society  (2003).

\bibitem{cox2009artifact}
B.~T. Cox and B.~E. Treeby, ``Artifact trapping during time reversal
  photoacoustic imaging for acoustically heterogeneous media,'' {\em IEEE
  transactions on medical imaging} {\bf 29}(2), 387--396  (2009).

\bibitem{Stefanov:2009tatvs}
P.~Stefanov and G.~Uhlmann, ``Thermoacoustic tomography with variable sound
  speed,'' {\em Inverse Problems} {\bf 25}(7), 075011  (2009).

\bibitem{Guo2010compressed}
Z.~Guo, C.~Li, L.~Song, {\em et~al.}, ``Compressed sensing in photoacoustic
  tomography in vivo,'' {\em Journal of biomedical optics} {\bf 15}(2), 021311
  (2010).

\bibitem{Wang2012modelbased}
K.~Wang, R.~Su, A.~A. Oraevsky, {\em et~al.}, ``Investigation of iterative
  image reconstruction in three-dimensional optoacoustic tomography,'' {\em
  Physics in Medicine \& Biology} {\bf 57}(17), 5399  (2012).

\bibitem{Haltmeier2017iterative}
M.~Haltmeier and L.~V. Nguyen, ``Analysis of iterative methods in photoacoustic
  tomography with variable sound speed,'' {\em SIAM Journal on Imaging
  Sciences} {\bf 10}(2), 751--781  (2017).

\bibitem{chambolle2016introduction}
A.~Chambolle and T.~Pock, ``An introduction to continuous optimization for
  imaging,'' {\em Acta Numerica} {\bf 25}, 161--319  (2016).

\bibitem{benning2018modern}
M.~Benning and M.~Burger, ``Modern regularization methods for inverse
  problems,'' {\em Acta Numerica} {\bf 27}, 1  (2018).

\bibitem{Rosenthal2010fast}
A.~Rosenthal, D.~Razansky, and V.~Ntziachristos, ``Fast semi-analytical
  model-based acoustic inversion for quantitative optoacoustic tomography,''
  {\em IEEE transactions on medical imaging} {\bf 29}(6), 1275--1285  (2010).

\bibitem{Paltauf2002iterative}
G.~Paltauf, J.~Viator, S.~Prahl, {\em et~al.}, ``Iterative reconstruction
  algorithm for optoacoustic imaging,'' {\em The Journal of the Acoustical
  Society of America} {\bf 112}(4), 1536--1544  (2002).

\bibitem{Bauer2011quantitative}
A.~Q. Bauer, R.~E. Nothdurft, J.~P. Culver, {\em et~al.}, ``Quantitative
  photoacoustic imaging: correcting for heterogeneous light fluence
  distributions using diffuse optical tomography,'' {\em Journal of Biomedical
  Optics} {\bf 16}(9), 096016  (2011).

\bibitem{Hussain2018photoacoustic}
A.~Hussain, E.~Hondebrink, J.~Staley, {\em et~al.}, ``Photoacoustic and
  acousto-optic tomography for quantitative and functional imaging,'' {\em
  Optica} {\bf 5}(12), 1579--1589  (2018).

\bibitem{Cox2006}
B.~T. Cox, S.~R. Arridge, K.~P. K.~o. stli, {\em et~al.}, ``{Two-dimensional
  quantitative photoacoustic image reconstruction of absorption distributions
  in scattering media by use of a simple iterative method},'' {\em Applied
  Optics} {\bf 45}(8), 1866--1875  (2006).

\bibitem{Buchmann2019three}
J.~Buchmann, B.~A. Kaplan, S.~Powell, {\em et~al.}, ``Three-dimensional
  quantitative photoacoustic tomography using an adjoint radiance monte carlo
  model and gradient descent,'' {\em Journal of biomedical optics} {\bf 24}(6),
  066001  (2019).

\bibitem{tensorflow2015-whitepaper}
M.~Abadi, A.~Agarwal, P.~Barham, {\em et~al.}, ``{TensorFlow}: Large-scale
  machine learning on heterogeneous systems,''  (2015).
\newblock Software available from tensorflow.org.

\bibitem{pytorch_NEURIPS2019}
A.~Paszke, S.~Gross, F.~Massa, {\em et~al.}, ``Pytorch: An imperative style,
  high-performance deep learning library,'' in {\em Advances in Neural
  Information Processing Systems 32},  8024--8035, Curran Associates, Inc.
  (2019).

\bibitem{lecun2015deep}
Y.~LeCun, Y.~Bengio, and G.~Hinton, ``Deep learning,'' {\em Nature} {\bf
  521}(7553), 436--444  (2015).

\bibitem{goodfellow2016deep}
I.~Goodfellow, Y.~Bengio, and A.~Courville, {\em Deep learning}, MIT press
  (2016).

\bibitem{schmidhuber2015deep}
J.~Schmidhuber, ``Deep learning in neural networks: An overview,'' {\em Neural
  networks} {\bf 61}, 85--117  (2015).

\bibitem{rumelhart1986learning}
D.~E. Rumelhart, G.~E. Hinton, and R.~J. Williams, ``Learning representations
  by back-propagating errors,'' {\em nature} {\bf 323}(6088), 533--536  (1986).

\bibitem{lecun1989backpropagation}
Y.~LeCun, B.~Boser, J.~S. Denker, {\em et~al.}, ``Backpropagation applied to
  handwritten zip code recognition,'' {\em Neural computation} {\bf 1}(4),
  541--551  (1989).

\bibitem{kingma2014adam}
D.~P. Kingma and J.~Ba, ``Adam: A method for stochastic optimization,'' {\em
  3rd International Conference for Learning Representations (ICLR), 2015}
  (2015).

\bibitem{zhu2017unpaired}
J.-Y. Zhu, T.~Park, P.~Isola, {\em et~al.}, ``Unpaired image-to-image
  translation using cycle-consistent adversarial networks,'' in {\em
  Proceedings of the IEEE international conference on computer vision},
  2223--2232  (2017).

\bibitem{antholzer2019deep}
S.~Antholzer, M.~Haltmeier, and J.~Schwab, ``Deep learning for photoacoustic
  tomography from sparse data,'' {\em Inverse problems in science and
  engineering} {\bf 27}(7), 987--1005  (2019).

\bibitem{Ronneberger2015}
O.~Ronneberger, P.~Fischer, and T.~Brox, ``U-net: Convolutional networks for
  biomedical image segmentation,'' in {\em International Conference on Medical
  Image Computing and Computer-Assisted Intervention},  234--241, Springer
  (2015).

\bibitem{daubechies1992ten}
I.~Daubechies, {\em Ten lectures on wavelets}, vol.~61, Siam  (1992).

\bibitem{mallat1989theory}
S.~G. Mallat, ``A theory for multiresolution signal decomposition: the wavelet
  representation,'' {\em IEEE transactions on pattern analysis and machine
  intelligence} {\bf 11}(7), 674--693  (1989).

\bibitem{hauptmann2018model}
A.~Hauptmann, F.~Lucka, M.~Betcke, {\em et~al.}, ``Model-based learning for
  accelerated, limited-view 3-d photoacoustic tomography,'' {\em IEEE
  transactions on medical imaging} {\bf 37}(6), 1382--1393  (2018).

\bibitem{Putzky2017recurrent}
P.~Putzky and M.~Welling, ``Recurrent inference machines for solving inverse
  problems,'' {\em arXiv preprint arXiv:1706.04008}   (2017).

\bibitem{he2016deep}
K.~He, X.~Zhang, S.~Ren, {\em et~al.}, ``Deep residual learning for image
  recognition,'' in {\em Proceedings of the IEEE conference on computer vision
  and pattern recognition},  770--778  (2016).

\bibitem{maier2019learning}
A.~K. Maier, C.~Syben, B.~Stimpel, {\em et~al.}, ``Learning with known
  operators reduces maximum error bounds,'' {\em Nature machine intelligence}
  {\bf 1}(8), 373--380  (2019).

\bibitem{scianna2013review}
M.~Scianna, C.~Bell, and L.~Preziosi, ``A review of mathematical models for the
  formation of vascular networks,'' {\em Journal of theoretical biology} {\bf
  333}, 174--209  (2013).

\bibitem{erhan2010does}
D.~Erhan, Y.~Bengio, A.~Courville, {\em et~al.}, ``Why does unsupervised
  pre-training help deep learning?,'' {\em Journal of Machine Learning
  Research} {\bf 11}(Feb), 625--660  (2010).

\bibitem{yosinski2014transferable}
J.~Yosinski, J.~Clune, Y.~Bengio, {\em et~al.}, ``How transferable are features
  in deep neural networks?,'' in {\em Advances in neural information processing
  systems},  3320--3328  (2014).

\bibitem{treeby2010kWave}
B.~E. Treeby and B.~T. Cox, ``k-wave: Matlab toolbox for the simulation and
  reconstruction of photoacoustic wave fields,'' {\em Journal of biomedical
  optics} {\bf 15}(2), 021314  (2010).

\bibitem{baguer2020computed}
D.~O. Baguer, J.~Leuschner, and M.~Schmidt, ``Computed tomography
  reconstruction using deep image prior and learned reconstruction methods,''
  {\em arXiv preprint arXiv:2003.04989}   (2020).

\bibitem{an2020application}
Y.~An, H.~Meng, Y.~Gao, {\em et~al.}, ``Application of machine learning method
  in optical molecular imaging: a review,'' {\em Science China Information
  Sciences} {\bf 63}(1), 111101  (2020).

\bibitem{zhang2019brief}
L.~Zhang, G.~Zhang, {\em et~al.}, ``Brief review on learning-based methods for
  optical tomography,'' {\em Journal of Innovative Optical Health Sciences}
  {\bf 12}(6), 1930011  (2019).

\bibitem{sivasubramanian2020deep}
K.~Sivasubramanian and L.~Xing, ``Deep learning for image processing and
  reconstruction to enhance led-based photoacoustic imaging,'' in {\em
  LED-Based Photoacoustic Imaging},  203--241, Springer  (2020).

\bibitem{antholzer2018deep}
S.~Antholzer, J.~Schwab, and M.~Haltmeier, ``Deep learning versus
  $\ell^{1}$-minimization for compressed sensing photoacoustic tomography,'' in
  {\em 2018 IEEE International Ultrasonics Symposium (IUS)},  206--212, IEEE
  (2018).

\bibitem{antholzer2018photoacoustic}
S.~Antholzer, M.~Haltmeier, R.~Nuster, {\em et~al.}, ``Photoacoustic image
  reconstruction via deep learning,'' in {\em Photons Plus Ultrasound: Imaging
  and Sensing 2018},   {\bf 10494}, 104944U, International Society for Optics
  and Photonics  (2018).

\bibitem{guan2019fully}
S.~Guan, A.~Khan, S.~Sikdar, {\em et~al.}, ``Fully dense unet for 2d sparse
  photoacoustic tomography artifact removal,'' {\em IEEE journal of biomedical
  and health informatics}   (2019).

\bibitem{deng2019machine}
H.~Deng, X.~Wang, C.~Cai, {\em et~al.}, ``Machine-learning enhanced
  photoacoustic computed tomography in a limited view configuration,'' in {\em
  Advanced Optical Imaging Technologies II},   {\bf 11186}, 111860J,
  International Society for Optics and Photonics  (2019).

\bibitem{shan2019accelerated}
H.~Shan, G.~Wang, and Y.~Yang, ``Accelerated correction of reflection artifacts
  by deep neural networks in photo-acoustic tomography,'' {\em Applied
  Sciences} {\bf 9}(13), 2615  (2019).

\bibitem{awasthi2019pa}
N.~Awasthi, K.~R. Prabhakar, S.~K. Kalva, {\em et~al.}, ``Pa-fuse: deep
  supervised approach for the fusion of photoacoustic images with distinct
  reconstruction characteristics,'' {\em Biomedical optics express} {\bf
  10}(5), 2227--2243  (2019).

\bibitem{waibel2018reconstruction}
D.~Waibel, J.~Gr{\"o}hl, F.~Isensee, {\em et~al.}, ``Reconstruction of initial
  pressure from limited view photoacoustic images using deep learning,'' in
  {\em Photons Plus Ultrasound: Imaging and Sensing 2018},   {\bf 10494},
  104942S, International Society for Optics and Photonics  (2018).

\bibitem{davoudi2019deep}
N.~Davoudi, X.~L. De{\'a}n-Ben, and D.~Razansky, ``Deep learning optoacoustic
  tomography with sparse data,'' {\em Nature Machine Intelligence} {\bf 1}(10),
  453--460  (2019).

\bibitem{hariri2020deep}
A.~Hariri, K.~Alipour, Y.~Mantri, {\em et~al.}, ``Deep learning improves
  contrast in low-fluence photoacoustic imaging,'' {\em Biomedical Optics
  Express} {\bf 11}(6), 3360--3373  (2020).

\bibitem{farnia2020high}
P.~Farnia, M.~Mohammadi, E.~Najafzadeh, {\em et~al.}, ``High-quality
  photoacoustic image reconstruction based on deep convolutional neural
  network: towards intra-operative photoacoustic imaging,'' {\em Biomedical
  Physics \& Engineering Express}   (2020).

\bibitem{zhang2020new}
H.~Zhang, L.~Hongyu, N.~Nyayapathi, {\em et~al.}, ``A new deep learning network
  for mitigating limited-view and under-sampling artifacts in ring-shaped
  photoacoustic tomography,'' {\em Computerized Medical Imaging and Graphics} ,
  101720  (2020).

\bibitem{schwab2018real}
J.~Schwab, S.~Antholzer, R.~Nuster, {\em et~al.}, ``Real-time photoacoustic
  projection imaging using deep learning,'' {\em arXiv preprint
  arXiv:1801.06693}   (2018).

\bibitem{schwab2019learned}
J.~Schwab, S.~Antholzer, and M.~Haltmeier, ``Learned backprojection for sparse
  and limited view photoacoustic tomography,'' in {\em Photons Plus Ultrasound:
  Imaging and Sensing 2019},   {\bf 10878}, 1087837, International Society for
  Optics and Photonics  (2019).

\bibitem{anas2018enabling}
E.~M.~A. Anas, H.~K. Zhang, J.~Kang, {\em et~al.}, ``Enabling fast and high
  quality led photoacoustic imaging: a recurrent neural networks based
  approach,'' {\em Biomedical Optics Express} {\bf 9}(8), 3852--3866  (2018).

\bibitem{hochreiter1997long}
S.~Hochreiter and J.~Schmidhuber, ``Long short-term memory,'' {\em Neural
  computation} {\bf 9}(8), 1735--1780  (1997).

\bibitem{xingjian2015convolutional}
S.~Xingjian, Z.~Chen, H.~Wang, {\em et~al.}, ``Convolutional lstm network: A
  machine learning approach for precipitation nowcasting,'' in {\em Advances in
  neural information processing systems},  802--810  (2015).

\bibitem{singh2020deep}
M.~K.~A. Singh, K.~Sivasubramanian, N.~Sato, {\em et~al.}, ``Deep
  learning-enhanced led-based photoacoustic imaging,'' in {\em Photons Plus
  Ultrasound: Imaging and Sensing 2020},   {\bf 11240}, 1124038, International
  Society for Optics and Photonics  (2020).

\bibitem{kim2020deep}
M.~W. Kim, G.-S. Jeng, I.~Pelivanov, {\em et~al.}, ``Deep-learning image
  reconstruction for real-time photoacoustic system,'' {\em IEEE Transactions
  on Medical Imaging}   (2020).

\bibitem{godefroy2020solving}
G.~Godefroy, B.~Arnal, and E.~Bossy, ``Solving the visibility problem in
  photoacoustic imaging with a deep learning approach providing prediction
  uncertainties,'' {\em arXiv preprint arXiv:2006.13096}   (2020).

\bibitem{vu2020generative}
T.~Vu, M.~Li, H.~Humayun, {\em et~al.}, ``A generative adversarial network for
  artifact removal in photoacoustic computed tomography with a linear-array
  transducer,'' {\em Experimental Biology and Medicine} , 1535370220914285
  (2020).

\bibitem{allman2018photoacoustic}
D.~Allman, A.~Reiter, and M.~A.~L. Bell, ``Photoacoustic source detection and
  reflection artifact removal enabled by deep learning,'' {\em IEEE
  transactions on medical imaging} {\bf 37}(6), 1464--1477  (2018).

\bibitem{allman2018exploring}
D.~Allman, A.~Reiter, and M.~Bell, ``Exploring the effects of transducer models
  when training convolutional neural networks to eliminate reflection artifacts
  in experimental photoacoustic images,'' in {\em Photons Plus Ultrasound:
  Imaging and Sensing 2018},   {\bf 10494}, 104945H, International Society for
  Optics and Photonics  (2018).

\bibitem{allman2017machine}
D.~Allman, A.~Reiter, and M.~A.~L. Bell, ``A machine learning method to
  identify and remove reflection artifacts in photoacoustic channel data,'' in
  {\em 2017 IEEE International Ultrasonics Symposium (IUS)},  1--4, IEEE
  (2017).

\bibitem{ren2015faster}
S.~Ren, K.~He, R.~Girshick, {\em et~al.}, ``Faster r-cnn: Towards real-time
  object detection with region proposal networks,'' in {\em Advances in neural
  information processing systems},  91--99  (2015).

\bibitem{allman2018deep}
D.~Allman, F.~Assis, J.~Chrispin, {\em et~al.}, ``Deep neural networks to
  remove photoacoustic reflection artifacts in ex vivo and in vivo tissue,'' in
  {\em 2018 IEEE International Ultrasonics Symposium (IUS)},  1--4, IEEE
  (2018).

\bibitem{awasthi2020sinogram}
N.~Awasthi, R.~Pardasani, S.~K. Kalva, {\em et~al.}, ``Sinogram
  super-resolution and denoising convolutional neural network (srcn) for
  limited data photoacoustic tomography,'' {\em arXiv preprint
  arXiv:2001.06434}   (2020).

\bibitem{awasthi2020deep}
N.~Awasthi, G.~Jain, S.~K. Kalva, {\em et~al.}, ``Deep neural network based
  sinogram super-resolution and bandwidth enhancement for limited-data
  photoacoustic tomography,'' {\em IEEE Transactions on Ultrasonics,
  Ferroelectrics, and Frequency Control}   (2020).

\bibitem{shang2020two}
R.~Shang, K.~Hoffer-Hawlik, and G.~P. Luke, ``A two-step-training deep learning
  framework for real-time computational imaging without physics priors,'' {\em
  arXiv preprint arXiv:2001.03493}   (2020).

\bibitem{grohl2018confidence}
J.~Gr{\"o}hl, T.~Kirchner, T.~Adler, {\em et~al.}, ``Confidence estimation for
  machine learning-based quantitative photoacoustics,'' {\em Journal of
  Imaging} {\bf 4}(12), 147  (2018).

\bibitem{lan2019deep}
H.~Lan, C.~Yang, D.~Jiang, {\em et~al.}, ``Deep learning approach to
  reconstruct the photoacoustic image using multi-frequency data,'' in {\em
  2019 IEEE International Ultrasonics Symposium (IUS)},  487--489, IEEE
  (2019).

\bibitem{lan2019reconstruct}
H.~Lan, C.~Yang, D.~Jiang, {\em et~al.}, ``Reconstruct the photoacoustic image
  based on deep learning with multi-frequency ring-shape transducer array,'' in
  {\em 2019 41st Annual International Conference of the IEEE Engineering in
  Medicine and Biology Society (EMBC)},  7115--7118, IEEE  (2019).

\bibitem{guan2020limited}
S.~Guan, A.~A. Khan, S.~Sikdar, {\em et~al.}, ``Limited-view and sparse
  photoacoustic tomography for neuroimaging with deep learning,'' {\em
  Scientific Reports} {\bf 10}(1), 1--12  (2020).

\bibitem{tong2020domain}
T.~Tong, W.~Huang, K.~Wang, {\em et~al.}, ``Domain transform network for
  photoacoustic tomography from limited-view and sparsely sampled data,'' {\em
  Photoacoustics} , 100190  (2020).

\bibitem{lan2020real}
H.~Lan, D.~Jiang, C.~Yang, {\em et~al.}, ``Real-time photoacoustic tomography
  system via single data acquisition channel,'' {\em arXiv preprint
  arXiv:2001.07454}   (2020).

\bibitem{reiter2017machine}
A.~Reiter and M.~A.~L. Bell, ``A machine learning approach to identifying point
  source locations in photoacoustic data,'' in {\em Photons Plus Ultrasound:
  Imaging and Sensing 2017},   {\bf 10064}, 100643J, International Society for
  Optics and Photonics  (2017).

\bibitem{johnstonbaugh2019novel}
K.~Johnstonbaugh, S.~Agrawal, D.~Abhishek, {\em et~al.}, ``Novel deep learning
  architecture for optical fluence dependent photoacoustic target
  localization,'' in {\em Photons Plus Ultrasound: Imaging and Sensing 2019},
  {\bf 10878}, 108781L, International Society for Optics and Photonics  (2019).

\bibitem{johnstonbaugh2020deep}
K.~Johnstonbaugh, S.~Agrawal, D.~A. Durairaj, {\em et~al.}, ``A deep learning
  approach to photoacoustic wavefront localization in deep-tissue medium,''
  {\em IEEE Transactions on Ultrasonics, Ferroelectrics, and Frequency Control}
    (2020).

\bibitem{shan2019simultaneous}
H.~Shan, G.~Wang, and Y.~Yang, ``Simultaneous reconstruction of the initial
  pressure and sound speed in photoacoustic tomography using a deep-learning
  approach,'' in {\em Novel Optical Systems, Methods, and Applications XXII},
  {\bf 11105}, 1110504, International Society for Optics and Photonics  (2019).

\bibitem{matthews2018parameterized}
T.~P. Matthews, J.~Poudel, L.~Li, {\em et~al.}, ``Parameterized joint
  reconstruction of the initial pressure and sound speed distributions for
  photoacoustic computed tomography,'' {\em SIAM journal on imaging sciences}
  {\bf 11}(2), 1560--1588  (2018).

\bibitem{boink2018sensitivity}
Y.~E. Boink, S.~A. Van~Gils, S.~Manohar, {\em et~al.}, ``Sensitivity of a
  partially learned model-based reconstruction algorithm,'' {\em PAMM} {\bf
  18}(1), e201800222  (2018).

\bibitem{boink2019partially}
Y.~E. Boink, S.~Manohar, and C.~Brune, ``A partially-learned algorithm for
  joint photo-acoustic reconstruction and segmentation,'' {\em IEEE
  transactions on medical imaging} {\bf 39}(1), 129--139  (2019).

\bibitem{boink2019robustness}
Y.~E. Boink, C.~Brune, and S.~Manohar, ``Robustness of a partially learned
  photoacoustic reconstruction algorithm,'' in {\em Photons Plus Ultrasound:
  Imaging and Sensing 2019},   {\bf 10878}, 108781D, International Society for
  Optics and Photonics  (2019).

\bibitem{Chambolle2011}
A.~Chambolle, S.~E. Levine, and B.~J. Lucier, ``An upwind finite-difference
  method for total variation-based image smoothing,'' {\em SIAM J. Imaging
  Sci.} {\bf 4}(1), 277--299  (2011).

\bibitem{hauptmann2018approximate}
A.~Hauptmann, B.~Cox, F.~Lucka, {\em et~al.}, ``Approximate k-space models and
  deep learning for fast photoacoustic reconstruction,'' in {\em International
  Workshop on Machine Learning for Medical Image Reconstruction},  103--111,
  Springer  (2018).

\bibitem{Cox:2005fwdkspace}
B.~T. Cox and P.~C. Beard, ``Fast calculation of pulsed photoacoustic fields in
  fluids using k-space methods,'' {\em The Journal of the Acoustical Society of
  America} {\bf 117}(6), 3616--3627  (2005).

\bibitem{yang2019accelerated}
C.~Yang, H.~Lan, and F.~Gao, ``Accelerated photoacoustic tomography
  reconstruction via recurrent inference machines,'' in {\em 2019 41st Annual
  International Conference of the IEEE Engineering in Medicine and Biology
  Society (EMBC)},  6371--6374, IEEE  (2019).

\bibitem{lonning2018recurrent}
K.~L{\o}nning, P.~Putzky, M.~W. Caan, {\em et~al.}, ``Recurrent inference
  machines for accelerated mri reconstruction,'' in {\em International
  Conference on Medical Imaging with Deep Learning (MIDL 2018)},   (2018).

\bibitem{lan2019hybrid}
H.~Lan, K.~Zhou, C.~Yang, {\em et~al.}, ``Hybrid neural network for
  photoacoustic imaging reconstruction,'' in {\em 2019 41st Annual
  International Conference of the IEEE Engineering in Medicine and Biology
  Society (EMBC)},  6367--6370, IEEE  (2019).

\bibitem{lan2019ki}
H.~Lan, K.~Zhou, C.~Yang, {\em et~al.}, ``Ki-gan: Knowledge infusion generative
  adversarial network for photoacoustic image reconstruction in vivo,'' in {\em
  International Conference on Medical Image Computing and Computer-Assisted
  Intervention},  273--281, Springer  (2019).

\bibitem{lan2019net}
H.~Lan, D.~Jiang, C.~Yang, {\em et~al.}, ``Y-net: A hybrid deep learning
  reconstruction framework for photoacoustic imaging in vivo,'' {\em arXiv
  preprint arXiv:1908.00975}   (2019).

\bibitem{li2020nett}
H.~Li, J.~Schwab, S.~Antholzer, {\em et~al.}, ``Nett: Solving inverse problems
  with deep neural networks,'' {\em Inverse Problems}   (2020).

\bibitem{antholzer2019nett}
S.~Antholzer, J.~Schwab, J.~Bauer-Marschallinger, {\em et~al.}, ``Nett
  regularization for compressed sensing photoacoustic tomography,'' in {\em
  Photons Plus Ultrasound: Imaging and Sensing 2019},   {\bf 10878}, 108783B,
  International Society for Optics and Photonics  (2019).

\bibitem{schwab2019deepSVD}
J.~Schwab, S.~Antholzer, R.~Nuster, {\em et~al.}, ``Deep learning of truncated
  singular values for limited view photoacoustic tomography,'' in {\em Photons
  Plus Ultrasound: Imaging and Sensing 2019},   {\bf 10878}, 1087836,
  International Society for Optics and Photonics  (2019).

\bibitem{kirchner2018context}
T.~Kirchner, J.~Gr{\"o}hl, and L.~Maier-Hein, ``Context encoding enables
  machine learning-based quantitative photoacoustics,'' {\em Journal of
  biomedical optics} {\bf 23}(5), 056008  (2018).

\bibitem{cai2018end}
C.~Cai, K.~Deng, C.~Ma, {\em et~al.}, ``End-to-end deep neural network for
  optical inversion in quantitative photoacoustic imaging,'' {\em Optics
  letters} {\bf 43}(12), 2752--2755  (2018).

\bibitem{yang2019quantitative}
C.~Yang, H.~Lan, H.~Zhong, {\em et~al.}, ``Quantitative photoacoustic blood
  oxygenation imaging using deep residual and recurrent neural network,'' in
  {\em 2019 IEEE 16th International Symposium on Biomedical Imaging (ISBI
  2019)},  741--744, IEEE  (2019).

\bibitem{chen2020deep}
T.~Chen, T.~Lu, S.~Song, {\em et~al.}, ``A deep learning method based on u-net
  for quantitative photoacoustic imaging,'' in {\em Photons Plus Ultrasound:
  Imaging and Sensing 2020},   {\bf 11240}, 112403V  (2020).

\bibitem{luke2019net}
G.~P. Luke, K.~Hoffer-Hawlik, A.~C. Van~Namen, {\em et~al.}, ``O-net: A
  convolutional neural network for quantitative photoacoustic image
  segmentation and oximetry,'' {\em arXiv preprint arXiv:1911.01935}   (2019).

\bibitem{bench2020towards}
C.~Bench, A.~Hauptmann, and B.~Cox, ``Towards accurate quantitative
  photoacoustic imaging: learning vascular blood oxygen saturation in 3d,''
  {\em arXiv preprint arXiv:2005.01089}   (2020).

\bibitem{yang2019eda}
C.~Yang and F.~Gao, ``Eda-net: Dense aggregation of deep and shallow
  information achieves quantitative photoacoustic blood oxygenation imaging
  deep in human breast,'' in {\em International Conference on Medical Image
  Computing and Computer-Assisted Intervention},  246--254, Springer  (2019).

\bibitem{yu2018deep}
F.~Yu, D.~Wang, E.~Shelhamer, {\em et~al.}, ``Deep layer aggregation,'' in {\em
  Proceedings of the IEEE conference on computer vision and pattern
  recognition},  2403--2412  (2018).

\bibitem{zhou2018unetpp}
Z.~Zhou, M.~M.~R. Siddiquee, N.~Tajbakhsh, {\em et~al.}, ``Unet++: A nested
  u-net architecture for medical image segmentation,'' in {\em Deep Learning in
  Medical Image Analysis and Multimodal Learning for Clinical Decision
  Support},  D.~Stoyanov, Z.~Taylor, G.~Carneiro, {\em et~al.}, Eds., 3--11,
  Springer  (2018).

\bibitem{Durairaj2020}
D.~Durairaj, S.~Agrawal, K.~Johnstonbaugh, {\em et~al.}, ``Unsupervised deep
  learning approach for photoacoustic spectral unmixing,'' in {\em Photons Plus
  Ultrasound: Imaging and Sensing 2020},   {\bf 11240}, 112403H  (2020).

\bibitem{grohl2019estimation}
J.~Gr{\"o}hl, T.~Kirchner, T.~Adler, {\em et~al.}, ``Estimation of blood
  oxygenation with learned spectral decoloring for quantitative photoacoustic
  imaging (lsd-qpai),'' {\em arXiv preprint arXiv:1902.05839}   (2019).

\bibitem{Cox2011}
B.~T. Cox, T.~Tarvainen, and S.~R. Arridge, ``{Multiple Illumination
  Quantitative Photoacoustic Tomography using Transport and Diffusion
  Models},'' {\em Contemporary Mathematics} {\bf 559}, 1--12  (2011).

\bibitem{Vogt2019SO2}
W.~C. Vogt, X.~Zhou, R.~Andriani, {\em et~al.}, ``Photoacoustic oximetry
  imaging performance evaluation using dynamic blood flow phantoms with tunable
  oxygen saturation,'' {\em Biomedical optics express} {\bf 10}(2), 449--464
  (2019).

\bibitem{fang2009monte}
Q.~Fang and D.~A. Boas, ``Monte carlo simulation of photon migration in 3d
  turbid media accelerated by graphics processing units,'' {\em Optics express}
  {\bf 17}(22), 20178--20190  (2009).

\bibitem{adler2019deepPosterior}
J.~Adler and O.~{\"O}ktem, ``Deep posterior sampling: Uncertainty
  quantification for large scale inverse problems,'' in {\em International
  Conference on Medical Imaging with Deep Learning},   (2019).

\bibitem{schlemper2018bayesian}
J.~Schlemper, D.~C. Castro, W.~Bai, {\em et~al.}, ``Bayesian deep learning for
  accelerated mr image reconstruction,'' in {\em International Workshop on
  Machine Learning for Medical Image Reconstruction},  64--71, Springer
  (2018).

\bibitem{denker2020conditional}
A.~Denker, M.~Schmidt, J.~Leuschner, {\em et~al.}, ``Conditional normalizing
  flows for low-dose computed tomography image reconstruction,'' {\em arXiv
  preprint arXiv:2006.06270}   (2020).

\bibitem{raissi2019physics}
M.~Raissi, P.~Perdikaris, and G.~E. Karniadakis, ``Physics-informed neural
  networks: A deep learning framework for solving forward and inverse problems
  involving nonlinear partial differential equations,'' {\em Journal of
  Computational Physics} {\bf 378}, 686--707  (2019).

\bibitem{etmann2020iunets}
C.~Etmann, R.~Ke, and C.-B. Sch{\"o}nlieb, ``iunets: Fully invertible u-nets
  with learnable up-and downsampling,'' {\em arXiv preprint arXiv:2005.05220}
  (2020).

\bibitem{putzky2019invert}
P.~Putzky and M.~Welling, ``Invert to learn to invert,'' in {\em Advances in
  Neural Information Processing Systems},  444--454  (2019).

\bibitem{hauptmann2020multi}
A.~Hauptmann, J.~Adler, S.~R. Arridge, {\em et~al.}, ``Multi-scale learned
  iterative reconstruction,'' {\em IEEE Transactions on Computational Imaging}
   (2020).

\bibitem{arridge2019networks}
S.~Arridge and A.~Hauptmann, ``Networks for nonlinear diffusion problems in
  imaging,'' {\em Journal of Mathematical Imaging and Vision} , 1--17  (2019).

\bibitem{lunz2020learned}
S.~Lunz, A.~Hauptmann, T.~Tarvainen, {\em et~al.}, ``On learned operator
  correction,'' {\em arXiv preprint arXiv:2005.07069}   (2020).

\bibitem{smyl2020learning}
D.~Smyl, T.~N. Tallman, J.~A. Black, {\em et~al.}, ``Learning and correcting
  non-gaussian model errors,'' {\em arXiv preprint arXiv:2005.14592}   (2020).

\bibitem{siahkoohi2019neural}
A.~Siahkoohi, M.~Louboutin, and F.~J. Herrmann, ``Neural network augmented
  wave-equation simulation,'' {\em arXiv preprint arXiv:1910.00925}   (2019).

\end{thebibliography}
\bibliographystyle{spiejour}

%%%%% Biographies of authors %%%%%

% \vspace{2ex}\noindent\textbf{First Author} is an assistant professor at the University of Optical Engineering. He received his BS and MS degrees in physics from the University of Optics in 1985 and 1987, respectively, and his PhD degree in optics from the Institute of Technology in 1991.  He is the author of more than 50 journal papers and has written three book chapters. His current research interests include optical interconnects, holography, and optoelectronic systems. He is a member of SPIE.

%\listoffigures
%\listoftables

%\end{spacing}
\end{document}